%% file: ncln1.tex
\documentstyle[12pt,epsfig]{article}
\input{defi}
\begin{document}
\begin{titlepage}

\begin{flushright}
 June  12  1997\\
TAUP 2432/97\\
%{\bf hep-ph/9705375}\\
\end{flushright}

%\noindent
\begin{center}
{\Large\bf{PARTON DENSITIES IN A NUCLEON}}\\[6ex]
{\large \bf { A. L.
Ayala  F$^{\underline{o}}$ ${}^{a)\,b)}$${}^*$\footnotetext{ ${}^*$ E-mail:
ayala@if.ufrgs.br},
 M. B. Gay  Ducati ${}^{a)}$$^{**}$\footnotetext{${}^{**}$
E-mail:gay@if.ufrgs.br}
  and 
 E. M. Levin ${}^{c)\,d)\,\dagger}$
\footnotetext{$^{\dagger}$ E-mail:  leving@ccsg.tau.ac.il} 
}} \\[1.5ex]

{\it ${}^{a)}$Instituto de F\'{\i}sica, Univ. Federal do Rio Grande do Sul}\\
{\it Caixa Postal 15051, 91501-970 Porto Alegre, RS, BRAZIL}\\[1.5ex]
{\it ${}^{b)}$Instituto de F\'{\i}sica e Matem\'atica, Univ. 
Federal de Pelotas}\\
{\it Campus Universit\'ario, Caixa Postal 354, 96010-900, Pelotas, RS,
BRAZIL}\\[1.5ex]
{\it ${}^{c)}$ HEP Department, School of Physics and Astronomy}\\
{\it Raymond and Beverly  Sackler Faculty of Exact Science}\\
{\it Tel Aviv University, Tel Aviv 69978, ISRAEL}
\\[1.5ex]
{\it$ {}^{d)}$ Theory Department, Petersburg Nuclear Physics Institute}\\
{\it 188350, Gatchina, St. Petersburg, RUSSIA}\\[3.5ex]
\end{center}
{\large \bf Abstract:}
In this paper we re-analyse the situation with the shadowing 
corrections (SC) in QCD for the proton deep inelastic structure functions.
We reconsider the  Glauber - Mueller approach for the SC in deep 
inelastic scattering (DIS) and suggest a new nonlinear evolution 
equation. We argue that this equation solves the problem of the SC in 
the wide kinematic region where $\as\, \kappa \,=\,\as\,\frac{3 \pi \as}{2
Q^2 
R^2}\,x G(x,Q^2) \,\,\leq\,1$. Using the new equation we estimate the 
value of the SC which turn out to be essential in the gluon deep 
inelastic structure function but  rather small in $F_2(x,Q^2)$.
We claim that the SC in $xG(x,Q^2)$ is so large that the BFKL Pomeron is 
hidden under the SC and cannot be seen even in such  ``hard" processes
that have been proposed to test it. We found that the gluon density
is proportional to $\ln(1/x)$ in the region of very small $x$. This 
result means that the gluon density does not reach saturation in the 
region of applicability of the new evolution equation. It 
should be confronted with the solution of the GLR equation which leads 
to saturation.   

\end{titlepage}

\section{Introduction.}
The main goal of this paper is to reconsider the whole issue of the shadowing 
corrections ( SC ) to the quark and gluon  densities in a nucleon.
The motivation to investigate the SC for parton distributions
comes from some inconsistencies in the interpretation of the present
HERA data, which we will discuss below.

 The experiment
 \cite{HERA} shows that the deep inelastic structure function $F_2(x,Q^2)$
has a steep behaviour in the small $x$ region ($10^{-2}\,>\,x\,>\,10^{-5}$), 
even for very small virtualities. Indeed, considering $F_2 \, \propto \, 
x^{- \lambda}$ for small $x$, the experimental data go from  $\l \, = \,0.15$
for $Q^2 \, = \, 0.85 \, GeV^2$  to  $\l \, = \,0.4$
for $Q^2 \, = \, 20 \, GeV^2$. This steep behaviour is well described
in perturbative QCD by the DGLAP evolution equations 
 \cite{MRS}-\cite{GRV}.  The phenomenological input, with
the quark and gluon distributions at
initial virtuality $Q^2 = Q^2_0$, can be chosen at sufficiently low
values of $Q^2$ using the backwards evolution of the experimental data in
the region of $Q^2 \,\approx\, 4 - 5 \, GeV^2$. The DGLAP evolution describes
the data down to $Q^2 \, = \, 1.5 \,GeV^2$.

From the above discussion  one can conclude that 
the parton cascade is a rather deluted system of partons with small parton -
parton interaction, which can be neglected in a first approximation.  
Therefore, no SC is needed to describe the experimental data.

Nevertheless, there is  a set of several facts that
does not fit into  this  scheme.

%{\bf 1.} The best parameterization of HERA data is not the solution 
%of the DGLAP
%equations but a simple formula \cite{BH}:
%$$
%F_2(x,Q^2)\,\,=\,\,a\,\,+\,\,m\,\log \frac{Q^2}{Q^2_0}\,\log \frac{x_0}{x}
%$$
%with $a $\,=\,0.078 ; $m$\,=\,0.364 ; $x_0$\,=\,0.074 ; $Q^2_0$\, =\,0.5
%$GeV^2$. It is clear that this simple formula cannot be a solution of the
%DGLAP evolution equations.  To make obvious this remark it is enough to
%recalculate the gluon structure function from the above expression as 
%it has been done in Ref.\cite{BH}. Indeed, $x G(x,Q^2)$ turns out to be 
%equal to
%$$ xG(x,Q^2) \,\, =\,\,3\,\,\log \frac{x_0}{x}$$
%without any $Q^2$ - dependence in direct contradiction with
% the DGLAP evolution.

1. We expect that the quark distribution will not grow indefinitely as $x$
goes to zero, since it would violate unitarity for some value of $x$ 
\cite{GLR}. In Ref.\cite{FRST} the unitarity bound (UB) for $F_2(x,Q^2)$
was established and it turns out
 that the
 $F_2$ structure function reaches the UB at $Q^2\, = \, 1 \, - \, 2 
\,GeV^2$ in HERA kinematic region. It means that  large shadowing  
corrections to the normal DGLAP evolution 
should take place in this kinematic region.

\begin{figure}[hbtp]
\centerline{\psfig{file=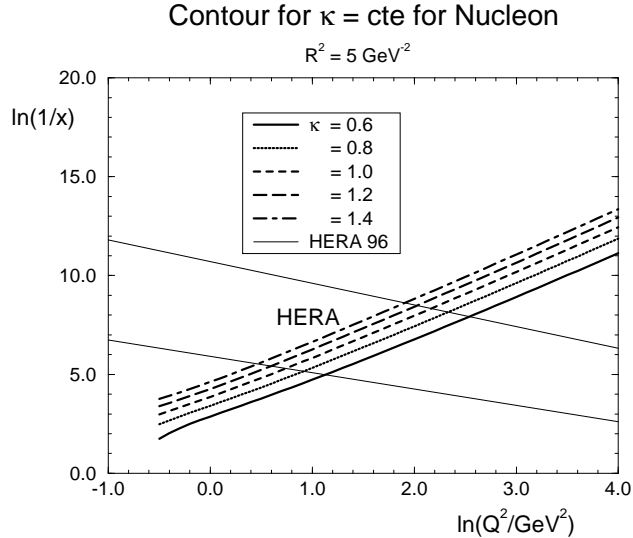,width=90mm}} 
\caption{{\em Contour plot for $\kappa= \, cte$ for the GRV
95\protect\cite{GRV}
distribution and $R^2= \, 5 \, GeV^{-2}$. The  solid lines show the HERA
kinematic region
 \protect\cite{HERA}.}}
\label{fig.1}
\end{figure}

2. From  HERA data we can evaluate also
the probability $\kappa$ of the parton - parton (gluon - gluon) interaction, 
which is given by \cite{GLR}, \cite{MUQI}
\beq 
\label{1}
\kappa\,\,=\,\,x G(x,Q^2) \frac{\s(GG)}{ Q^2 \, \pi R^2}\,\,=\,\,
\frac{ 3 \,\pi\,\as}{ Q^2\,R^2}\,\,xG(x,Q^2)\,\,,
\eeq
where $xG(x,Q^2)$ is the number of partons ( gluons) in the parton cascade 
and $R^2$ is the radius 
 of the area populated by gluons in a nucleon. $\s (GG)$ is the gluon
cross section inside the parton cascade  and 
was evaluated in \cite{MUQI}. 
  Using HERA 
data on photoproduction of J/$\Psi$ meson
\cite{HERA}  the 
value of $R^2$ was  estimated as $R^2 \,\,\leq\,\,5 \, GeV^{-2}$ 
\cite{FRST}. Using the GRV parameterization \cite{GRV} for the gluon 
structure function and the value of $R^2 = 5 \, GeV^{-2}$,
we obtain  that $\kappa$ reaches 1 at HERA 
kinematic region ( see Fig.\ref{fig.1} ), meaning shadowing corrections  
should not  be neglected (see Ref. \cite{FRST} for details).

\vspace{0.5 true cm}

\begin{figure}[htbp]
\centerline{\psfig{file=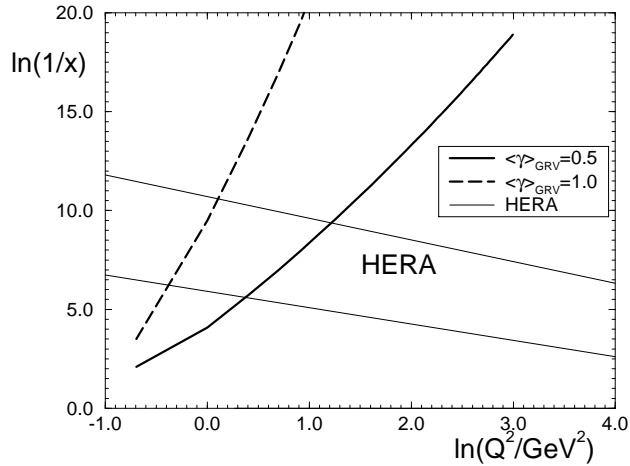,width=90mm}}
\caption{{\em Contours for $ <\gamma > $ = 1 and $<\gamma >$ = 1/2 for
the GRV95 gluon density  and HERA kinematic region.}}
\label{Fig.3}
\end{figure}

3. The situation looks even more controversial if we plot the average 
value of the anomalous dimension, 
 $ < \gamma  >\,\,=\,\,\partial \ln( x G(x,Q^2))/\partial \ln Q^2$, in the
GRV parameterization.
 Fig.\ref{Fig.3} shows two 
remarkable lines: the line  $< \gamma > =1$, where the deep inelastic 
cross section
reaches the value compatible with the geometrical size of the proton; and
the line
$ < \gamma > \,=\,1/2$, which is the characteristic line in which vicinity both
the BFKL Pomeron \cite{BFKL}   and the GLR equation \cite{GLR} 
should take over the DGLAP evolution equations. HERA data passed
 over the second line and even for sufficiently small values of $Q^2$ they 
crossed the first one without any indication of a strange behaviour
near these lines.

From the above discussion we conclude that the physical interpretation
of HERA data looks 
controversial and the statement that the DGLAP evolution works is first but 
not the last  outcome of HERA data. We expect also that the SC will
 be essential in HERA kinematic region. However, to treat the SC we 
need a more general approach than the GLR one \cite{GLR}, \cite{MUQI}, 
which gave a theoretical approach to the SC only to the right of the 
critical line $ < \gamma > \,=\,\frac{1}{2}$, at large values of $Q^2$.
  In 
this paper  we develop a general approach to the SC, that will allow us
to give reliable estimates for the SC in the kinematic region to the 
left of the
line $< \gamma >\,=\,1/2$, i.e., for large $ln( 1/x)$
and small $Q^2$.

This paper is organized as follows.
 In section 2  we present the Glauber or Eikonal 
approach for high energy scattering off nucleon. We discuss the
 SC for the gluon distribution in a nucleon and for the $F_2$
deep inelastic structure function, concentrating mostly on the theory
status of this approach and on the predicted scale for the SC. We analyse
also
the effect of the SC on the  $b_t$-dependence of the gluon  distributions
and $F_2$, and the consequent approach to the ``soft'' physics.
%theory status and the predicted scale for the SC.
After a brief discussion, in section 3, of 
the first correction to the Glauber (Eikonal) approach,
we present, in section 4, a new evolution equation that sums
all SC diagrams  which are of order $ (\as \,\ln (1/x)\,\ln Q^2 )^n$
and $\kappa^n$. 
This equation is solved  for $\as$ constant in the semiclassical approach.
The solution is compared  with 
that of the GLR equation  \cite{GLR} and with the Glauber
approach. Finally, 
section 5  summarizes our results.

\section{ The Eikonal  approach in QCD .}
\label{eaqcd}
\subsection{ The Mueller formula.}
%\newline
In this section we  present the general idea for
the Glauber formula in QCD  both for the gluon
distribution and the $F_2$ structure function of DIS.
These Glauber formulas will take into account the SC from the 
eikonalization of the parton-nucleon cross section.
The ideas were  originally formulated in 
Refs.\cite{LR87}-\cite{AGL}.

Let us consider a high energy virtual particle that probes the 
nucleon gluon distribution ( for example, such a probe could be a virtual 
graviton or a heavy Higgs boson). In the space-time picture of 
the process, the virtual probe decays in a gluon-gluon ($GG$) pair, with 
 transverse separation $r_{\perp}$ and the fraction of energy of the probe 
$z$ and $ 1 - z$, which interacts with the nucleon through a gluon ladder 
exchange. In high energy (small $x$) limit,
 $r_{\perp}$ is constant during the interaction  for $x\,\,\ll\,\,{1}/(2\,m\,R)$,
where $R$ is the target nucleon size. Therefore, 
the transverse distance $r_{\perp}$ is a good degree of freedom for our 
problem.

The  cross section of the absorption
of gluon($G^*$) with virtuality $Q^2$ and Bjorken $x$ can be written 
in the form
\beq \label{5}
\sigma (\,G^*\,)\,\,=\,\,
\int^1_0 d z \,\,\int \,\frac{d^2 r_{\perp}}{ \pi}\,\,
\int\frac{d^2 b_t}{ \pi}
| \, \Psi^{G^*}_{\perp} (Q^2, r_{\perp},x,z)\,  |^2\,\,\sigma_{tot} (GG \, + \, 
nucleon) 
 \,\, ,
\eeq
where $z$ is the fraction of energy 
carried by the gluon, $b_t$ is the impact parameter
and  $\Psi^{G^*}_{\perp}$ is the wave 
function of the transverse polarized gluon in the virtual probe.
The
cross section of the interaction of the $GG$ pair with the nucleon
is $\sigma_{tot} (GG \, + \, nucleon) = 
\sigma (x,r^2_{\perp})$. It is
important to note that this description is valid in 
leading log ( $1 / x$ ) approximation (LL($1/x$)A), in which only 
contributions of the order of $ (\, \as \ln(\frac{1}{x}\,))^n$ were taken 
into account. In this approximation we can neglect the change of $z$ 
during the interaction and describe the cross section $\sigma_{tot} (
GG\,+\,nucleon)$ as a function of the variable $x$ \cite{GLR}, 
\cite{RY}, \cite{FIVE}.

The cross section $\sigma (x,r^2_{\perp})$ can be written in the form
\beq \label{CS}
\sigma (x,r^2_{\perp})\,\,=\,\,2\,\,\int\,\,d^2 b_t 
\,\,Im\,a(x,r_{\perp},b_t)\,\,,
\eeq
where $a$ is the elastic amplitude for which we have the $s$ - channel 
unitarity constraint
\beq \label{UN}
2\,\,Im\,a(x,r_{\perp},b_t)\,\,=\,\, |a(x,r_{\perp},b_t )
|^2\,\,+\,\,G_{in}(x,r_{\perp},b_t)\,\,,
\eeq
where $G_{in}$ is the contribution of all inelastic processes. Eq.(\ref{UN})
gives the relation between two unknowns $a $ and $G_{in}$ and has a 
general solution in the limit of small  real part of the elastic
amplitude at $x\,\rightarrow 0$

\beq \label{EIK}
a (x,r_{\perp},b_t)\,\,=
\,\,i\,\,\{\,\,1\,\,-\,\,e^{-\,\frac{1}{2}\,\Omega(x, r_{\perp},b_t)}\,\,
\}\,\, ,
\eeq
$$
G_{in} (x,r_{\perp},b_t)\,\,=\,\,1\,\,-\,\,e^{-\,\Omega(x, 
r_{\perp},b_t)}\,\,  
\,\, .
$$

The opacity function  $\Omega$ has a simple physical meaning, namely
$e^{- \Omega}$ is the probability that a $GG$ pair has no inelastic
 interaction during the passage through the target. 
$\Omega$ is an arbitrary real function, which can be specified
only in a more detailed theory or approach than the unitarity constraint.
One of such specific model is the Glauber approach or the Eikonal model. 

However,
before we  discuss this model let us make one important remark on the
strategy for the approach to the SC: being able to calculate the
opacity $\Omega$ we will have the theory
or model for all inelastic processes. Indeed, using AGK cutting rules \cite{AGK}
we can calculate any inelastic process, if we know $\Omega$, in accordance
 with the $s$-channel unitarity. It is worthwhile mentioning that the inverse
procedure does not work. If we know the SC in all details for a particular
inelastic process, say for the inclusive production, we cannot reconstruct
all other process and the total cross section in particular.

Now, let us built the Glauber approach. First, let us assume that $\Omega$ is
 small ($\Omega\,\,\ll\,\,1$)
 and its $b_t$ dependence can be factorized as $\Omega\,=\,\widetilde{
\Omega}(x,r_{\perp}) \,S(b_t)$ with the normalization $\int \,d^2 b_t \,S(b_t)
\,=\,1$. The factorization was proven for the DGLAP evolution equations 
\cite{GLR} and, therefore, all our further calculations will be valid for
the DGLAP evolution equations \cite{DGLAP} in the region of small $x$ or,
in other 
words, in the Double Log Approximation (DLA)  of perturbative QCD (pQCD). 
 Expanding \eq{EIK} and substituting it in \eq{5}, one can obtain
\beq \label{9}
\s_{tot} (x,r_{\perp})\,\,=\,\,\widetilde{\Omega}(x,r_{\perp})\, .
\eeq
The substitution of \eq{9} in \eq{5} gives
\beq \label{8a}
\sigma (\,G^*\,)\,\,=\,\,
\int^1_0 d z \,\,\int \,\frac{d^2 r_{\perp}}{ \pi}\,\,
| \, \Psi^{G^*}_{\perp} (Q^2, r_{\perp},x,z) \, |^2
 \widetilde{\Omega} (x,r_{\perp} )\,\, .
\eeq
Using \eq{8a} and $\sigma( G^* )\,\,=\,\,\frac{4 \pi^2 \as}{Q^2}\,xG(x,Q^2)$
as well as the expression for the  wave function of the $ GG$  pair 
in the virtual gluon probe ( see Ref.\cite{MU90} ), one can express 
 $ \widetilde{\Omega} 
(x,r_{\perp} )$  through the gluon structure function \cite{LR87},\cite{MU90} 
\bea    \label{10}
\widetilde{\Omega}\,\,=\,\,\s^{GG}_{N} (x,\frac{r^2_{\perp}}{4} )\,\,=
\,\,\frac{ 3\, \pi^2 \, \as}{4}
\,\,r^2_{\perp}\,xG(x,\frac{4}{r^2_{\perp}})\,\,.
\eea

The Glauber (eikonal ) approach is the assumption that
 $\Omega \,=\,\widetilde{\Omega}\,S(b_t)$ with $\widetilde{\Omega}$ 
given by
\eq{10} in the whole kinematical region. 
From the point of view of the structure of the final state,
 this assumption means that the typical rich inelastic event was modeled
as a sum of the diffractive dissociation of the $GG$  pair and 
 uniform in rapidity
distribution of produced gluons. For example, we neglected in the Glauber
    approach all the rich structure of the large rapidity gaps events including
 the diffractive dissociation in the region of large mass.

Substituting  \eq{10} in  \eq{9} and the result in \eq{5},
and using the wave function $ \Psi^{G^*}_{\perp} (Q^2, r_{\perp},x,z)$
calculated in Ref.\cite{MU90} we obtain the Glauber (Mueller) formula
for the gluon structure function
\beq \label{MF}
x G(x,Q^2) = \frac{4}{\pi^2} \int_{x}^{1} \frac{d x'}{x'} 
\int_{\frac{4}{Q^2}}^{\infty} \frac{d^2 r_{\perp}}{\pi r_{\perp}^{4}} 
\int_{0}^{\infty} \frac{d^2 b_t}{\pi}  2
\left\{ 1 -
e^{- \frac{1}{2} \, \s^{GG}_N ( x^{\prime},\frac{r^2_{\perp}}{4} ) 
S(b_t) } \right\} \, .
\eeq
 
The first term in the expansion of \eq{MF}
with respect to $\s^{GG}_N$ gives the DGLAP equation in the region 
of small $x$.

To simplify our calculations we will use
the Gaussian parameterization for  the profile function $S(b_t)$, namely
\beq \label{13}
S(b^2_t)\,\,=\,\,\frac{1}{\pi\,R^2}\,e^{ -\, \frac{b^2_t}{R^2} }\,\,.
\eeq
Using this profile function and  integrating over $b_t$, we obtain  ($N_c = N_f = 3$)
\bea
x G(x,Q^2) = \frac{2 R^{2}}{\pi^2} \int_{x}^{1} \frac{d x'}{x'} 
\int^{\frac{1}{Q^2_0}}_{\frac{1}{Q^2}}  \frac{d r_{\perp}^2}{ r_{\perp}^{4}} 
\left\{\,\, C \,\,+ \,\,ln(\kappa_{G} ( x', r_{\perp}^{2})) \,\,+\,\,
E_1 (\kappa_{G} ( x', r_{\perp}^{2}))  \right\} \,\, ,
\label{14}
\eea
where $C$ is the Euler constant and $E_1$ is the exponential integral
(see Ref.\cite{r20} Eq. {\bf 5.1.11}) and
\beq
\kappa_{G} ( x', r_{\perp}^{2}) = \frac{3 \as(\frac{1}{r^2_{\perp}})  \pi 
r^2_{\perp}}{2 R^{2}} x' G^{DGLAP} (x', \frac{1}{r^2_{\perp}} ) \,\, .
\label{kapa}
\eeq
The \eq{MF} is the master equation of this section and it gives a way to
estimate the value of the SC.

A similar Glauber (eikonal ) formula may be obtained for the deep
inelastic structure function  $F_2(x,Q^2) $. In this case, the virtual
probe decays into a quark anti-quark pair which interacts with the nucleon
trough a gluon ladder exchange. Taking into account $N_f$ quark flavours and integrating
the quark pair wave function over $z$, we obtain a Glauber formula for
$F_2$
( see Refs. \cite{FRST}  \cite{LR87} \cite{MU90} )
\beq \label{F2}
F_2(x,Q^2)\,\,=\,\,\frac{ N_c}{6\,\pi^3}\,\sum^{N_f}_{1}\,\,Z^2_f\,\, 
 \,\int^{\infty}_{\frac{1}{Q^2}} \,\frac{ d r^2_{\perp}}{r^4_{\perp}}
\, \int\,d^2 b_{t}\,\,\{\,
1\,\,-\,\,e^{-\,\frac{1}{2} \,
\O_{q \bar q}( x, r_{\perp}, b_{t} )}\,\}\,\,,
\eeq 
where $\O_{q \bar q }\,\,=\,\s^{q \bar q}_{N} \, 
S(b_t^2 )\, = \, \frac{4}{9} \O $, and $Z_f$ 
is the quark charge fraction. 
After integration over $b_t$ using the  Gaussian 
profile function $S(b_t)$ one can reduce \eq{F2} to  
( $N_c = 3 $ )
\beq \label{F2INT}
F_2 (x, Q^2 )\,\,=\,\,\frac{R^2}{2 \pi^2} \, \sum^{N_{f}}_{1} Z^2_f \,
\int^{\frac{1}{Q^2_0}}_{\frac{1}{Q^2}}
\,\,\frac{ d\, r^2_{\perp}}{ r^4_{\perp}}\left\{\,\, C \,\,+ \,\,ln(\kappa_{q} ( x', r_{\perp}^{2})) \,\,+\,\,
E_1 (\kappa_{q} ( x', r_{\perp}^{2}))  \right\}\,\,,
\eeq
with $\kappa_q\,\,=\,\,\frac{4}{9}\,\kappa_G$.

\subsection{  Theory status of the Mueller formula.}

In this section we shall recall the main assumptions that have been made 
to obtain
the Mueller formula.

1. The gluon energy ($x$) should be  high (small) enough to satisfy 
$x\,\ll\,\frac{1}{m R}$  and  \,\,\,\,$\as ln(1/x) \leq 1$. The last 
condition means that we have to assume
the leading $ln (1/x)$ approximation of perturbative QCD for the nucleon gluon
structure function.

2. The DGLAP evolution equations hold in the region of small $x$ or, in other
words, $\as ln(1/r^2_{\perp}) \leq 1$. One of the lessons from HERA data is the
fact that the DGLAP evolution can describe the experimental data.

These two assumptions mean that we describe the gluon  emission in
 Double Log Approximation ( DLA) of perturbative QCD. In other words, we
extract from each Feynman diagram of order $\as^n$ the contribution
of order $(\,\as\,\ln1/x\,\ln Q^2/Q^2_0\,)^n$, neglecting all other
contributions of the same diagram. In terms of the DGLAP evolution, we
have to assume that the DGLAP evolution equations describe the gluon
 emission in the region of small $x$. However, the first assumption
is very important for the whole picture, since it
 allows us to treat successive rescatterings as independent
 and simplifies all formulae 
 reducing the problem to an eikonal picture of the
classical propagation of a relativistic particle with high energy ($ E R
\,
\gg 
 \,1$) through 
the target.

3. Only the fastest partons ($GG$ pairs) interact with the target.
This assumption is an artifact of the Glauber approach, which looks strange in
 the parton picture of the interaction. Indeed, in the parton model  
we rather  expect that all partons, not only the fastest one, should interact
 with the target. 
In the next
 section we will show that corrections to the Glauber approach 
due to the interaction of slower partons are essential in QCD.

4.  There are
no correlations (interaction) between partons from the different parton
 cascades.
This assumption means that even the interaction of 
the fastest $GG$-pair was taken into account in the Mueller formula only
approximately and we have to assume that we are dealing with large number 
of colours to trust the Mueller formula. Indeed, it has been proven that
correlations between partons from different parton cascades lead to 
corrections to the Mueller formula of the order of $1/N^2_c$, where $N_c$ is the
number of colours 
(\cite{HT}). 

\begin{figure}[htbp]
\begin{tabular}{c c}
\psfig{file=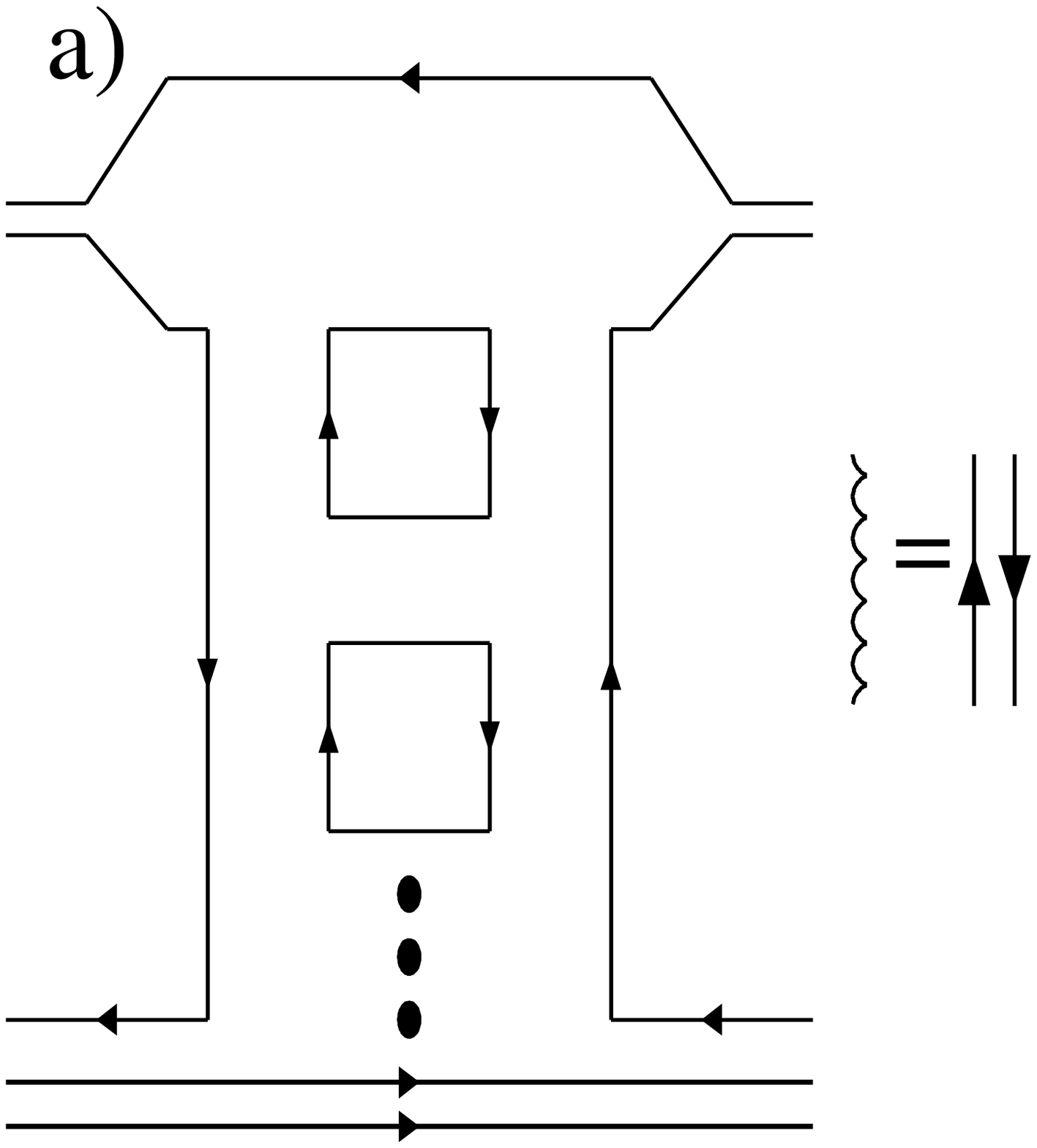,width=70mm} & \psfig{file=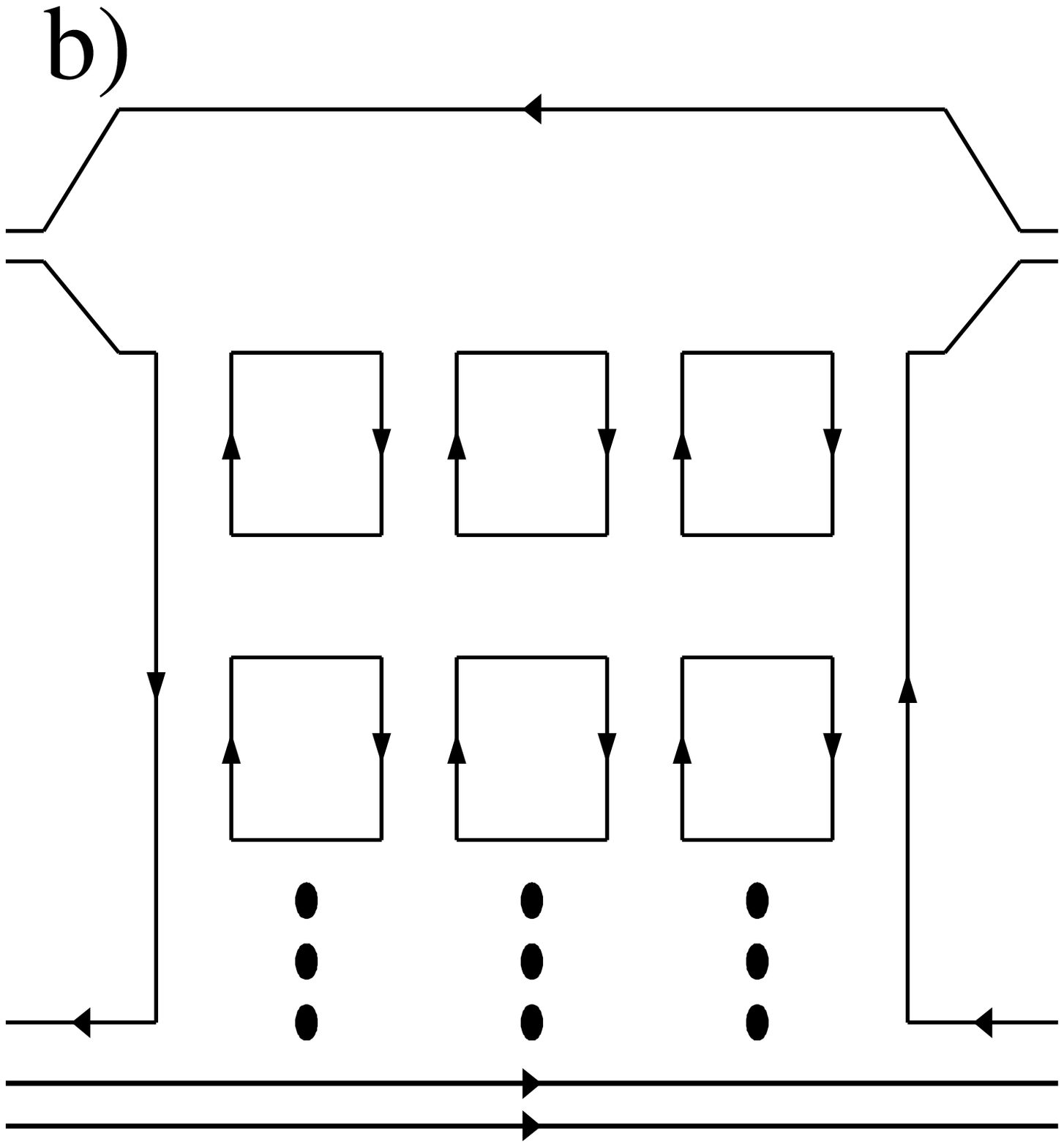,width=70mm} \\
\psfig{file=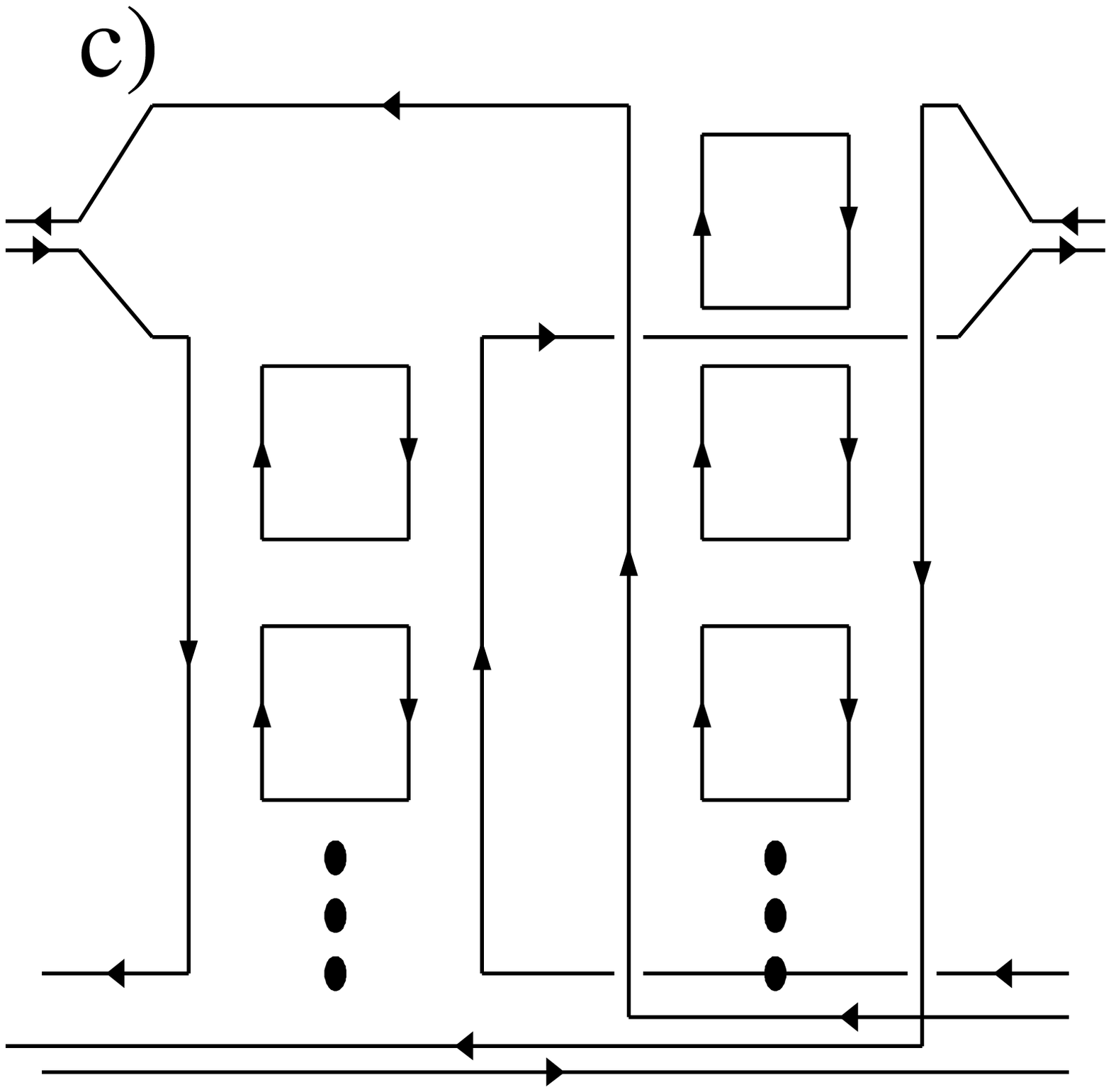,width=70mm} & \psfig{file=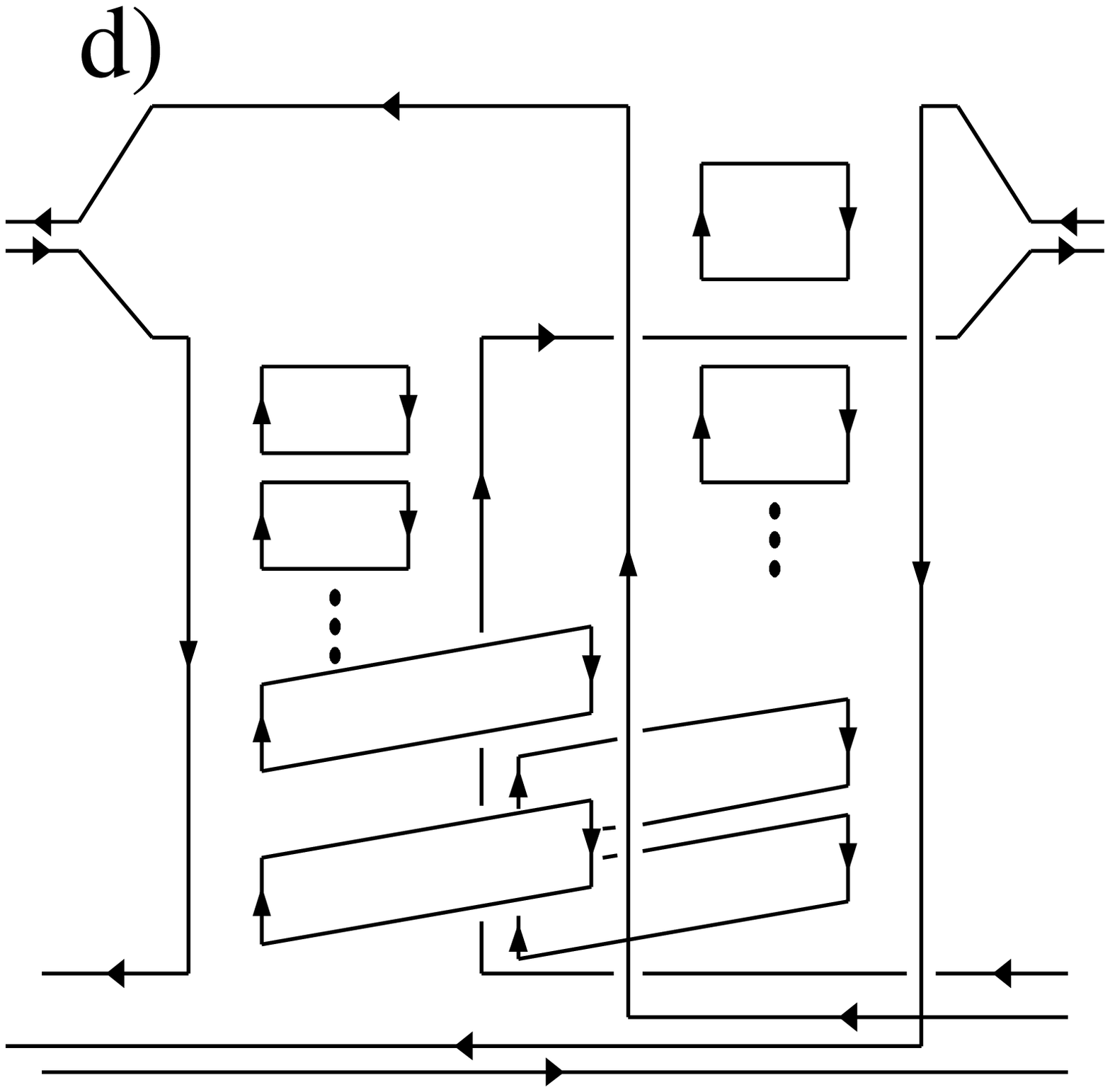,width=70mm} \\
\end{tabular}
\caption{{\em Topological configurations for large $N_c$ approach: a)
ladder diagram; b) general planar  diagram; c) Glauber-Mueller  approach;
d) general
two sheet diagram.}}
\label{NC}
\end{figure}

Let us discuss the large $N_c$
approximation in more detail. 
 The main principle, used in our large $N_c$ approach, has
been formulated by Veneziano et al. \cite{VENEZ}, namely, we sum leading
$N_c$ diagrams considering $N_c \,\as \,\approx\,1$ while
$\as\,\,<\,\,1$, separately in each topological configuration. For example,
at high
energy ( low $x$ ) the main contribution gives the planar diagrams ( see 
Fig.\ref{NC}a )which lead to the DGLAP evolution equations in leading
$\ln(1/x)$ approximation of pQCD. In fact, it has not been proven
yet that more complicated planar diagrams such as of Fig.\ref{NC}b could
be neglected at high energies, but several simple examples have been
considered ( see Ref.\cite{HT} ) which show that this hypothesis is very
likely ( see review \cite{LIP} for a detailed discussion on the subject).
The two sheet configuration (see Fig.\ref{NC}c) is suppressed since it is
proportional to an extra $\as$. The main contribution at high energy in this
configuration are the diagrams that has been summed in the Glauber -
Mueller approach (see Ref.\cite{HT}) in which the smallness of the order
of $\as$ is compensated by extra power of $xG(x,Q^2)$. The actual
parameter of the Glauber - Mueller approach is $\kappa$ which could be of
the order of 1. However we neglected the contribution of the diagrams of
Fig.\ref{NC}d type in which one can see a  transition from one two sheet
topology to another. 
As has been shown \cite{HT} such transition is
suppressed by a factor $\frac{1}{N^2_c}$. The use of the $\frac{1}{N_c}$
expansion can be suspicious in this case, because such transition gives a 
 $\frac{1}{N_c^2}$ corrections to the anomalous dimension. However,
it was shown \cite{LALE} that  $\frac{1}{N_c}$ - expansion works quite
well, justifying our approach. To be safe we assume that
$\frac{\as}{N^2_c}\,\ln(1/x)\,<\,1$.

5. There are no correlations between different  gluons inside the nucleon,
 of about the same energy,
 except the fact that they are distributed in the area with radius $R$.
This is the main assumption of the Eikonal model which makes this model 
simple and, of course, it is a great simplification of the unknown 
( nonperturbative ) structure of the nucleon. $R$ is the 
correlation radius for such gluons and the estimates from the 
HERA data have been discussed in the introduction.

\subsection{ The modified Mueller formula.}

 The next step of our approach is to give an estimate of the SC using 
the Mueller formula. However, before doing so, we have to study how well 
the DLA of perturbative QCD works, as it was heavily used in the derivation of
the Mueller formula. Let us consider the small $x$ limit of the DGLAP
evolution equation, which takes a simple form
\beq \label{DLA}
\frac{\partial^2 x G(x,Q^2)}{\partial \ln(1/x)\,\partial \ln Q^2}\,\,=\,\,
\frac{\as\,N_c}{\pi}\,xG(x,Q^2)\,\,.
\eeq
This expression may be obtained from the full DGLAP equation taking only
the DLA limit of the anomalous dimension, namely, 
$
\g^{DLA}\,\,=\,\,\frac{\as N_c}{\pi} \frac{1}{\o}
$.

In order to estimate numericaly how well the DLA works, 
we use the GRV parameterization \cite{GRV} for the nucleon gluon distribution.
This parameterization  describes all available experimental data quite well, 
including recent
HERA data at low $x$ and low $Q^2$ ($Q^2 \, \geq \, 1.5 \, GeV^2$).
  Moreover, the  GRV parameterization is 
suited for our purpose
 because
(i) the initial virtuality for the DGLAP evolution is small
 ($Q_0^2 \approx 0.25\, GeV^2$) and we can discuss the contribution of the 
large distances in the MF
having some support from experimental data;
 (ii) in this parameterization the most
essential contribution comes from the region where $\alpha _s ln Q^2 \approx 1$
and $\alpha _s ln 1/ x \approx 1$. This allows the use of the double 
leading log
approximation of pQCD, where the MF is proven \cite{MU90}. It should be also
stressed  that we consider the GRV parameterization as a solution of
 the  DGLAP
evolution equations, disregarding how much of the SC has been taken into
 account in this parameterization in the form of the initial gluon distribution.

\begin{figure}[htbp]
\centerline{\psfig{figure=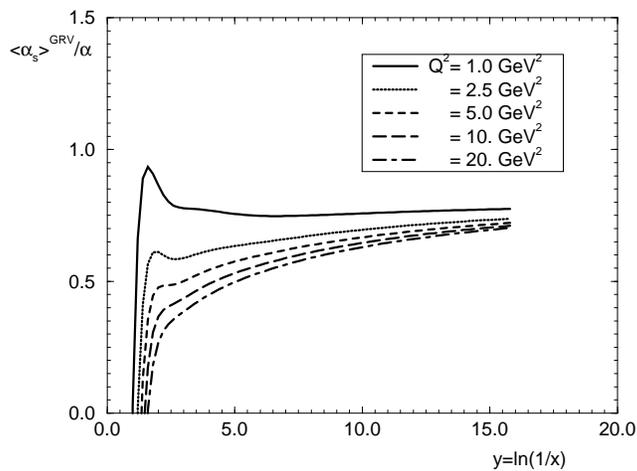,width=90mm}}
\caption{{\em The ration $\frac{< \as >^{GRV}}{\as}$ for different
 values of $Q^2$
in the GRV95 parameterization.}}
\label{alphaav}
\end{figure}

 However, in spite of the fact that the DGLAP evolution in the GRV
 parameterization starts from very low virtuality 
( $Q^2_0 \,\sim\,0.25 \, GeV^2$)
 it does not reach the DLA in the accessible
kinematic region ($Q^2\,> \, 1 \, GeV^2, \, x \,>\,10^{-5}$). 
To illustrate this statement we plot in Fig.\ref{alphaav} the ratio:
$$
\frac{< {\as} >^{GRV} }{ \as }\,\,=\,\,\frac{
\frac{\partial^2 x G^{GRV}(x,Q^2)}{\partial \ln(1/x)\,\partial \ln Q^2}
}{ \frac{\as N_c}{\pi}  xG^{GRV}(x,Q^2)}\,\,.
$$
This ratio is equal to 1 if the DLA holds. From Fig.\ref{alphaav} 
one can see that this 
ratio is rather around 1/2 even at large values of $Q^2$.

We can understand why the corrections to the DLA are so big modeling the 
small $x$ anomalous dimension $\g$ by a simple formula \cite{EKL}
\beq \label{APRGAM}
\g(\o)\,\,=\,\,\frac{\as N_c}{\pi}  \,\{\,\,\frac{1}{\o} \,\,-\,\,1\,\,\}
\,\, ,
\eeq
which has the correct DLA limit at small $\o$ and satisfies 
the momentum conservation constraint  $\g(\o=1)\,=\,0$. 
The typical values of $\o$
in all available parameterizations, even in the GRV, which is the closest
 one to the DLA, is $< \o > \,\approx\, 0.5 $. Therefore, there is about 50\%
correction to the DLA, meaning it cannot provide  reliable estimates
for the gluon structure function.

On the other hand, our
master equation (see \eq{MF}) is proven in  DLA. Willing to develop a
realistic approach in the region of not ultra small $x$ ($x \,>\,10^{-4})$
we have to change the master equation ( \eq{MF} ). We suggest to 
substitute  the full DGLAP kernel
in the first term of the r.h.s, which gives
\bea
x G(x,Q^2)\,\, & = &\,\, x G(x, Q^2)(\,\eq{MF}\,) \,\,+\,\,
x G^{DGLAP}(x,Q^2)\,\,
\nonumber \\
& - &\,\,\frac{\as N_c}{\pi} \,\int^1_x \,\int^{Q^2}_{Q^2_0} \,\,
\frac{d x'}{x'}\,
\frac{ d Q'^2}{Q'^2} \,x' G^{DGLAP}(x',Q'^2)\,\,.
 \label{FINANS}
\eea
 The above equation includes $ x G^{DGLAP}(x, Q^2_0)$ as the initial
condition for the gluon distribution and gives $ x G^{DGLAP}(x, Q^2)$ as the
 first term of the expansion with respect to $\kappa_G$. Therefore, this
equation is an attempt to include the full expression for the anomalous
 dimension for the scattering off each nucleon, while we use the DLA to 
take into account all SC. Our hope, which we will confirm by numerical
calculation, is that the SC  are small enough for $x \,>\,10^{-3}$ and
we can be not so careful in the accuracy of their calculation in this kinematic
 region. Going to smaller $x$, the DLA becomes better and \eq{FINANS} tends
 to the master equation (\ref{MF}).

We suggest an analogous improvement in the calculation of the SC 
correction for the structure
function $F_2$.  We subtract the Born term of  \eq{F2INT} and add the 
DGLAP evolved  structure function $F_2^{DGLAP}$. The result reads
\bea
F_2 ( x,Q^2) = F_2^{SC} - F_2^{BORN} + F_2^{DGLAP}\, ,
\label{f21}
\eea
where $F_2^{SC} $ is given by \eq{F2INT}, and the Born term is given by
\bea
F_2^{BORN}(x, Q^2 ) = \frac{1}{2 \pi}\, \sum^{N_f}_{1}\,\,  \as \,Z^2_f\,\,
\int^{\ln Q^2}_{\ln Q^2_0} \,\frac{2}{3}
d(\ln Q'^2) x G^{DGLAP}(x, Q'^2)\,\, .
\label{f2b}
\eea
As discussed above, $F_2^{DGLAP}(x, Q^2_0)$ is the initial condition
for $F_2$ and $F_2^{DGLAP}(x, Q^2)$ is the first term of the
expansion with respect to $\kappa_q$. Now, we are able to evaluate the
amount of SC in the nucleon gluon distribution and in the $F_2$ structure
function predicted by the eikonal approach.   

\subsection{ The gluon structure function for nucleon.}

 Before performing calculations we would like to discuss which observable
we are able to calculate using the Mueller formula. One can see from
Eq.(9) that the large distances enter into this formula. In spite of the
fact that this formula provides the infrared safe answer, or in other
words, the integral is finite in the region of the large distances, we
cannot trust our calculations in this region since we have no arguments
to apply pQCD in this kinematic region. Therefore, first we need
to indroduce a scale of distances for which we can trust the pQCD
calculations. We decided to choose the value  of $
r^2_{\perp\,0}\,\,=\,\, 1\,GeV^{- 2} $ as  such scale. The argument
comes from the new HERA data on the deep inelastic structure function
$F_2(x,Q^2)$ \cite{HERA} which show that the GRV parameterization
 works quite well starting from $Q^2 \,\,\geq\,\,1\,GeV^2$. Therefore, we
consider the GRV structure functions at $Q^2 \,=\,1\,GeV^2$ both for
gluons and quarks, as a lucky
guess of the initial conditions for the QCD evolution that takes into
account the unknown  nonperturbative large distance physics, disregarding
all physical motivation for such a guess. We believe that such a choice 
give us a reliable estimate of the SC which stem from small
distances,
being on the theoretical control of pQCD.  In the  following calculations
we will use a fixed value for the QCD running coupling constant, $\as\, = \, 0.25$.

In order to evaluate  quantitatively the influence of the SC on the evolution
of gluon structure function we calculate three functions
\bea
R_1 \, = \, \frac{xG(x, Q^2)}{xG^{GRV}(x,Q^2)}\,\, ,
\label{r1}
\eea
\beq
< \omega >\,\,=\,\,\frac{\partial \ln (x G (x,Q^2))}{\partial \ln(1/x)}\,\, ,
\label{omega}
\eeq
and
\beq 
< \gamma >\,\,=
\,\,\frac{\partial \ln (x G (x,Q^2))}{\partial \ln (Q^2/Q^2_0)}\,\,.
\label{gama}
\eeq 

We chose these functions because,
in the semiclassical approach ( see Ref. \cite{GLR} ),
the gluon structure function has a general form
\beq \label{SEMICLAS}
x G (x,Q^2)\,\,=\,\,C \,\{Q^2\}^{< \gamma >}\,\,
\{\frac{1}{x}\}^{ < \omega >}\,\, ,
\eeq
where we expect the exponents to be smooth functions 
of $ln (\frac{1}{x})$ and $ln (Q^2)$. 
We will see {\em a posteriori} that, indeed, $ < \o > $ and $<\gamma> $ 
turn out to be smooth functions in the region of small $x$.
All functions in \eq{r1} - \eq{gama} have transparent physical meaning.
Indeed,
 the function $< \gamma>$ is the average anomalous dimension in DIS,
 the function 
$< \omega>$ is the intercept of so called ``hard" Pomeron while 
 function $R_1$ is equal to
$1/N_{Pomeron}$,
where $N_{Pomeron}$ is the average number of ``hard" Pomerons taking part
in the rescatterings .
``Hard" Pomeron in our approach is
$xG(x,\frac{1}{r^2_{\perp}})$ in \eq{MF} ($\, ( x
G(x,\frac{1}{r^2_{\perp}}))^n$ corresponds to $n$ - Pomeron exchange).
Indeed, due to the AGK cutting rules \cite{AGK}, which have been proven in
 QCD in  Ref. \cite{BARY}, the inclusive cross
section is proportional to the exchange of one Pomeron or, in other words,
only the first term in \eq{MF} contributes to the inclusive cross section
of particles in the central region of rapidity ($\s_{incl} = \frac{d N}{d
y}$). On the other hand, $\s_{incl} \,=\,N\,\s_{tot}$, where $N$ is
the average multiplicity and $\s_{tot}$ is the total cross section given by
\eq{MF}. $N$ is equal to $N_{Pomeron} N (one \,\,Pomeron
\,\,exchange)$, where $N(one\,\,Pomeron\,\,exchange
)\,\,=\,\,\frac{1}{\s_{Pomeron}} \frac{d \s_{Pomeron}}{d y} $. Noticing
that $\s_{Pomeron}$ is the first term in \eq{MF}, we obtain that
$N_{Pomeron}\,=\,\frac{1}{R_1}$.

\begin{figure}
%\centerline{\psfig{figure=r1n2.eps,width=120mm}}
\begin{tabular}{c c}
\psfig{file=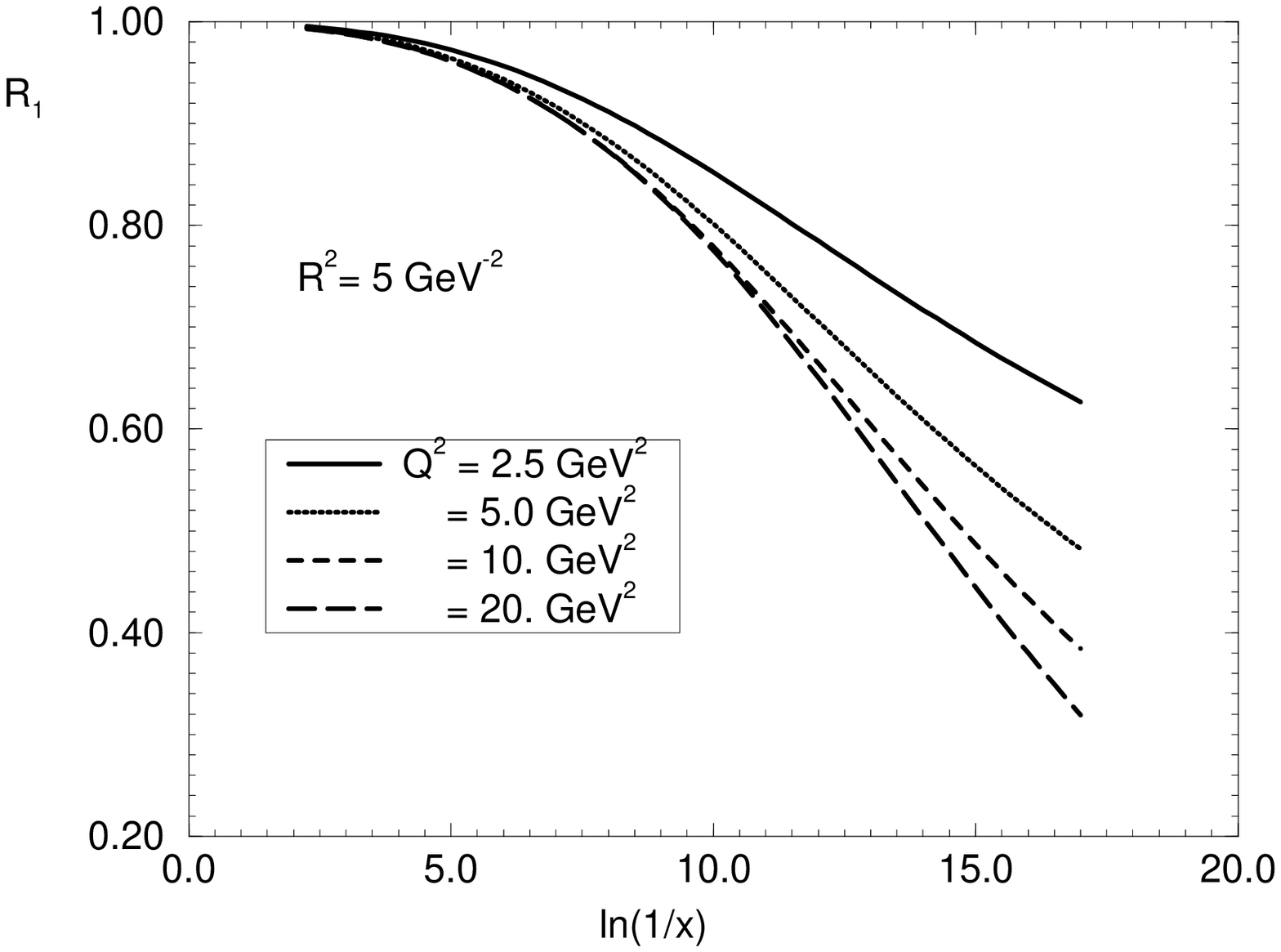,width=80mm} & \psfig{file=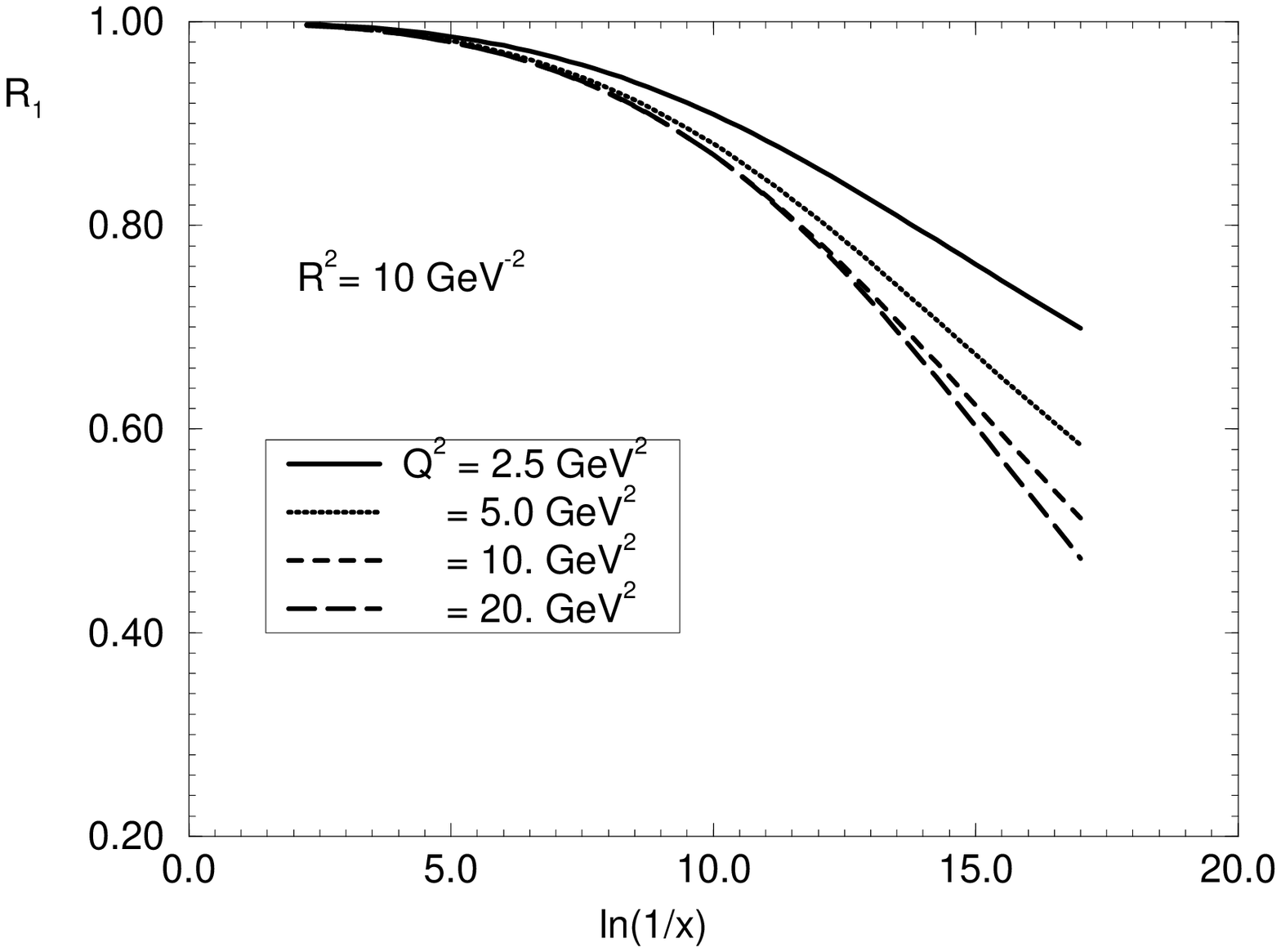,width=80mm}\\
\end{tabular}
\caption{\it $R_1$ ratio as a function of $y=ln(1/x)$ and $Q^2$ for two
values of $R^2$.}
\label{r1n}
\end{figure}

 In Fig.\ref{r1n}, we show the values
of $R_1$ as a function
of $y=ln(1/x)$ for different values of the virtuality $Q^2$ and two values of
$R^2$. The figure illustrates the general feature of the SC in
 the eikonal model:
the distribution $xG$ is strongly suppressed as $y$ increases  ( $x$  
tends to zero). 
In HERA kinematic region,  $2 \, < \, y \, < \, 10$
and $Q^2 = 2.5 \,\, GeV^2$, the modified Mueller formula for
 $R^2 = 10 \,\, GeV^{-2}$
predicts a suppression that
varies from less then $1 \%$ up to $10 \%$. For $Q^2 = 20\,\, GeV^2$, 
the suppression varies
from less then $1 \%$ up to $13 \% $. This effect is even bigger  for smaller 
value of $R^2$.  For example,
for $R^2 = 5 \,\, GeV^{-2}$ and $Q^2 = 2.5\,\, GeV^2$, the suppression goes from
less then $1 \%$ for $y=2$ to almost $15 \%$ for $y=10$. For $Q^2 = 20\,\,
 GeV^2$, the
suppression goes from less then $1 \%$ for $y=2$ to $22 \%$ for $y=10$. 

We can interpretate the result shown in Fig.\ref{r1n} in a different way
saying that the average number of the ``hard" Pomerons participating in
rescattering changes from 1 at $x = 10^{-2}$ to 1.25 ($R^2 = 10 \,GeV^{-2}$ and
$Q^2 \, = \, 10 \,\, GeV^2$) or
1.43 ($R^2 = 5 \, GeV^{-2}$) at $x \rightarrow 10^{-5}$. 

\begin{figure}[htbp]
\centerline{\psfig{figure=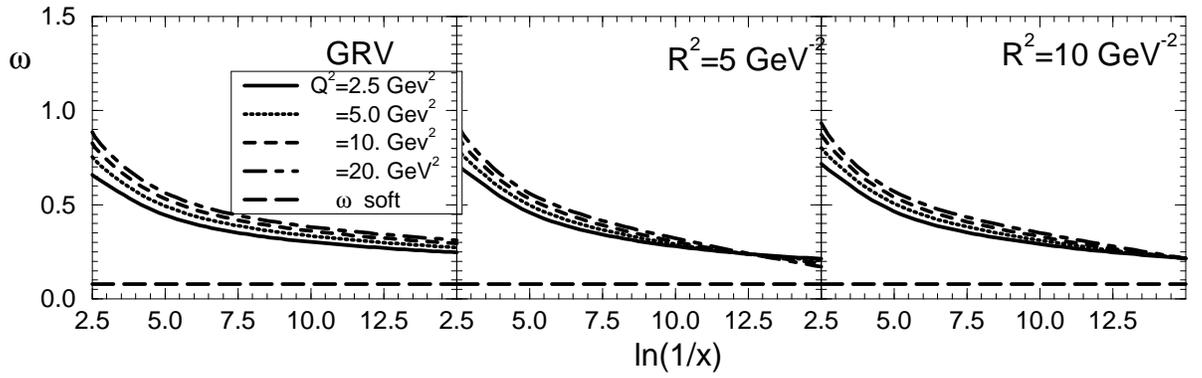,width=150mm}}
\caption{\it  The values of $<\omega >$ as a function of $y=ln(1/x)$ and $Q^2$ 
for GRV and
two values of $R^2$. It is also plotted the soft Pomeron
 intercept $\o \, = \, 0.08$.}
\label{omn}
\end{figure}

In Fig.\ref{omn} are shown the results for $<\omega >$ as a function of $y$ for
 several values of
$Q^2$ and two values of $R^2$. As we can see, the values
 of $\omega$ are softened
by the SC when compared with the GRV values.

\begin{figure}[hbtp]
\centerline{\psfig{figure=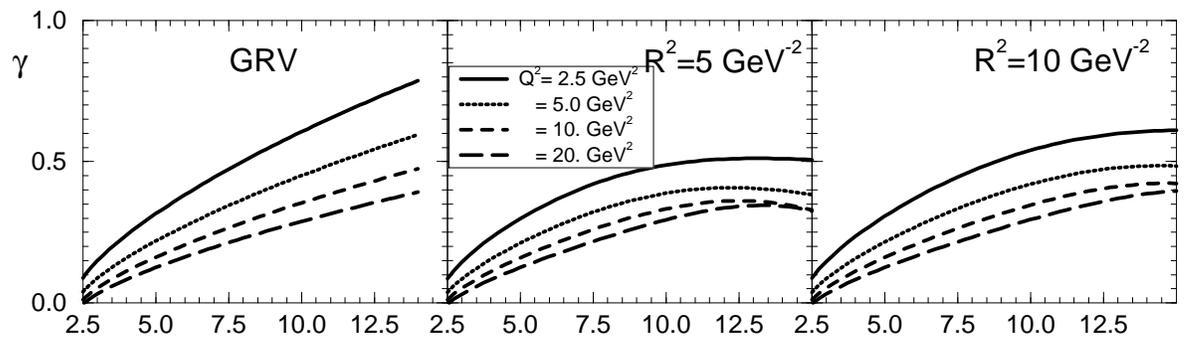,width=150mm}}
\caption{\it The values of $<\gamma >$ as a function of $y=ln(1/x)$ and $Q^2$ for GRV 
and two values of $R^2$.}
\label{gmn}
\end{figure}
In Fig.\ref{gmn} are  shown the results for $< \gamma >$ calculated from 
expression (\ref{gama}).
The values of $< \gamma >$ suffer a strong suppression as 
compared to the GRV ones.
It means  that the $Q^2$ evolution of
 the gluon distribution is significantly modified
by the SC. 

 The figures \ref{r1n}, \ref{omn} and \ref{gmn} also illustrate the 
 dependence of SC on the value of $R^2$.
 For example, for $R^2 \,\,=\,\,10\, GeV^{-2}$, the  SC, generated by 
the  eikonal approach, suppress the value of $\o$, but the value of the SC 
is  not enough  to provide a match between the
``soft" and ``hard" processes. To illustrate
 this point we put in Fig.\ref{omn} the
 intercept of the so called ``soft" Pomeron 
 $ \o_{soft}\,=\,0.08 $  \cite{SOFTPOMERON}. For  $R^2 \,\,=\,\,5\, GeV^{-2}$,
the suppression is bigger, but $\o$ does not reach the ``soft''
 Pomeron intercept.
However, for $Q^2 \, = \, 20 \, GeV^2$
sufficiently small values of $x$ ($y\approx 20$) and $Q^2 \, = \, 20 \, GeV^2$,
the SC reduces the values
of $\o$ reaching the ``soft'' Pomeron intercept. One can see that 
$< \o >$ turns out to be bigger 
than this intercept at  small values of $Q^2$. It occurs because
 these values of
$Q^2$ are close to the initial condition and do not generate enough shadowing.

 As we have discussed the SC in the eikonal approach  considerably
 reduce the value of $< \gamma >$.  For $R^2 \, =\, 5\,\, GeV^{-2}$,
the mean value of the anomalous dimension  $ < \gamma >$ reaches the value
 $\frac{1}{2}$ only for $Q^2 \, = \, 2.5 \, GeV^2$ in the  range 
$10\, < \, y \, < \, 16$. For $Q^2 \, > \, 2.5 \, GeV^2$,  $< \gamma >$
is always less then  $\frac{1}{2}$.
Thus we expect that the BFKL  Pomeron  will not be  seen in the deep inelastic
 gluon  structure function
at least in HERA kinematic region. For
$R^2 \, =\, 10 \,\, GeV^{-2}$, however,
the strength of the SC is such that   $ < \gamma >\,\,>\,\,\frac{1}{2} $ in the
accessible kinematic region.  It  means that,
 for  $Q^2 \,\leq\,2.5 \,\, GeV^2 $ and 
$7.5 \, < \, y$, 
the BFKL Pomeron could give a visible contribution.
 The above discussion illustrates
the role of SC and the $R^2$ parameter on the
transition to the BFKL dynamics which we are going to discuss a bit later in
 more details.

\begin{figure}
\begin{center}
\begin{tabular}{c c}
\psfig{file=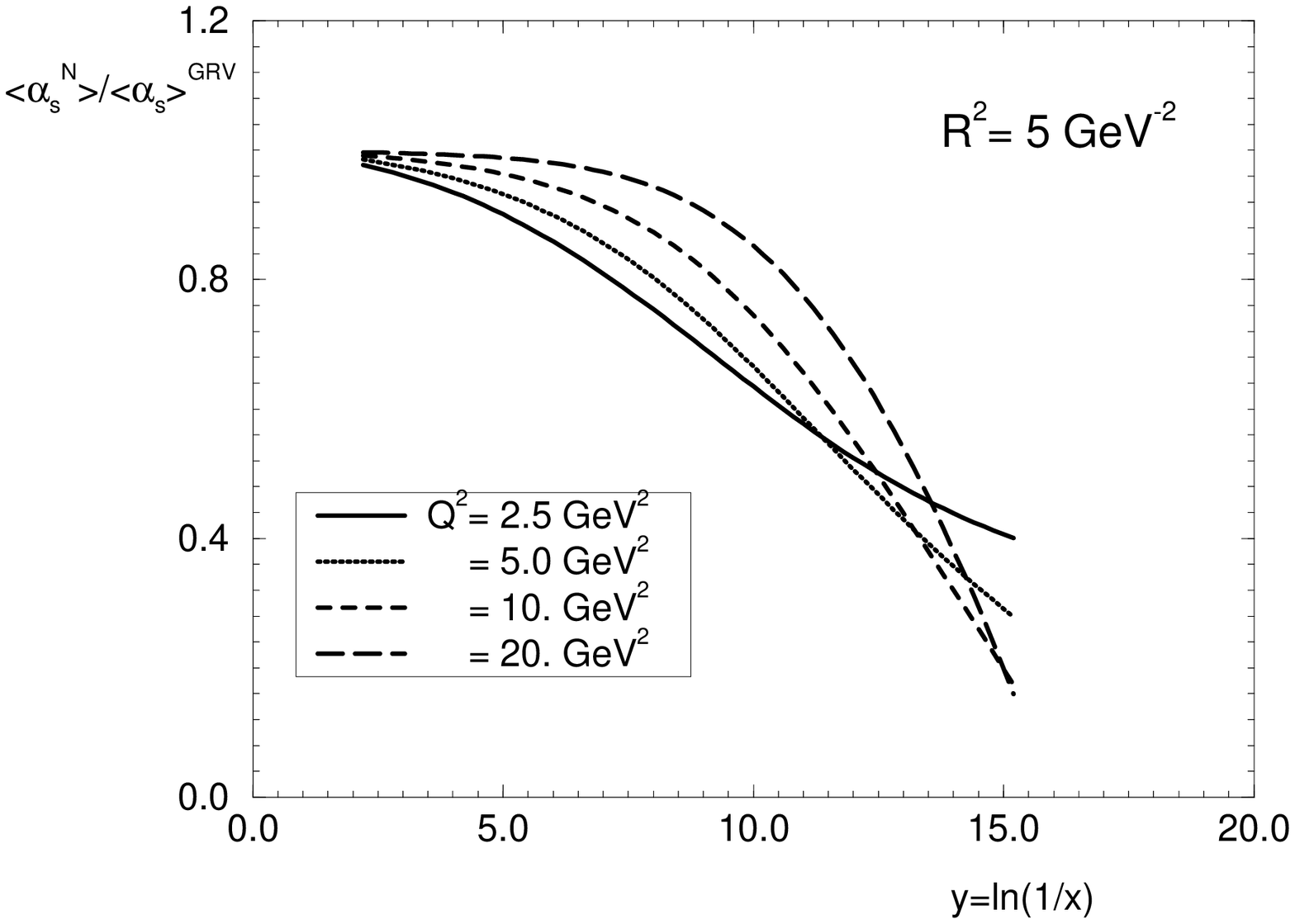,width=80mm} & \psfig{file=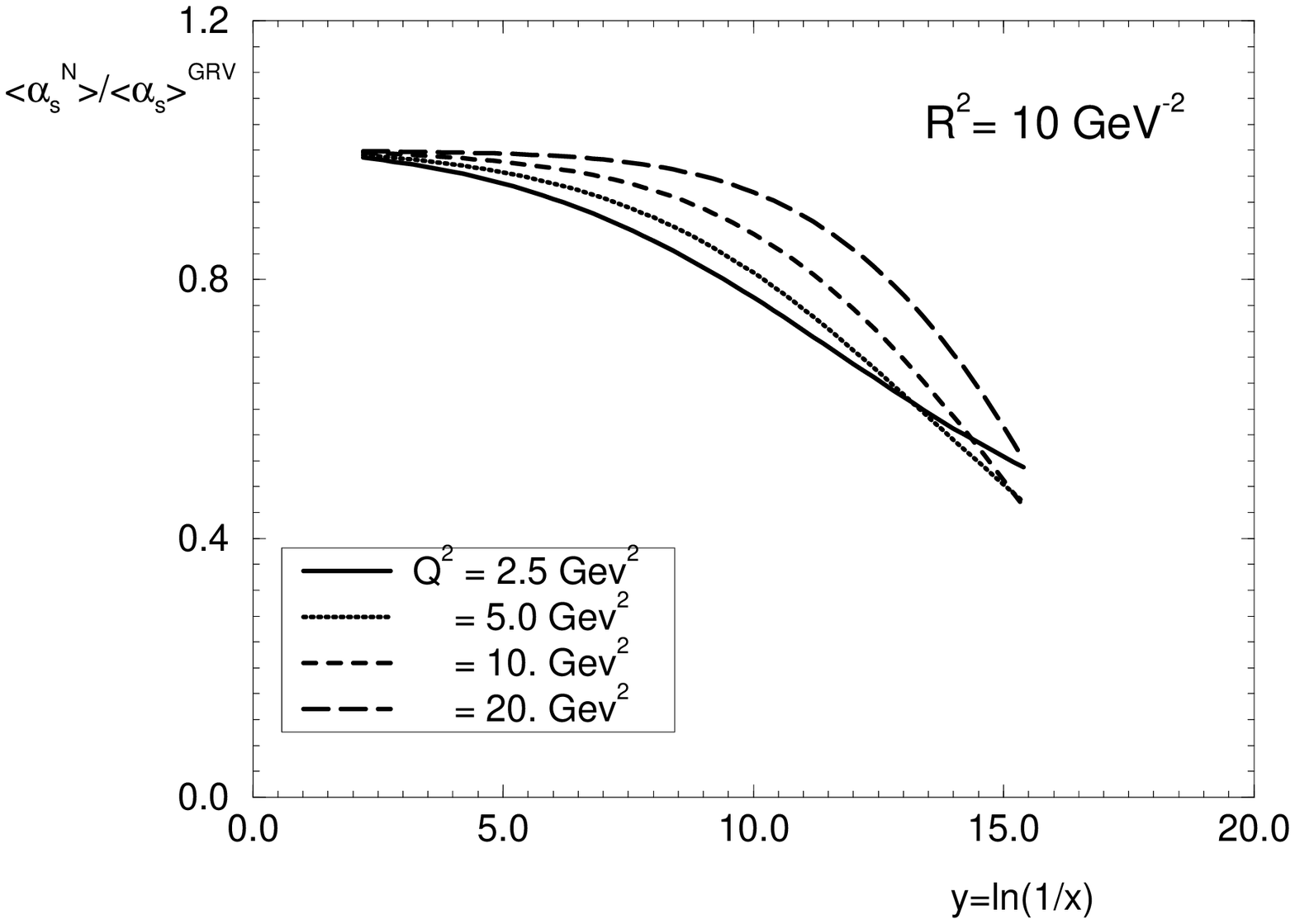,width=80mm}\\
\end{tabular}
\caption{\label{aba}
The values of the effective $\as$ in the parton cascade with SC. }
\end{center}
\end{figure}

In order to complete the discussion of the SC from  Mueller formula, we
plot in Fig.\ref{aba} the expression
\bea
\frac{< \as^N >}{< \as>^{GRV}} =
\frac{xG^{GRV}(x, Q^2)}{xG_N (x, Q^2 )}\frac{ \left[ 
\frac{\pa }{\pa \ln\frac{1}{x}\pa \ln Q^2} (xG_{N}(x,Q^{2})) \right]}{
 \left[ 
\frac{\pa }{\pa \ln\frac{1}{x}\pa \ln Q^2} (xG^{GRV}(x,Q^{2})) \right]}
 \, ,
\label{alphas}
\eea
which is the ratio of the effective $\as$ in DLA with and without the SC.
 In some sense this is the effective value of the QCD coupling 
constant in the parton cascade with SC,
 and characterizes the amount of shadowing taken
into account in the
DGLAP evolution of the gluon distribution. The results present a weakening
of the effective coupling constant which accepts  a simple interpretation.
In the small-$x$ region, the gluon
density increases and in a typical parton cascade cell there are many partons
with different colors.  The color average gives smaller mean color charge
at large gluon densities and $<\as >$ is reduced.
From the figure we can see that SC modify
the DGLAP evolution for a large kinematic region, even for $Q^2 \ge 10 \,\, GeV^2,$
though including a substantial part of HERA data.

\subsection{The DIS structure function $F_2$.}                  
\label{disf2}
In this section we evaluate the SC predicted by the Glauber approach for
$F_2$. We calculate $F_2$ from  the modified Mueller formula (\ref{f21})
$$
F_2 ( x,Q^2) = F_2^{SC} - F_2^{BORN} + F_2^{DGLAP}\, ,
$$
where $F_2^{SC} $ and $F_2^{BORN}$ are given by \eq{F2INT} and \eq{f2b} 
respectively. 
We calculate  $F_2^{DGLAP}$ from the GRV 95 \cite{GRV95} 
distribution $u$, $d$, $s$ and $xG$ 
using the expression
\bea
F_2^{DGLAP}(x, Q^2) = \sum_{u,d,s} \, \eps_q^2 \, x[q(x, Q^2) + \bar q (x, Q^2 ) ] + 
F_2^c (x, Q^2, m_c^2 )\, ,
\label{f2grv}
\eea
where $q$ and $\bar q$ represent the light quarks distributions. 
$F_2^c$ describes the heavy flavour contribution to $F_2$ 
(disregarding bottom contribution).
 In this scheme the heavy quarks are not
considered as intrinsic partons in the Renormalization Group evolution, 
but are produced, in LO of
perturbation theory, by the gluon-$\gamma^*$ fusion process. Thus, the 
charm component $F_2^c$ will be generated perturbatively from the intrinsic
gluon
distribution.

\begin{figure}[htbp]
\centerline{\psfig{figure=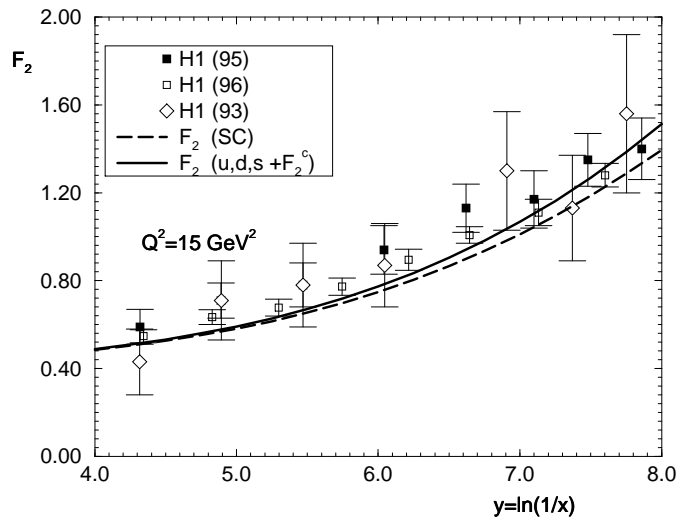,width=100mm}}
\caption{\it $F_2$ calculated from (\protect\ref{f21})
for $Q^2_0 \, = \, 1.0 \,\, GeV^2$.}
\label{f2nuc6}
\end{figure}

In Fig.\ref{f2nuc6} we plot $F_2$ using the full
DGLAP evolution, given by \eq{f2grv}, 
and the modified Mueller formula (\ref{f21})
to take into account the SC. 
The $F_2^c$ component was calculated from the gluon distribution in LO, as described
in Ref.\cite{GRV95}. The SC are not included in the gluon distribution that
enter in  the $F_2^c$ calculation. We expect that it would give a correction
for $F_2$ smaller then the
correction given by \eq{f21}, at least by a factor $\as (4 m_c^2) $.

%Furthermore, any correction to the gluon distribution in the
%charm component is
%inconsistent, since we have taken only light quarks to calculate the SC for $F_2$ in \eq{f21}.

\begin{figure}[hbtp]
\centerline{\psfig{figure=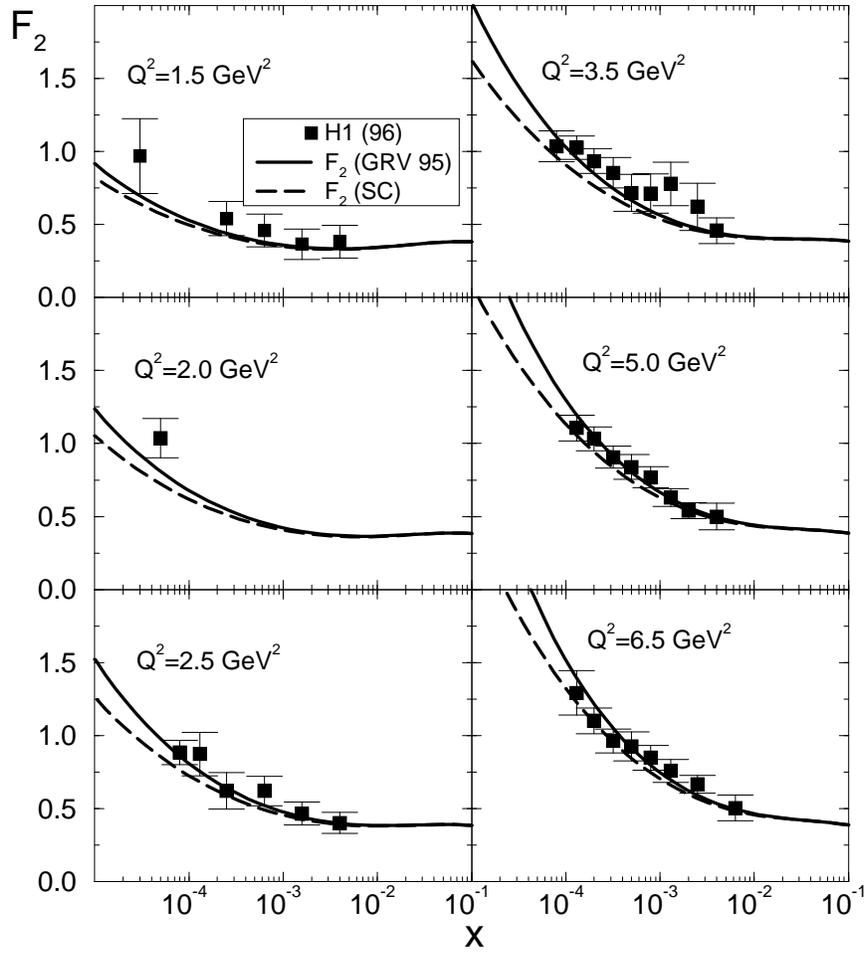,width=120mm}}
\caption{\it $F_2$ calculated from (\protect\ref{f21})
for $Q^2_0 \, = \, 1.0 \,\, GeV^2$
compared with H196 data\protect\cite{HERA}.}
\label{f2nuc8}
\end{figure}

\begin{figure}[hbtp]
\centerline{\psfig{figure=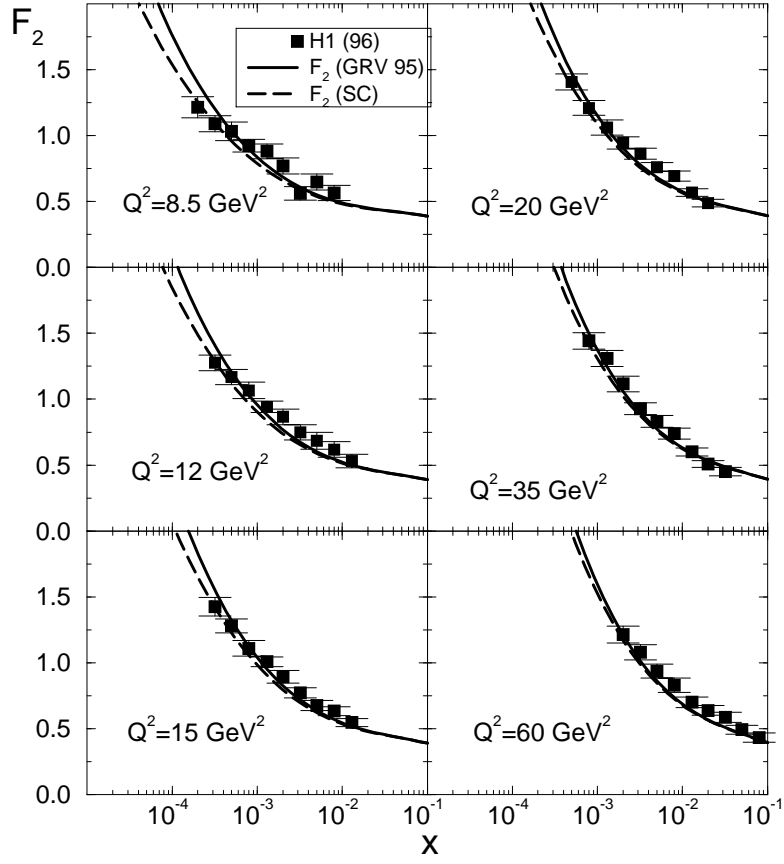,width=120mm}}
\caption{\it $F_2$ calculated from (\protect\ref{f21})
for $Q^2_0 \, = \, 1.0 \,\, GeV^2$ compared
with H196 data\protect\cite{HERA}.}
\label{f2nuc9}
\end{figure}

In figs.(\ref{f2nuc8}) and (\ref{f2nuc9}) we present the calculation of
$F_2^{DGLAP}$ and $F_2$ given by \eq{f21}
as a function of $x$ for different values 
of $Q^2$.  The results are compared with H1 96 data. As we can see
from the figures, the SC are small for $Q^2 \,=\,1.5\, GeV^2$ since
it is very close to the initial value  $Q^2_0 \,=\,1\, GeV^2$. When 
 $Q^2$ grows from  $Q^2 \,=\,1.5\, GeV^2$ to  $Q^2 \,=\,6\, GeV^2$,
the SC increase. For  $Q^2 \,>\, 6 \, GeV^2$, the SC start to fall down
again.   
In fact, the figures show the amount of shadowing
that we would expect to be present in the experimental data, if the Glauber
approach is supposed to be correct.
 We conclude that the contribution of the  SC to $F_2 (x, Q^2)$,
 predicted from the eikonal approach 
are rather small in HERA kinematic 
region unlike the case of the gluon structure function that has been
 discussed in the previous section. This is an explanation why we are able to
 describe the HERA experimental data on $F_2(x,Q^2)$ in the GLAP evolution, 
without taking into account SC in spite of the fact that parameter
$\kappa$
 is rather big. It turns out the story of the SC contribution is the story 
about the gluon densities for which we have no independent experimental 
information. The value of $R_1$ for $ xG(x,Q^2)$ is rather big, as we have
 discussed, but it is still smaller than uncertainties in the values
 of $xG(x,Q^2)$ from different attempts ( GRV, MRS and CTEQ) to
 describe the experimental HERA  data in the framework of the DGLAP evolution,
that  exist  on the market.

\subsection{ The $b_t$ dependence of the gluon distribution.}

 In the next two subsections we will investigate how the SC work at
different values of the impact parameter $b_t$. We hope, that the impact
parameter distribution will help us to understand better the matching
between ``soft" and ``hard" interactions.
In order to do this
we introduce the function $xG(x,Q^2,b_t)$ defined by  the normalization
\beq \label{BGNORM}
\int \,\,d^2 \,b_t\,\,x G(x,Q^2,b_t)\,\,=\,\,x G(x,Q^2)\,\,,
\eeq
where $x G(x,Q^2)$ is the gluon distribution in perturbative QCD. We introduce
also the
function  $xG(x,Q^2, t)$ given by
 \beq  \label{xgff} 
 xG(x,Q^2, t)\,\,=\,\,\int \,d^2 b_t e^{i\,\vec{q}_t \cdot \vec{b}_t } \,\,
 xG(x,Q^2, b_t) \,\,.
\eeq 
where $t$ is the momentum transfer $t = - q^2_t$.

In the DGLAP evolution equation, the $b_t$ dependence can be factorized out
in the form \cite{GLR}
\beq \label{BAP}
x G(x,Q^2,b_t)\,\,=\,\,x G(x, Q^2) \,S(b_t)\,\,,
\eeq
where  the profile function $S(b_t)$ is related to the two gluon nucleon 
form factor $F(t)$ as 
 \beq  \label{FF} 
F(t)\,\,=\,\,\int \,d^2 b_t e^{i\,\vec{q}_t \cdot \vec{b}_t } \,\,S(b_t)\,\,.
\eeq 
If we consider the exponential parameterization of the profile function $S(b_t)$,
 Eq.(~\ref{BAP}) and (\ref{FF}) give also an exponential $t$- dependence  for
 the gluon structure function, which can be obtained by the ratio
\beq
\frac{xG(x,Q^2, t)}{x G(x,Q^2,t = 0 )}\,\,=\,\,F(t)\,\,=\,\,e^{ - 
\frac{R^2}{4}\,|t|}\,\,.
\eeq
Thus the DGLAP evolution predicts a slope $B\,\,=\,\,\frac{R^2}{4}$, that 
does not depend on $x$ and $Q^2$. 

However, as we will show, the SC will lead to an $x$ and $Q^2$- dependence 
of the $B$ slope, as well as a non factorizable $b_t$ - dependence of the gluon
structure function.

We introduce the  impact parameter $b_t$ dependent
 function $xG(x,Q^2,b_t)$ defined 
by \eq{BGNORM},
where $x G(x,Q^2)$ is the gluon distribution given by the Glauber formula
(\ref{MF}).
From (\ref{BGNORM}),  $x G(x,Q^2,b_t)$ can be written as
\beq \label{BMF}
x G(x,Q^2,b_t)\,\,=\,\,\frac{2}{\pi^3}\,\,\int^1_x\,\frac{d\, x'}{ x'}
\,\,\int^{\infty}_{\frac{1}{Q^2}}
\,\,\frac{d\,r^2_{\perp}}{r^4_{\perp}}\,\,\{\,\,1\,\,-\,\,
e^{-\,\frac{1}{2} \s^{GG}_N\,
S(b_t) }\,\,\}\,\,,
\eeq
where  $\s^{GG}_N$ was defined in \eq{10}.  Following the discussion 
before \eq{FINANS} we will modify \eq{BMF} in order to use the full
DGLAP kernel to describe the gluon-nucleon interaction and the DLA
to take into account the SC. We add a term proportional to the 
gluon distribution and  subtract the  Born term
of (\ref{BMF}).
This procedure gives 
\bea
x G(x,Q^2,b_t)\,\,=\,& & \,  x G^{DGLAP}(x,Q^2)S(b_t) + 
x G(x,Q^2,b_t)(\ref{BMF}) \nonumber \\
& - & \frac{2}{\pi^3}\,\,\int^1_x\,\frac{d\, x'}{ x'}
\,\,\int^{\infty}_{\frac{1}{Q^2}}
\,\,\frac{d\,r^2_{\perp}}{r^4_{\perp}}\,\,\,\frac{1}{2} \s^{GG}_N\,
S(b_t)  \, .
\label{finalbt}
\eea 
This expression keeps the same $b_t$ 
dependence in the Born term and gives the correct match with \eq{BGNORM}.
In all calculations we use a fixed value for the coupling constant 
$\as= 0.25$ and 
put $Q_0^2 \, = \, 1 \, GeV^2$. 

\begin{figure}[hptb]
\begin{center}
\begin{tabular}{c   c}
\psfig{file=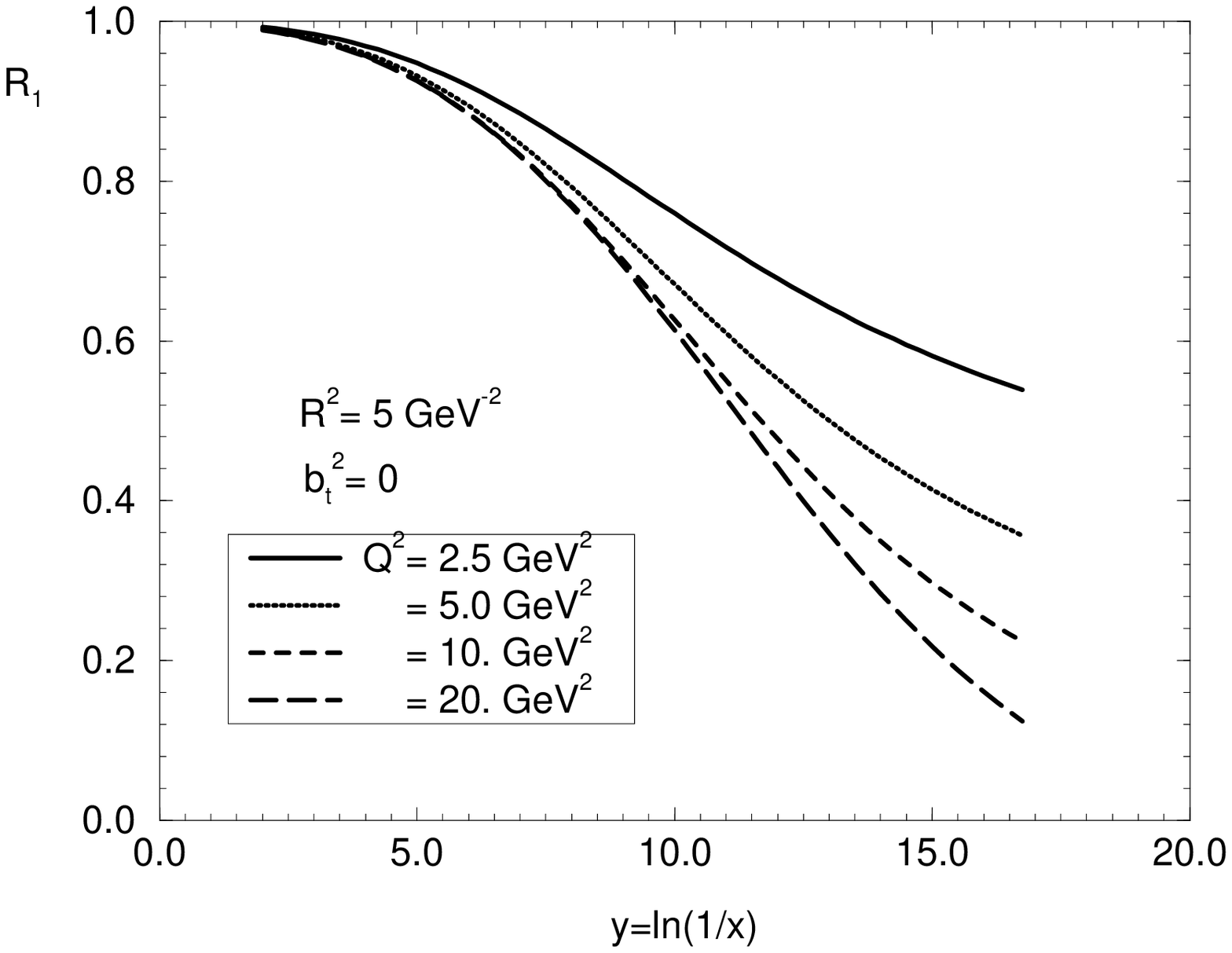,width=70mm} & \psfig{file=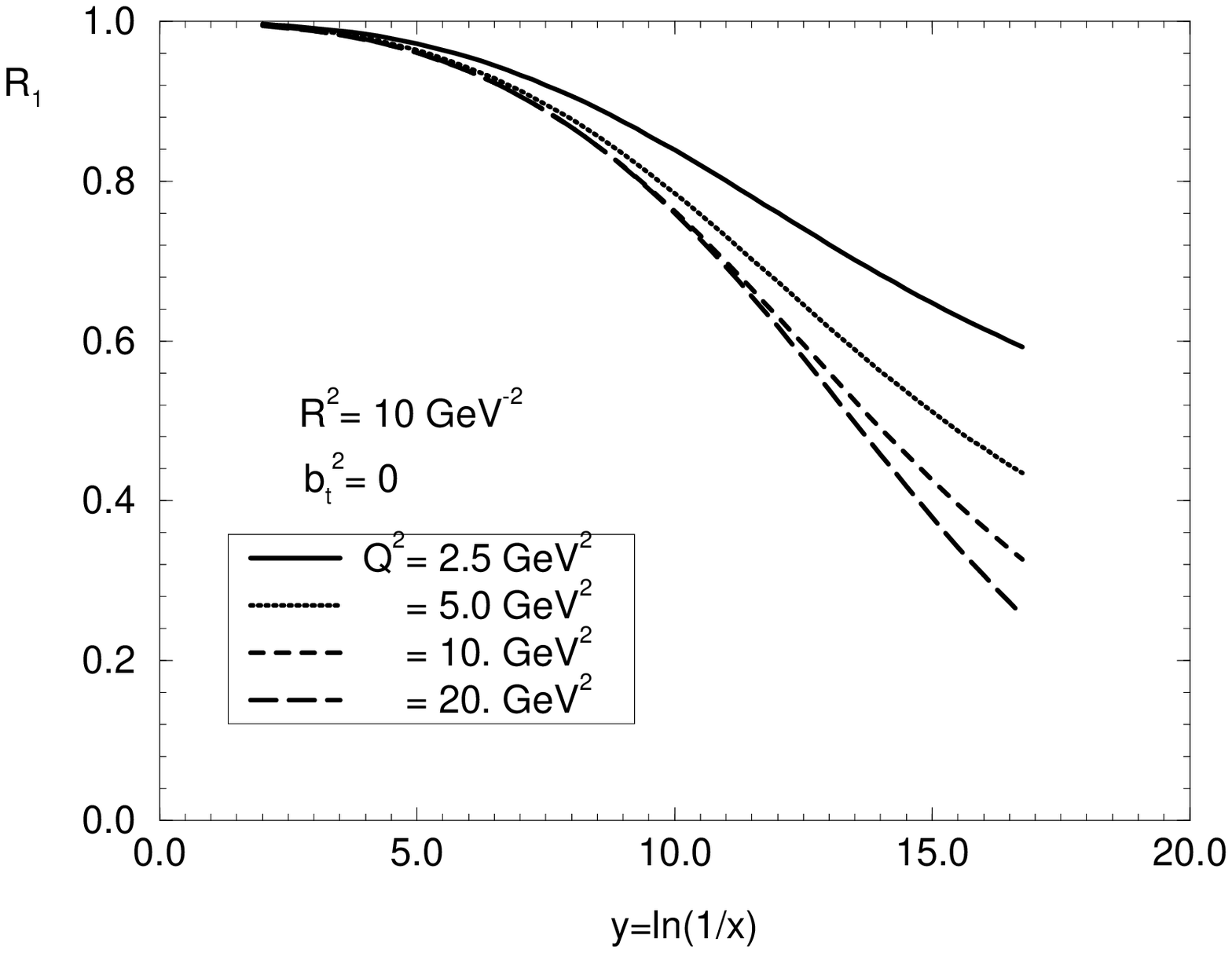,width=70mm}\\
\psfig{file=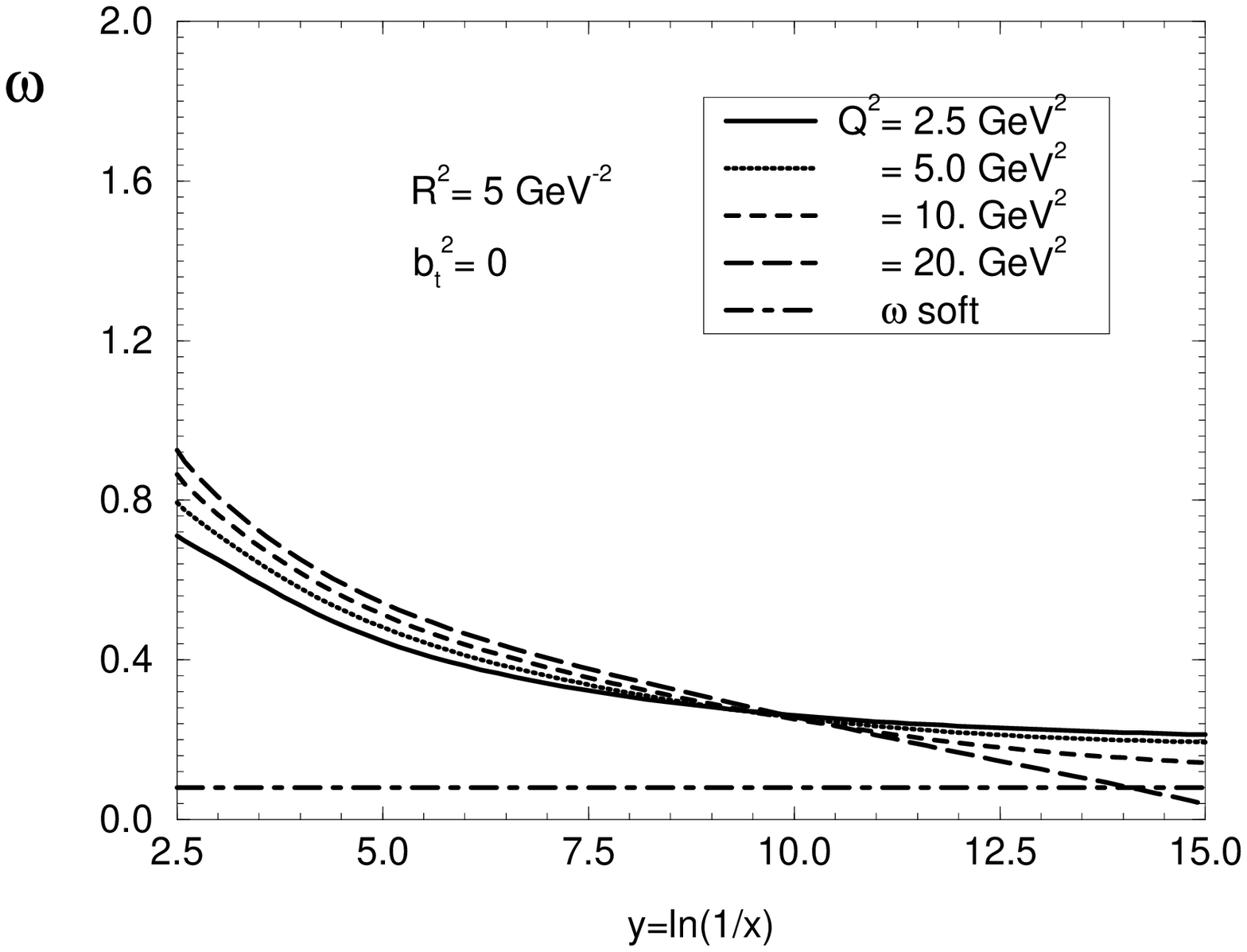,width=70mm} & \psfig{file=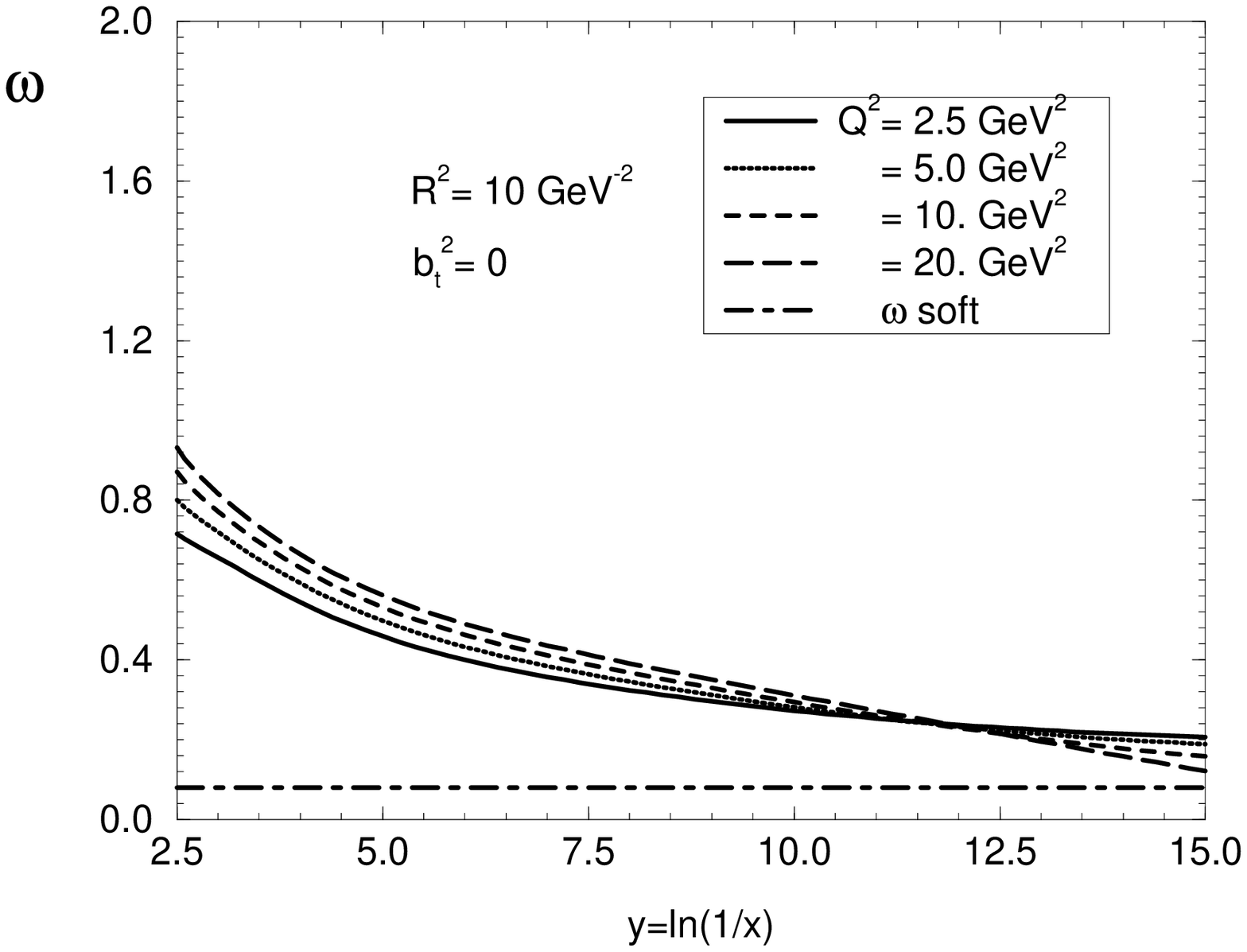,width=70mm}  \\
\psfig{file=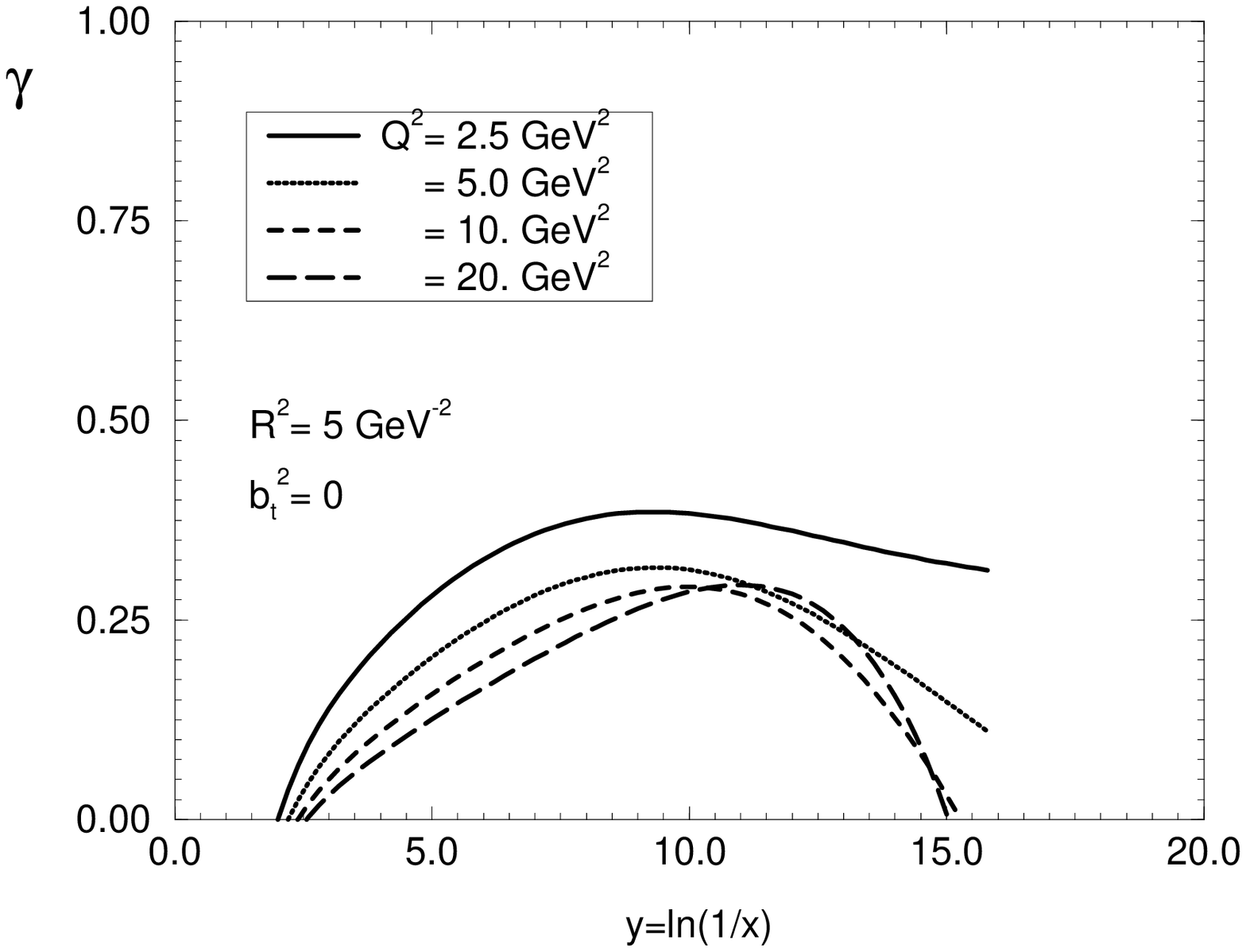,width=70mm} & \psfig{file=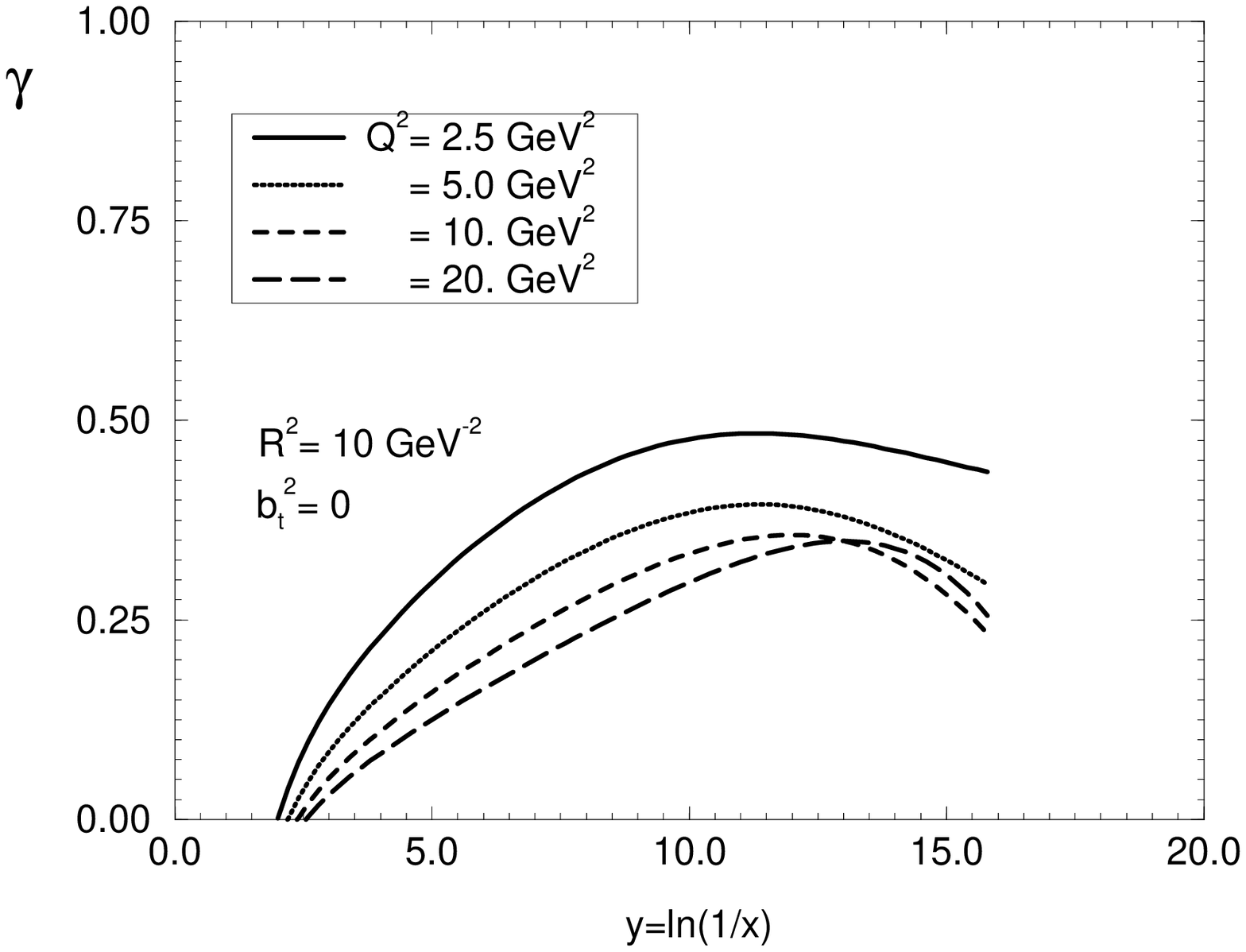,width=70mm} \\
\end{tabular}
\end{center}
\caption{\em  $R_1$( \protect\eq{r1}), $\omega$ 
(\protect\eq{omega}) and $\gamma$(\protect\eq{gama}) 
function calculated from expression (\protect\ref{finalbt}) with 
$b_t^2 \, = \, 0$. The results are plotted as functions of
$y$ for different values of $Q^2$.}
\label{rb0}
\end{figure}

In order to investigate the general properties of the $b_t$ dependent 
distribution,
we  
calculate all
the three 
functions $R_1$ (\eq{r1}), $<\omega >$ (\eq{omega}) and 
$<\gamma >$ (\eq{gama})
 using the above expression for $x G(x,Q^2,b_t)$ instead of 
$x G(x,Q^2)$.  The
results are plotted in Fig. \ref{rb0} for $b_t\,=\, 0$.   
We can see from the figure that the SC are bigger for the $b_t^2 \, = \, 0$
case when compared to the $b_t^2$ integrated results. The correction term
in (\ref{finalbt}) will be proportional to 
$\{1 -\, \kappa_G - exp(-\kappa_G)  \}$, which is bigger then the nonlinear
term of \eq{FINANS}.

It is interesting to notice that at $b_t$ = 0 the SC give $< \omega >$
 approaching $\omega_{soft}$ = 0.08  for $y \, \approx \, 14$, which
corresponds to $x \, = \, 10^{-7}$. Thus we may expect eventually a
 matching of the DIS data with
the ``soft" processes at this very small value of $x$. 

\begin{figure}[hptb]
\begin{center}
\begin{tabular}{c   c}
\psfig{file=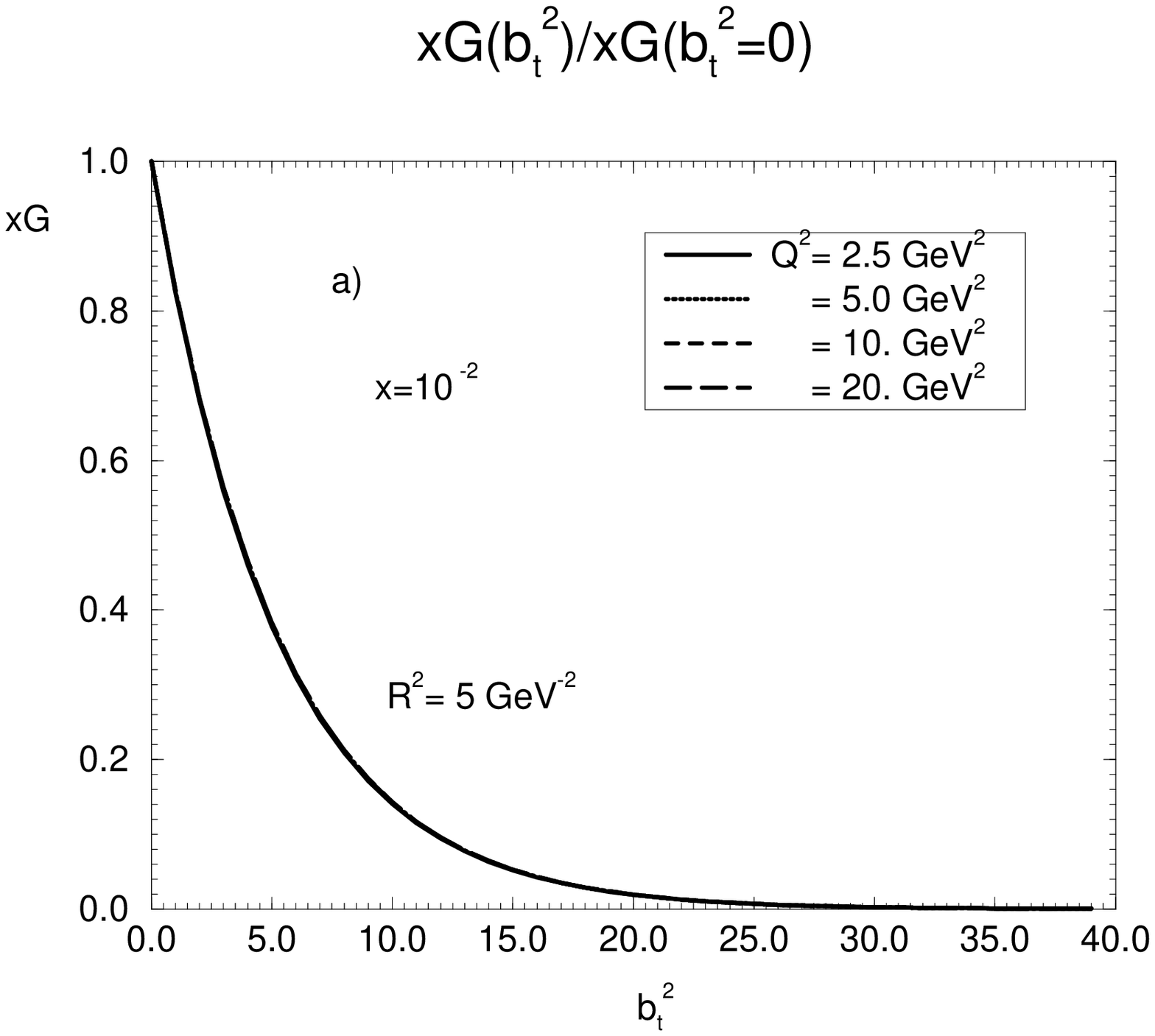,width=70mm} & \psfig{file=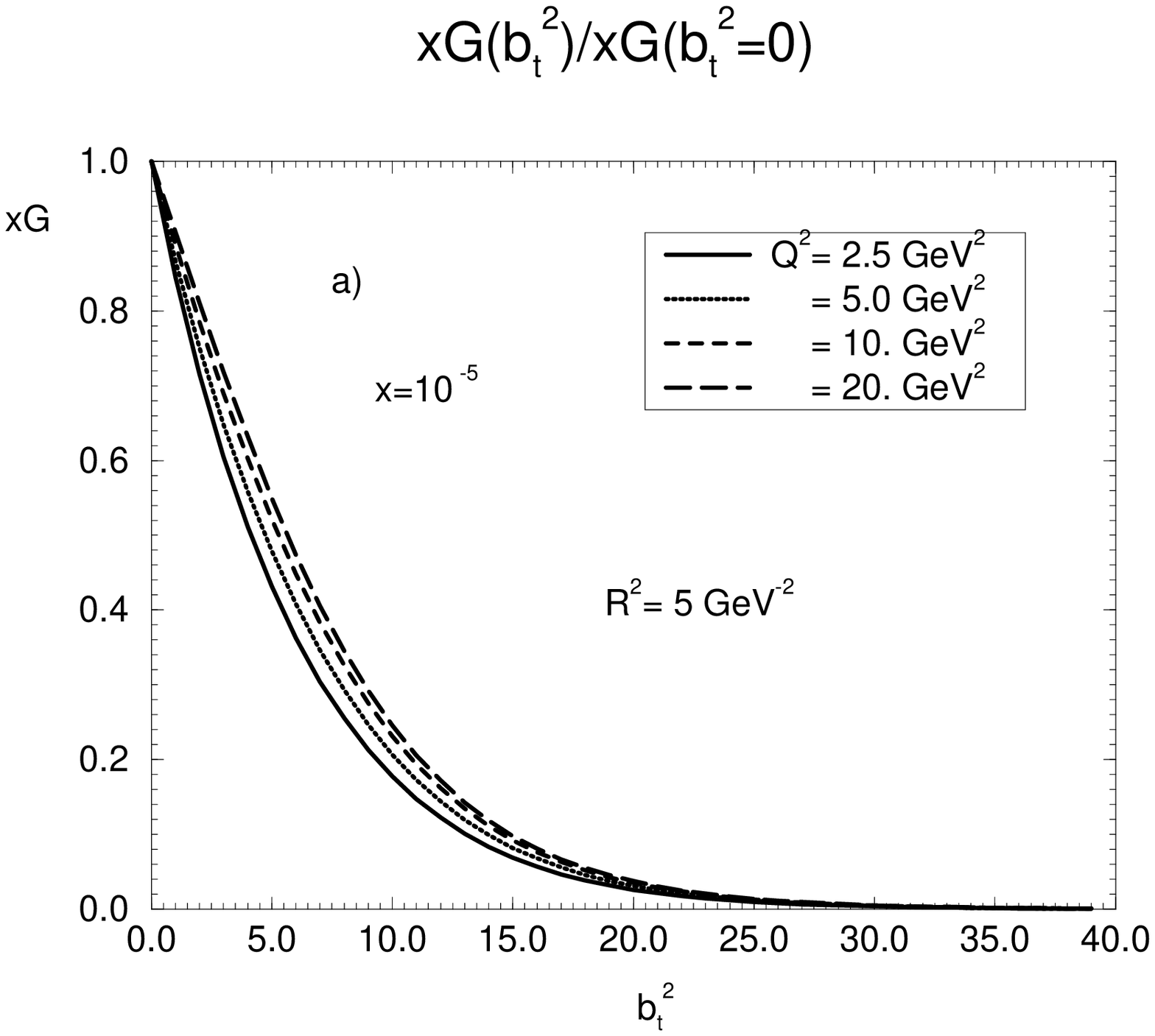,width=70mm}\\
\end{tabular}
\end{center}
\caption{\em  The values of $x G(x,Q^2,b_t)$ normalized at $b_t^2 \, =
\, 0$ as functions of $b_t$ for different values of $x$, $Q^2$ and $R^2\, =\, 5 \, GeV^{-2}$.}
\label{xbt}
\end{figure}

\begin{figure}[hptb]
\begin{center}
\begin{tabular}{c   c}
\psfig{file=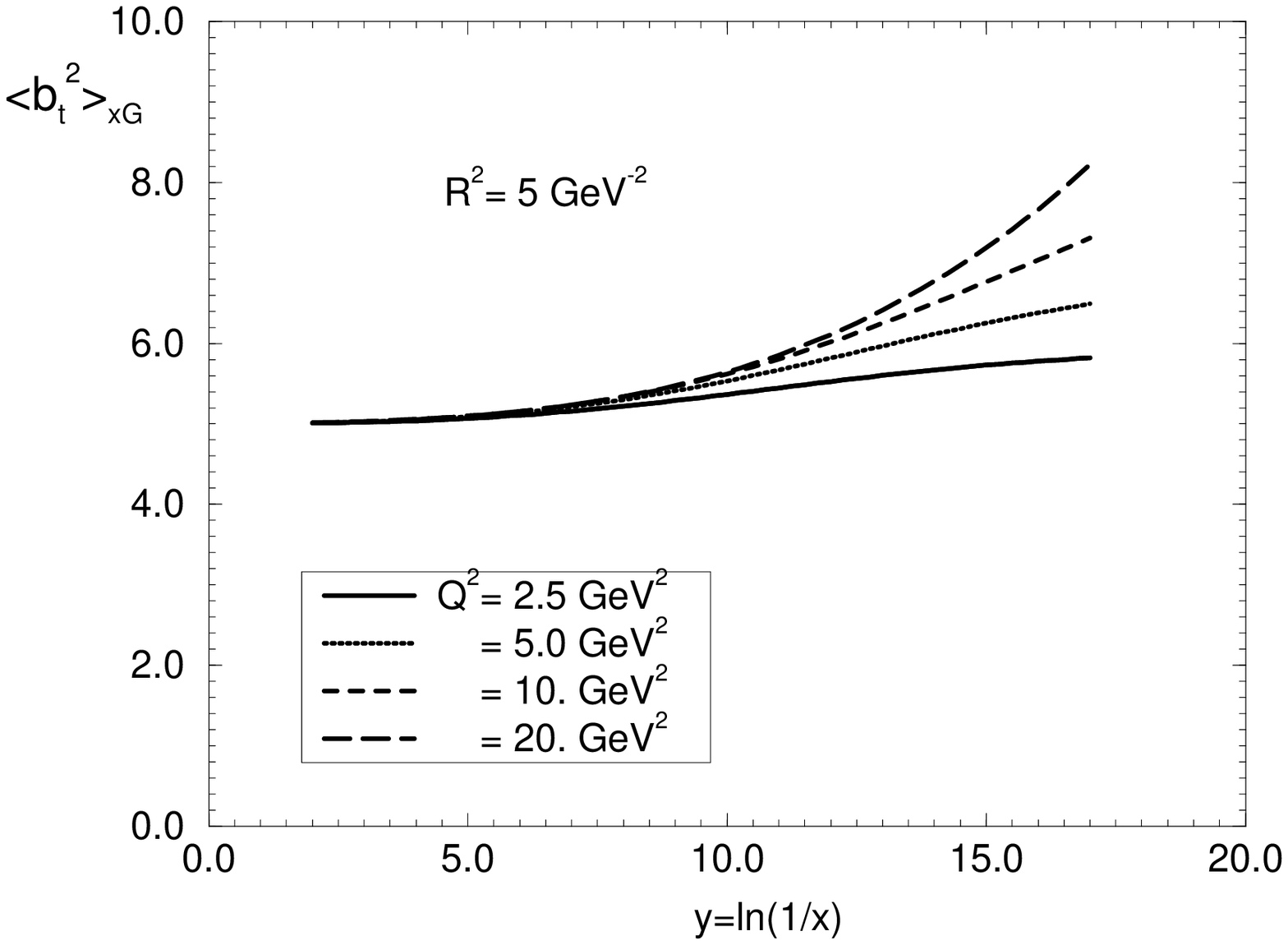,width=70mm} & \psfig{file=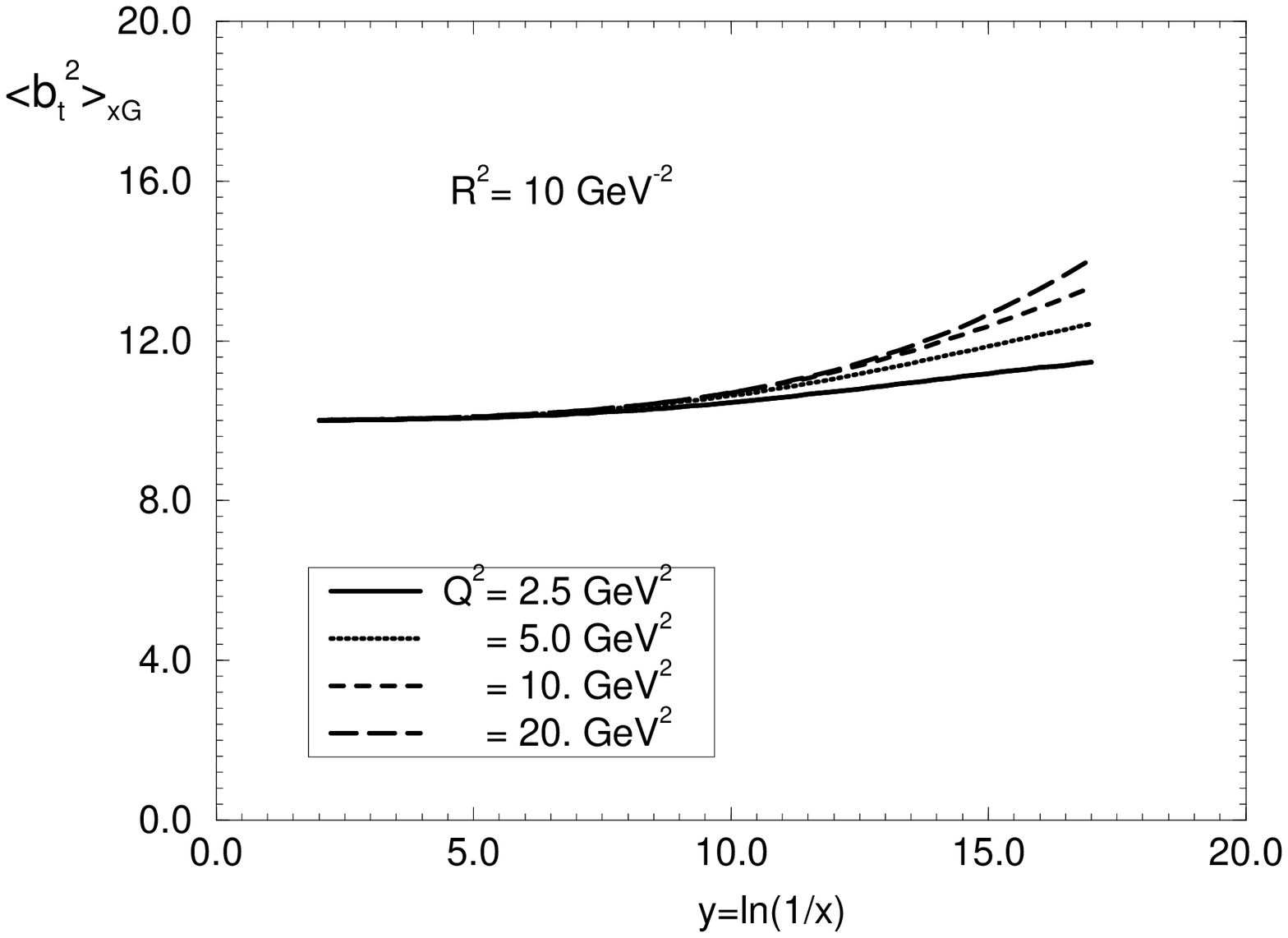,width=70mm}\\
\end{tabular}
\end{center}
\caption{\em  The mean radius of the interaction $< b_t^2 >$ for $xG$ 
(\protect\eq{b2g}).}
\label{b2}
\end{figure}

The  $b_t$ dependence of the gluon distribution is illustrated also
in 
Fig.\ref{xbt}, which shows $x G(x,Q^2,b_t)$ 
normalized to $b_t \, = \, 0$ at different values of $x$,
$Q^2$ and $R^2 \, = \, 5 \, GeV^{-2}$.
  Due to the SC, the width of the  $b_t$ - distribution
becomes bigger with respect to  the exponential
profile function, and presents 
an $x$ and $Q^2$ dependence. Comparing Fig.(\ref{xbt}a)
 and Fig.(\ref{xbt}b), we can
see that for smaller values of $x$, where SC are bigger, the $Q^2$
dependence
of the width is stronger, which leads to a broadening of the $b_t$ profile.
This effect reflects the $x$ and $Q^2$ dependence of 
the mean squared radius of the interaction $< b^2_t >$, defined by
\beq \label{b2g}
< b^2_t >\,\,=\,\,\frac{\int \,\,b^2_t\,\,d^2\,b_t\,\,xG(x,Q^2,b_t)}{\int\,\,d^2
\,b_t\,\,xG(x,Q^2,b_t)}\,\,=\,\,
\frac{\int \,\pi \,b^2_t\,\,d\,b^2_t\,\,xG(x,Q^2,b_t)}{xG(x,Q^2)} \, .
\eeq
 In Fig.\ref{b2}$, < b^2_t >$ is plotted  for  different values of $x$,
$Q^2$ and $R^2$.
Due to the SC, $< b^2_t >$  increases  as SC increase, i.e.,  
for $x\,\rightarrow\,0 $.  In HERA
kinematic region, even for 
$R^2 \, = \, 5 \, GeV^{-2}$, $< b^2_t >$ increases smoothly. 

%However, for $y \, > \, 10$ and 
%$R^2 \, = \, 5 \, GeV^{-2}$, 
 %$< b^2_t >$ increases very fast. For $R^2 \, = \, 10 \, GeV^{-2}$,
%this effect is less pronounced.

This result should be compared with the expectations for the soft 
Pomeron exchange.  For the soft Pomeron, the slope $B$ is equal to $B \,\,=\,\,
B_0 \,\,+\,\,\alpha'_P\,\,\ln(1/x)$, where $\alpha'_P$ does not depend 
on $Q^2$ being $\alpha'_P\,=\,$ 0.25 $GeV^{-2}$ \cite{SOFTPOMERON}.
We parameterize the behaviour of $< b^2_t >$ 
as a straightline for $5 \, < \, y \, < \, 10$,
$ < b^2_t > \,\,=\,\,4B( y\,=\, 5 ) \,\,+\,\,4\alpha'_{eff}\,\,\ln(1/x)$.
From  Fig.\ref{b2}, we
can conclude that $\alpha'_{eff}$ shows a $Q^2$ dependence   unlike
for 
the soft Pomeron and its value is rather small both for $R^2 \, = \, 5\, 
GeV^{-2}$ and $R^2 \, = \, 10\, 
GeV^{-2}$. Indeed, $\alpha'_{eff}\,\,\approx\,\,0.015 
\,GeV^{-2}\,\,\approx\,\,\frac{1}{16}\,\alpha'_P$ 
for $Q^2$ = 2.5 $GeV^2$ and  $\alpha'_{eff}\,\,\approx\,\,0.028 
\,GeV^{-2}\,\,\approx\,\,\frac{1}{9}\,\alpha'_P$ 
for $Q^2$ = 20 $GeV^2$ and    $R^2$ = 5 $ GeV^{-2}$. This 
straight line approximation
is not valid if we go beyond HERA kinematic region.

\subsection{ The $b_t$ dependence of the eikonal $F_2$ structure function.}

Here we will investigate the $x$ and $Q^2$ dependence of the slope $B$
as in the previous subsection, 
 for the $F_2$ structure function.
We define the modified $b_t$ dependent $F_2$ structure 
function as \bea
F_2 ( x,Q^2,b_t) = F_2^{SC} ( x,Q^2,b_t) - F_2^{BORN} ( x,Q^2,b_t)
 + F_2^{DGLAP} ( x,Q^2,b_t)\, .
\label{f21b2}
\eea
Each term in the above expression obeys the normalization condition
\bea
F_2 ( x,Q^2)= \int d^2 \, b_t \, F_2 ( x,Q^2,b_t) \, .
\label{f2f2bt}
\eea
The first term in Eq.(\ref{f21b2}) is the $b_t$ dependent Mueller formula
\bea
 F_2^{SC} ( x,Q^2,b_t) =\frac{1}{3 \pi^3}\int^{Q^2}_{Q_0^2}
d Q'^2 \{ 1 - e^{- \, \frac{1}{2} \, \Omega(x, Q'^2, b_t )} \} \, ,
\eea
where \cite{FRST}
\bea
\Omega(x, Q'^2, b_t ) = \frac{4 \as  \pi^2}{3 Q'^2}
xG (x, Q'^2 ) S(b_t ) \, ,
\eea
and $S(b_t )$ is the profile function.
The Born term for the  SC is written as
\bea
F_2^{BORN}(x, Q^2, b_t ) = \frac{2}{9 \pi}\, \sum^{N_f}_{1}\,\as\,Z^2_f\,\,
\int^{\ln Q^2}_{\ln Q^2_0} \,
d(\ln Q'^2)   x G^{DGLAP}(x, Q'^2)\,S(b_t ) \,\, .
\label{f2bbt}
\eea
The $b_t$ dependent DGLAP evolved $F_2$ is given by (see \eq{BAP})
\bea
 F_2^{DGLAP} ( x,Q^2,b_t)= F_2^{DGLAP} (x, Q^2) S(b_t ) \, ,
\eea
where $F_2^{DGLAP} (x, Q^2)$ is calculated from the GRV distributions for
valence, sea, and charm components as
already discussed in the $F_2$ section (\eq{f2grv}). All the above
expressions for $F_2$ fullfill the normalization (\ref{f2f2bt}).

\begin{figure}[hptb]
\begin{center}
\begin{tabular}{c   c}
\psfig{file=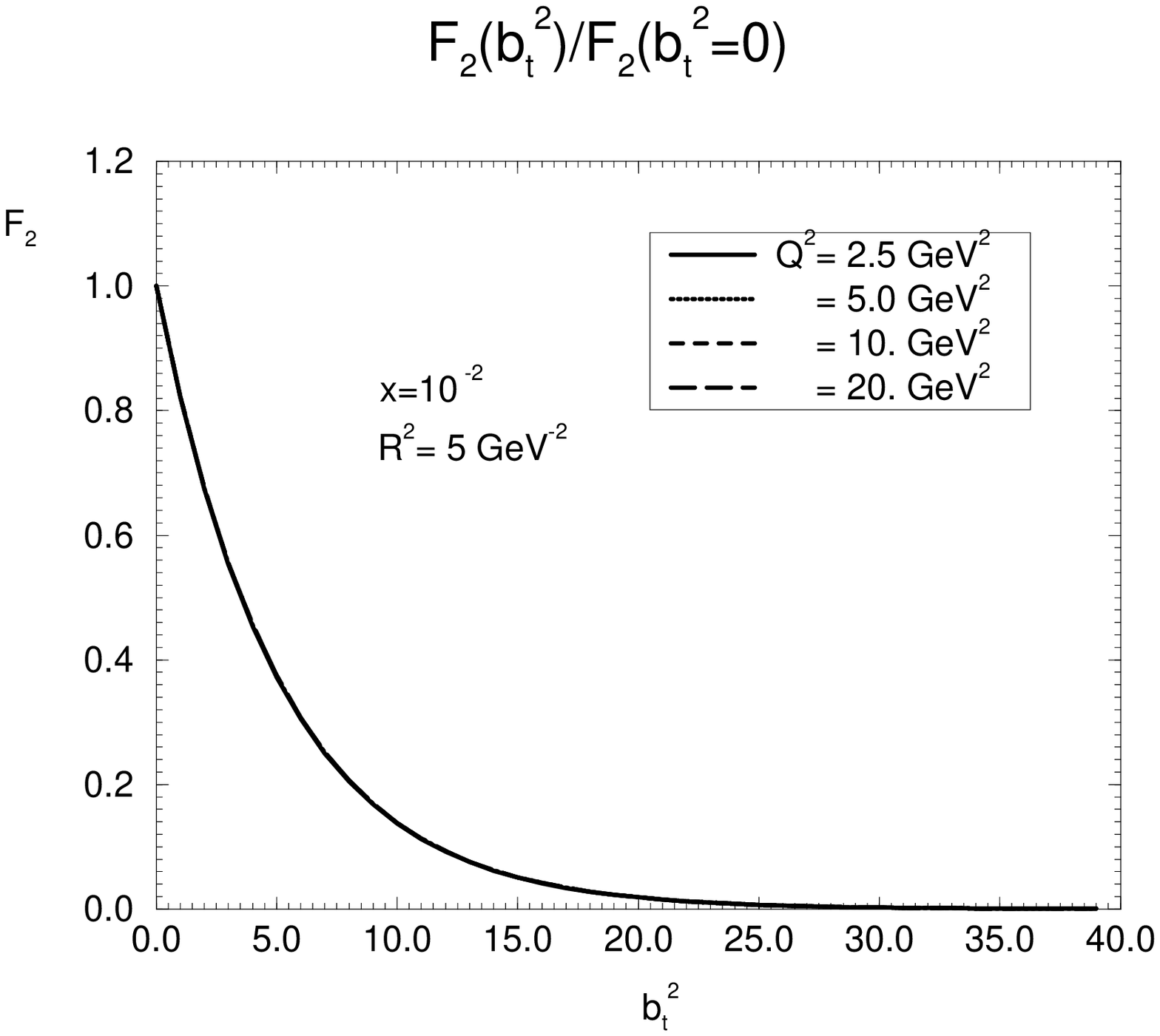,width=70mm} & \psfig{file=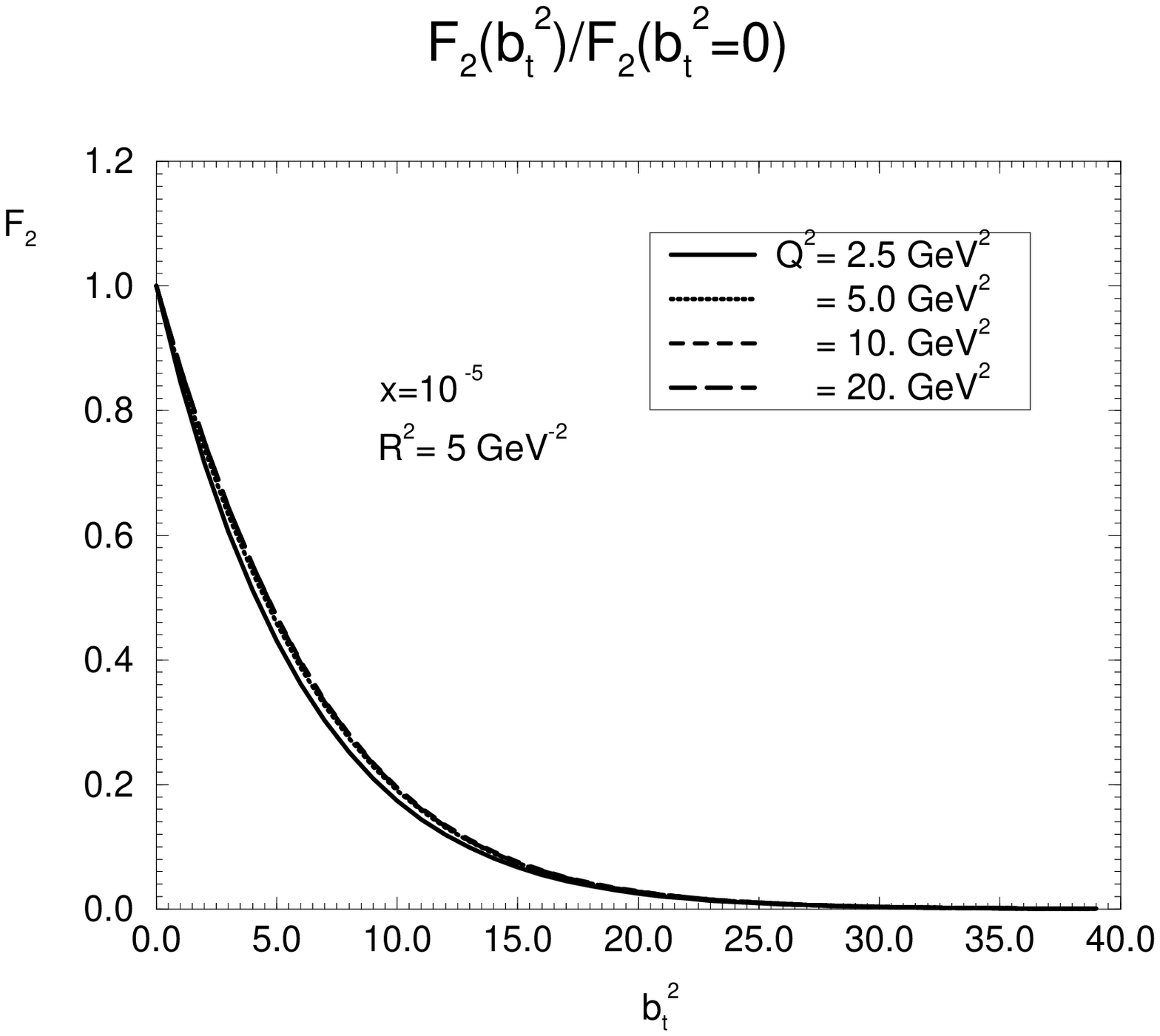,width=70mm}\\
\end{tabular}
\end{center}
\caption{\em  The values of $F_2 (b_t^2 )$ normalized at $b_t^2 \, =
\, 0$.}
\label{f2bt}
\end{figure}

In Fig.\ref{f2bt} we show the $b_t$ dependence of $F_2 (x,Q^2,b_t)$ 
normalized to $b_t \, = \, 0$ at different values of $x$,
$Q^2$ and $R^2 = 5 \, GeV^{-2}$. At low values of $x$, the SC lead to
a broadening of the $b_t$ profile, but the effect is less pronounced
when compared to the $b_t$ profile of the gluon distribution. 

\begin{figure}[hptb]
\begin{center}
\begin{tabular}{c   c}
\psfig{file=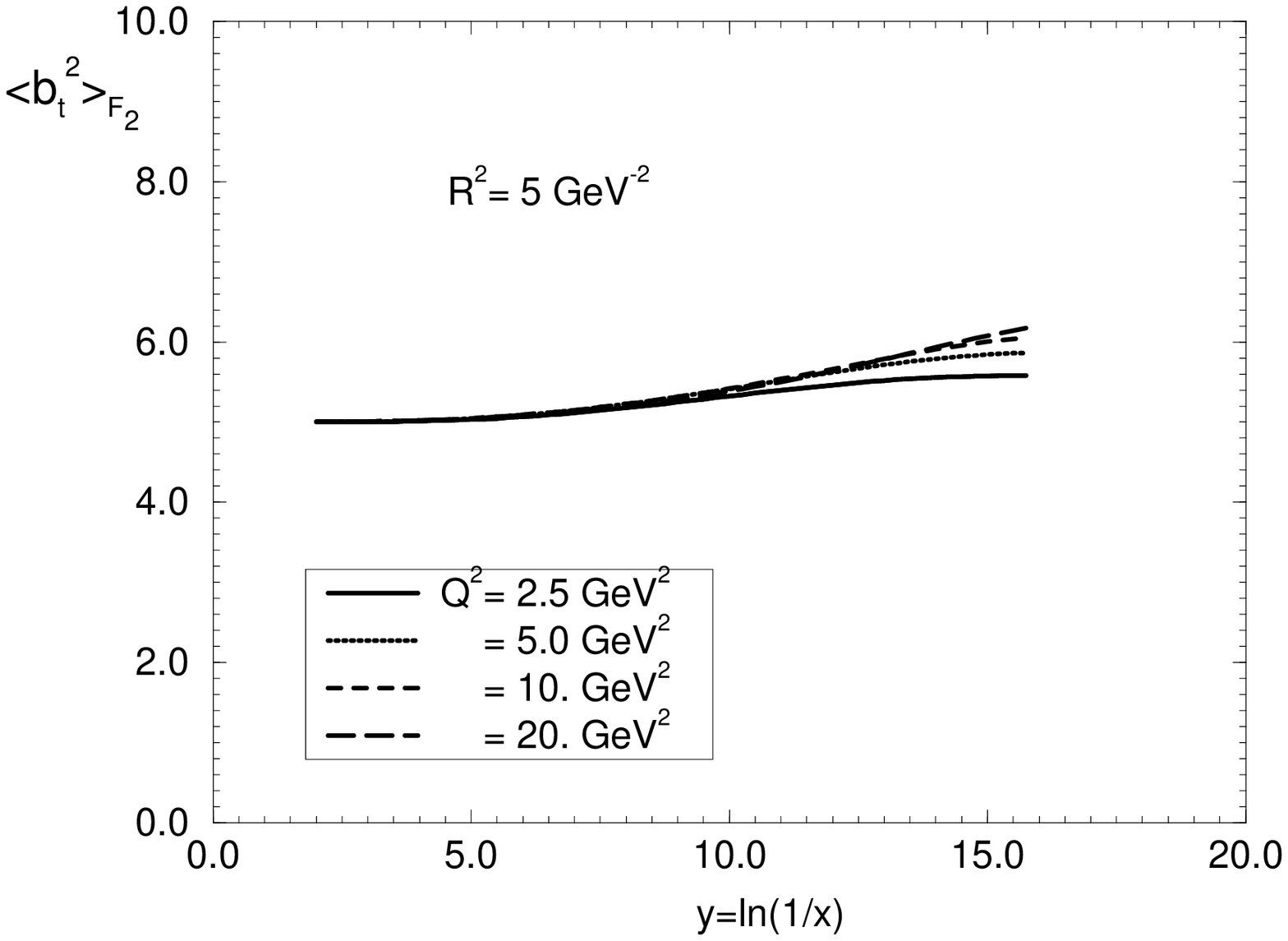,width=70mm} & \psfig{file=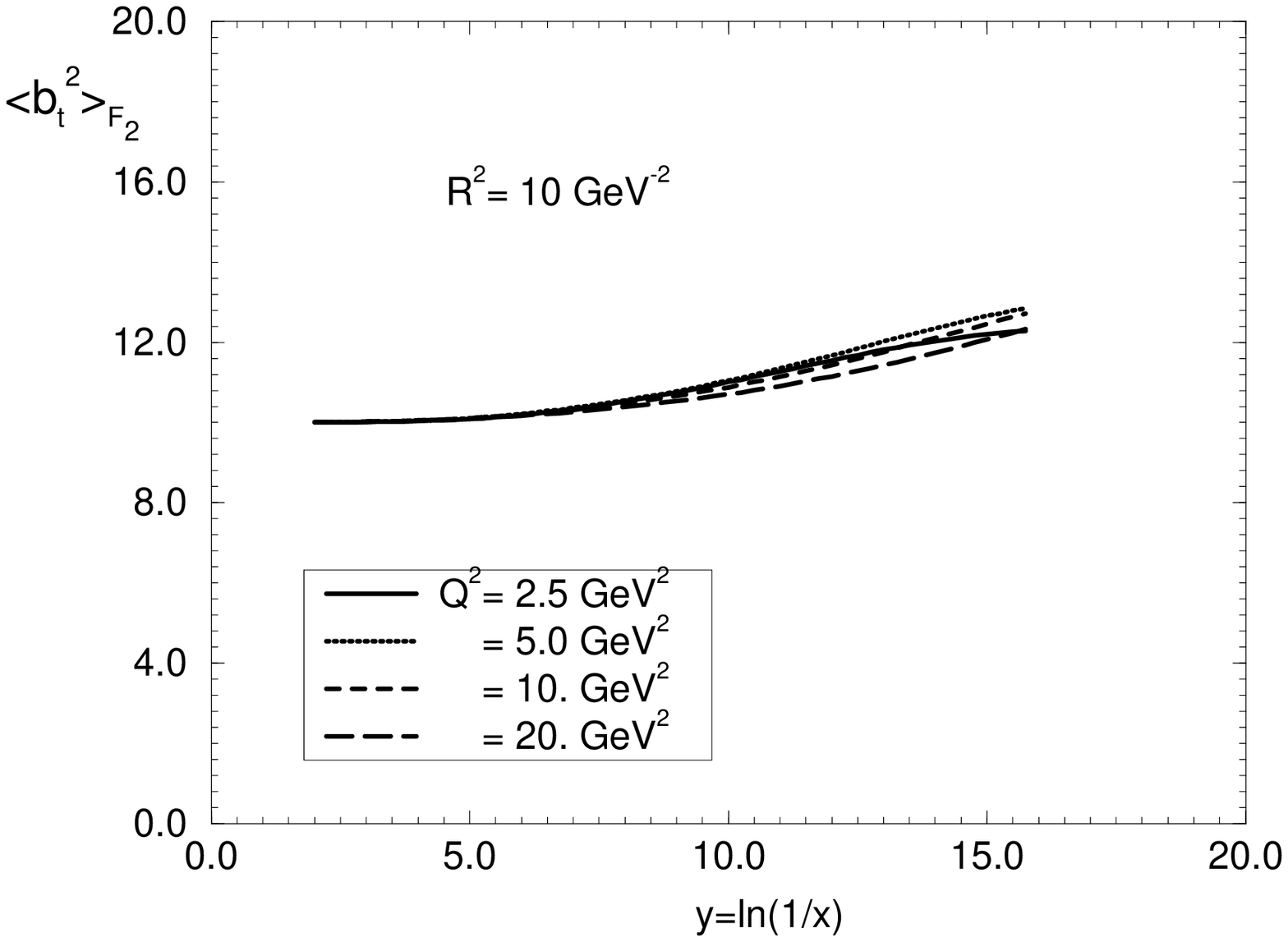,width=70mm}\\
\end{tabular}
\end{center}
\caption{\em  The mean radius of the interaction $< b_t^2 >$ calculated using
$F_2$(\protect\eq{f21b2} as a weight function 
( \protect\eq{btm}).}
\label{f2btm}
\end{figure}

We calculate also the mean radius $< \, b_t^2 \, >$ 
for the $F_2$ structure function,
which is given by  the expression
\bea
< \, b_t^2 \, > = \frac{\int  d \, b_t^2 \,\, b_t^2 \,\,
 F_2 ( x,Q^2,b_t^2)}
{\int d\, b_t^2 \,\, F_2 ( x,Q^2,b_t^2)} \, .
\label{btm}
\eea
 The results are  plotted in Fig.\ref{f2btm}. 
One can see
   that the effect of the SC
on the value of $< b^2_t >$ is smaller
in the  $F_2$ case then in  the $xG$ case.  
From Fig. 
\ref{f2btm} one can calculate the value of $\alpha'_{eff}$ which is equal
$\alpha'_{eff}\,\,\approx\,\,0.015 \,GeV^{-2}\,\,\approx \,\,\frac{1}{16} 
\alpha'_P$ for $Q^2 \,=\,2.5\, GeV^2$ and  
$\alpha'_{eff}\,\,\approx\,\,0.017 \,GeV^{-2}$ for  $Q^2 \,=\,20\, GeV^2$.
Comparing with $\alpha'_{eff}$ calculated from the gluon distribution, we
conclude that $\alpha'_{eff}$ for $F_2$ presents a less pronouced $Q^2$
dependence.  

\subsection{Energy sum rules.}
It is well known that the deep inelastic structure functions should obey
the energy sum rules which look as follows
\beq \label{E1}
\int^1_0 \,d x \{\,x \Sigma(x,Q^2) \,\,+\,\,xG(x,Q^2)\,\}\,\,=\,\,1\,\,,
\eeq
where $xG$ is the gluon structure function and $\Sigma(x,Q^2) \,\,=\,\,
\sum_f\,[\, q_i(x,Q^2)\,\,+\,\,\bar q(x,Q^2)\,]$ where $q $ ( $\bar q $ )
denotes quark ( antiquark) structure function.

\begin{figure}[hptb]
\begin{center}
\psfig{file=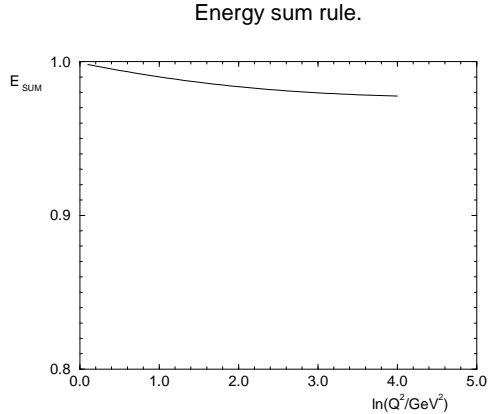,width=70mm}
\end{center}
\caption{\em The energy sum rule calculated from expression  \protect\eq{E1},
with the quark and gluon distribution given by the modified Mueller formula.}
\label{E}
\end{figure}

The SC in our approach violate the energy sum rules, since we heavily used
in the Mueller formula the leading $ln (1/x)$ approximation of perturbative
QCD in which the recoiled energy was neglected. However, the main
contribution to \eq{E1} comes from the value of $x$ which are not small
($x\,\,\approx\,0.5$) where we expect that the SC are
negligible. Since we took into account the full DGLAP evolution equations in the
first term of the modified Mueller formula we expect that this formula
will give us a small correction to \eq{E1}. In Fig.\ref{E}
we plot the energy sum rules as a function of $Q^2$ with quark and gluon
distributions calculated from the modified Mueller formula. One can see that 
the corrections to the energy sum rules is really small and can be
neglected. The corrections go from zero to $2 \, \%$ as $ln( Q^2/GeV^2)$ goes 
from the initial value $ln (Q_0)$ from $1$ to $4$.

\subsection{Where is the BFKL Pomeron?}

In this section  we address the question why we do not see 
any manifestation of the BFKL Pomeron \cite{BFKL}, since   the value of 
the anomalous dimension ($\g$)  has reached the
value $\frac{1}{2}$. First, let us summarize what we have learned about
the BFKL dynamics during the last two  decades.

1.In the BFKL equation all terms of the order of $( \frac{\as}{\o})^n$
have been taken into account. They generate the BFKL anomalous dimension
of the form \cite{BFKL},\cite{JAR}:
\beq \label{BFKL1}
\g^{BFKL}(\o)\,=\,\frac{\as N_c}{\pi} \frac{1}{\o}\,+\,\frac{2 \as^4
N^4_c \zeta(3)}{\pi^4} \frac{1}{\o^4}\,+\,\sum^{\infty}_{n=5}\,c_n
(\frac{\as}{\o})^n\,\,|_{\o\,\rightarrow\,\o_L}\,\rightarrow\,\frac{1}{2}\,
+\,\sqrt{\frac{\o_L - \o}{\Delta}}\,\,,
\eeq
where $\g^{BFKL}(\o = \o_l)\,=\,\frac{1}{2}$. From \eq{BFKL1} one can see
that the anomalous dimension does not exceed 
 much the value $\frac{1}{2}$ and the
second term in \eq{BFKL1} gives rise to the diffusion in log of
transverse momenta.

2. The Monte Carlo simulation of the BFKL dynamics as well as the
semiclassical solution of the evolution equation (see Ref.\cite{MW})
shows that we can see the difference between the BFKL and the DGLAP
evolutions only for the photon virtualities $Q^2$ which are close to the
initial one ( $Q^2_0$).

3. The BFKL diffusion leads to the breakdown of the operator product
expansion \cite{MU96} in the region of small $x$. We can trust the BFKL
equation only in a limited region of $x$, namely \cite{MU96}
\beq \label{BFKL2}
\ln\frac{x_0}{x}\,\,\leq\,c\as(Q^2_0)^{-3}
\eeq
with sufficiently small numeric coefficient $c$ (see Ref.\cite{MU96} for
details ). The attempts to take into account the running $\as$ in the BFKL
equation give even more restrictive bound \cite{LEREN}, namely
$$
\ln\frac{x_0}{x}\,\,\leq\,c\as(Q^2_0)^{-\frac{5}{3}}.
$$
Therefore, we expect that the BFKL evolution can be visible in a limited
range of $x$ at $Q^2\,\rightarrow \,Q^2_0$.

On the other hand the SC break the operator product expansion at any
value of $Q^2$ in the wide region of $x$, namely \cite{GLR}
\beq \label{BFKL3}
\ln\frac{x_0}{x}\,\,\geq\,c'\as(Q^2)^{-2}\,\,.
\eeq
This inequality can be derived  from the Mueller formula and
corresponds to  $\kappa\,\geq\,1$.

Our answer to the question formulated in the beginning of this section is
as follows. The SC turn out to be more important than the BFKL
resummation in the whole kinematic region at $x\,<\,10^{-2}$. Actually, one 
can conclude this just looking at Fig.1 and Fig.2,  since $<\g > \,\rightarrow\,\frac{1}{2}$
in the kinematic region where $\kappa\,>\,$ 1.

However, to come to more definite conclusion it is very instructive to notice
 ( see  Fig. \ref{gmn}), that the SC in the eikonal
approach reduce  considerably the value of $< \g >$. Even at $R^2 = 10 \, \,
GeV^{-2}$,  $<\g>$ becomes smaller than $\frac{1}{2}$ for any value of $x <
10^{-2}$ at $Q^2 \,>\,2.5\, GeV^2$.
 The situation becomes more pronounced at $b_t $= 0 ( see
Fig.\ref{rb0}). Indeed, for $b_t$ = 0 $, <\g>$ turns out to be smaller
 than $\frac{1}{2}$ for any value of $Q^2$ and $x$.
 However, we would like to recall that our calculation of
the SC in Glauber - Mueller approach depends on the hypothesis that we made
on the large distance contribution to the Mueller formula ( see \eq{MF}).
In Figs. \ref{gmn} and \ref{rb0} we calculated the SC originated from
distances $r_{\perp}\,\leq\,1 GeV^{-1}$.  To illustrate that our conclusion
on the BFKL contribution does not depend on large distances contribution
we calculate the function that gives also the average anomalous
dimension when the SC are small,
\beq
<\gamma > = \frac{1}{xG^{GRV}} \frac{\pa xG^{MF}}{\pa ln(Q^2)}\,\, .
\label{gams}
\eeq

\begin{figure}[hptb]
\begin{center}
\psfig{file=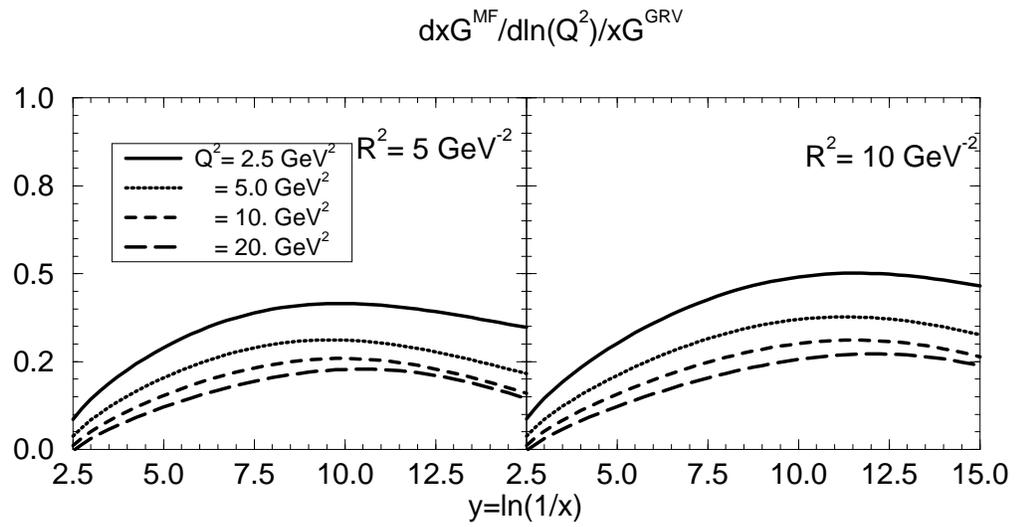,width=140mm}
\end{center}
\caption{\em The average anomalous dimension for small SC
(\protect\eq{gams}).}
\label{bfk1}
\end{figure}

The SC contribute to this
function only at small distances ($r_{\perp} \,\approx \,\frac{1}{Q}$)
( see \eq{MF} and the discussion in section 4.1) and, therefore, we
can calculate them  without any uncertainty related to the unknown ``soft"
processes. One can see in Figs. \ref{bfk1} that this
function turns out to be smaller then $\frac{1}{2}$. This fact supports our
conclusion that the BFKL Pomeron cannot be seen and it 
is hidden under large SC.  
 We want also to add to the previous
discussion that we do not believe  that $R^2 = 10\,\, GeV^{-2}$ is a correct
radius for a proton as we have discussed in the introduction.

It should be stressed that the SC to the  deep inelastic structure
functions are the smallest ones since in other processes we expect smaller
values for the radius. For example, in the inclusive production of high
transverse jet in DIS ( so called hot spots hunting \cite{HOTSPOT})  with
transverse momentum of the jet of the order of Q, $
R\,\approx\,\frac{1}{Q}$ and we expect that the SC will be more important
than the BFKL emission. Therefore, we think that the chance to
see the BFKL Pomeron is ruled out, following this formalism,
even in this specially suited process for the BFKL
contribution. We have to look for other processes to recover the BFKL
dynamics.

\section{ First corrections to  the Glauber-Mueller Approach.}

\label{cgf}
In this section we discuss  the corrections to the Glauber approach
 (the  Mueller formula of \eq{FINANS}.
To understand how big could be the corrections to the Glauber approach
 we calculate the second iteration of the Mueller formula of \eq{FINANS}. As
has been discussed, \eq{FINANS} describes the rescattering of the fastest gluon
( gluon - gluon pair ) during the passage through a nucleon. 
In the second iteration we take into account also
the rescattering of the next to the fastest gluon. This is a well defined task
 due to the strong ordering in the parton fractions of energy in the
parton cascade in leading $ln (1/x)$ approximation of pQCD that we are 
dealing with. Namely
\beq \label{95}
x_B\,\,<\,\,x_n\,\,<\,\,...\,\,<x_1\,\,<\,\,1\,\, ,
\eeq
where 1 corresponds to the fastest parton in the cascade.

In the first iteration, we evaluate the SC inserting the gluon distribution 
$x G^{DGLAP}(x,Q^2)$ into  the
 modified Mueller's formula, \eq{FINANS}. For the second iteration we substitute the correction term 
of the first iteration
\beq \label{96}
x \,G\,\,=
\,\,x G^1(x, Q^2) (\eq{FINANS} )\,\,-\,\,x\, G^{DGLAP}( x,Q^2)\,\,;
\eeq
in \eq{FINANS} and add the first iteration to the result.

%\begin{figure}[htbp]
%\centerline{\epsfig{figure=Fig5np.eps,height=100mm}}
%\caption{\em  The interaction with nucleon that is  taken into
%account in the second iteration of 
%M ueller formula. }
%\label{Fig.5}
%\end{figure}
The substraction of $ x G^{DGLAP} $ in \eq{96} means that, doing 
the second iteration, we take into account only two or more  
interactions of the next to the fastest gluon with the target nucleon. 
We have to do such a substraction to avoid double counting since the one 
iteration of all gluons with the target has been included in the DGLAP 
evolution or in other words in the
Born term of our approach ( see Ref.\cite{AGL} for details).

\begin{figure}
\begin{center}
\begin{tabular}{c c}
\epsfig{file=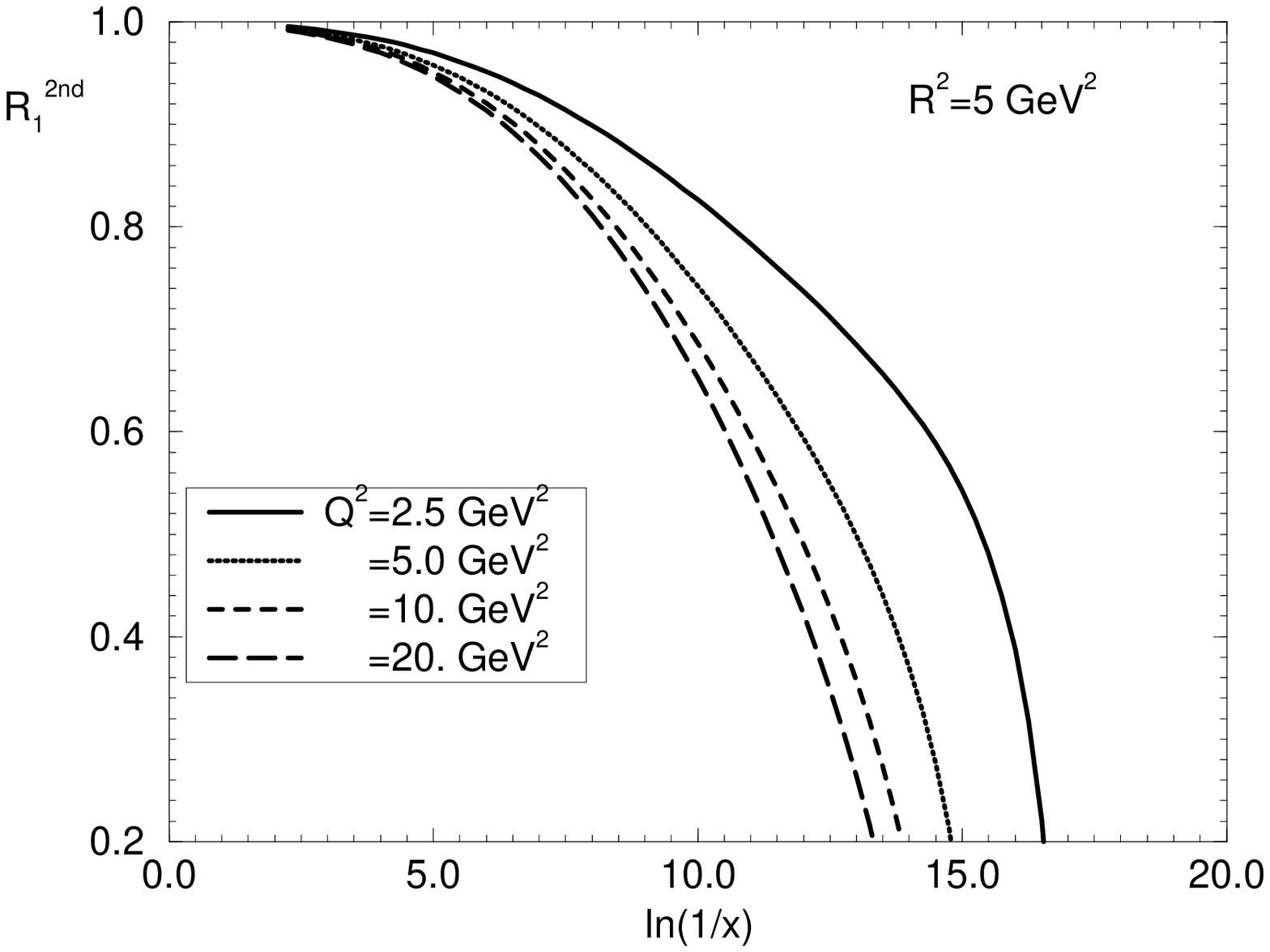,width=80mm}& \epsfig{file=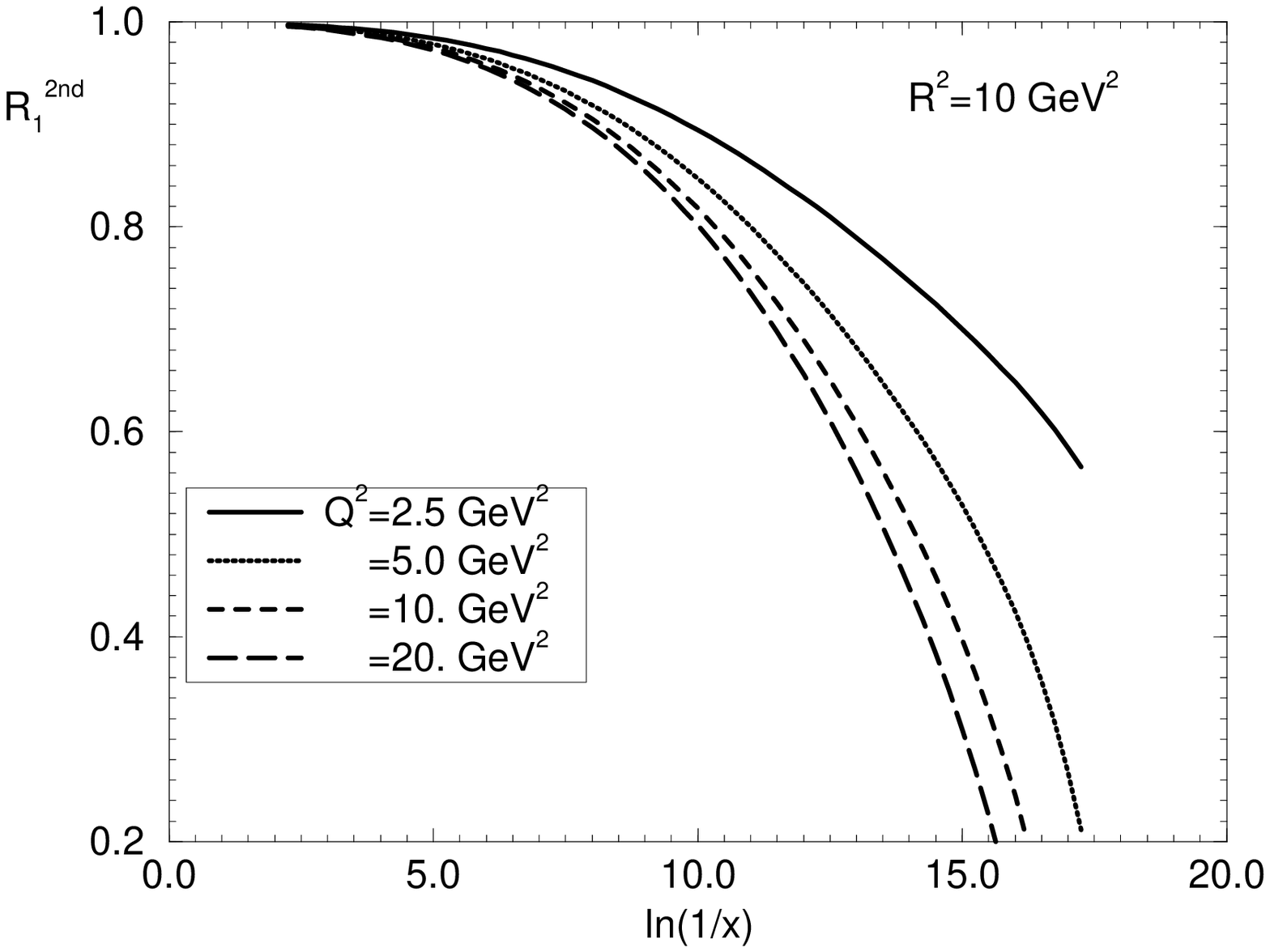,width=80mm}\\
\epsfig{file=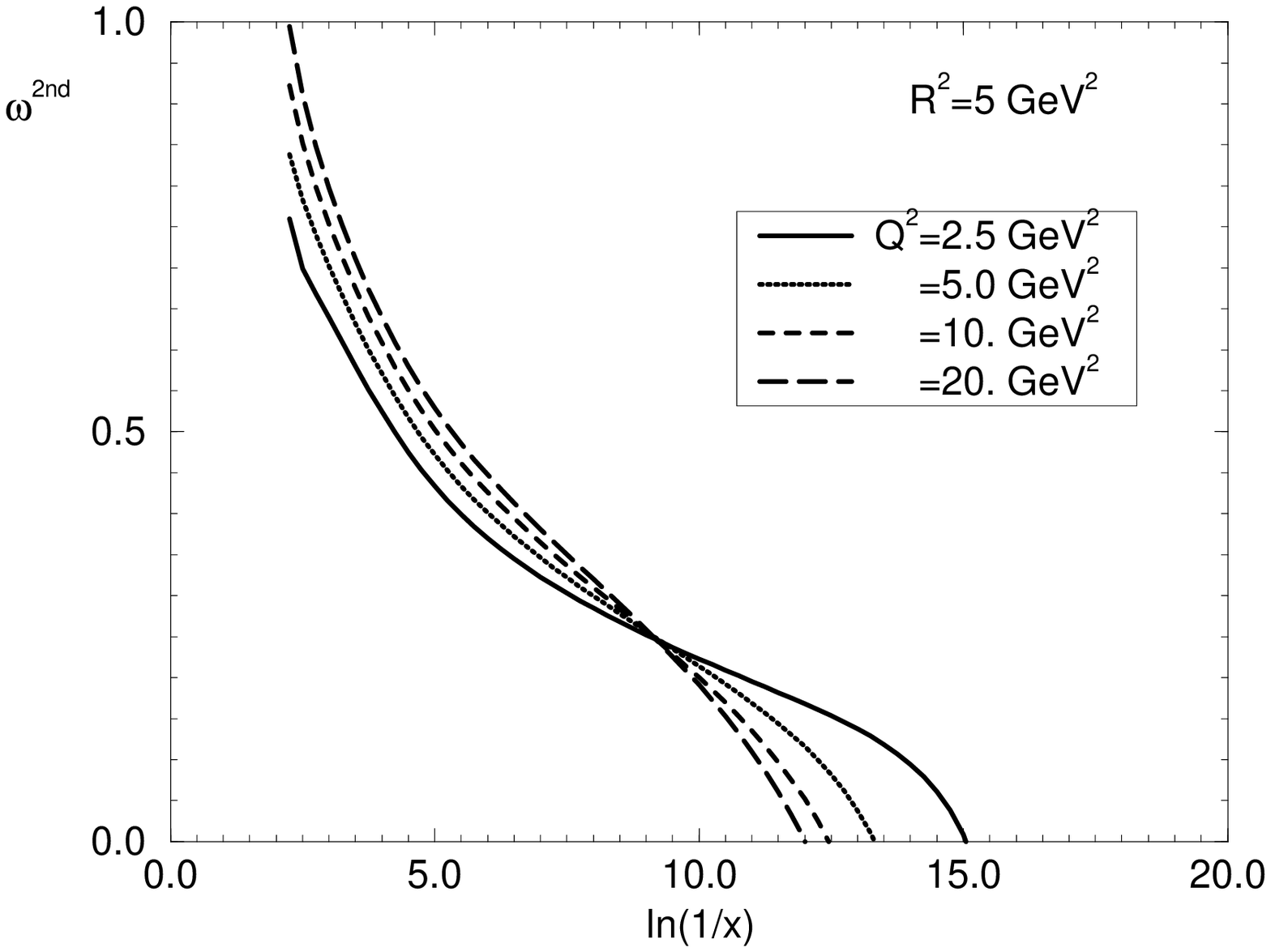,width=80mm} &\epsfig{file=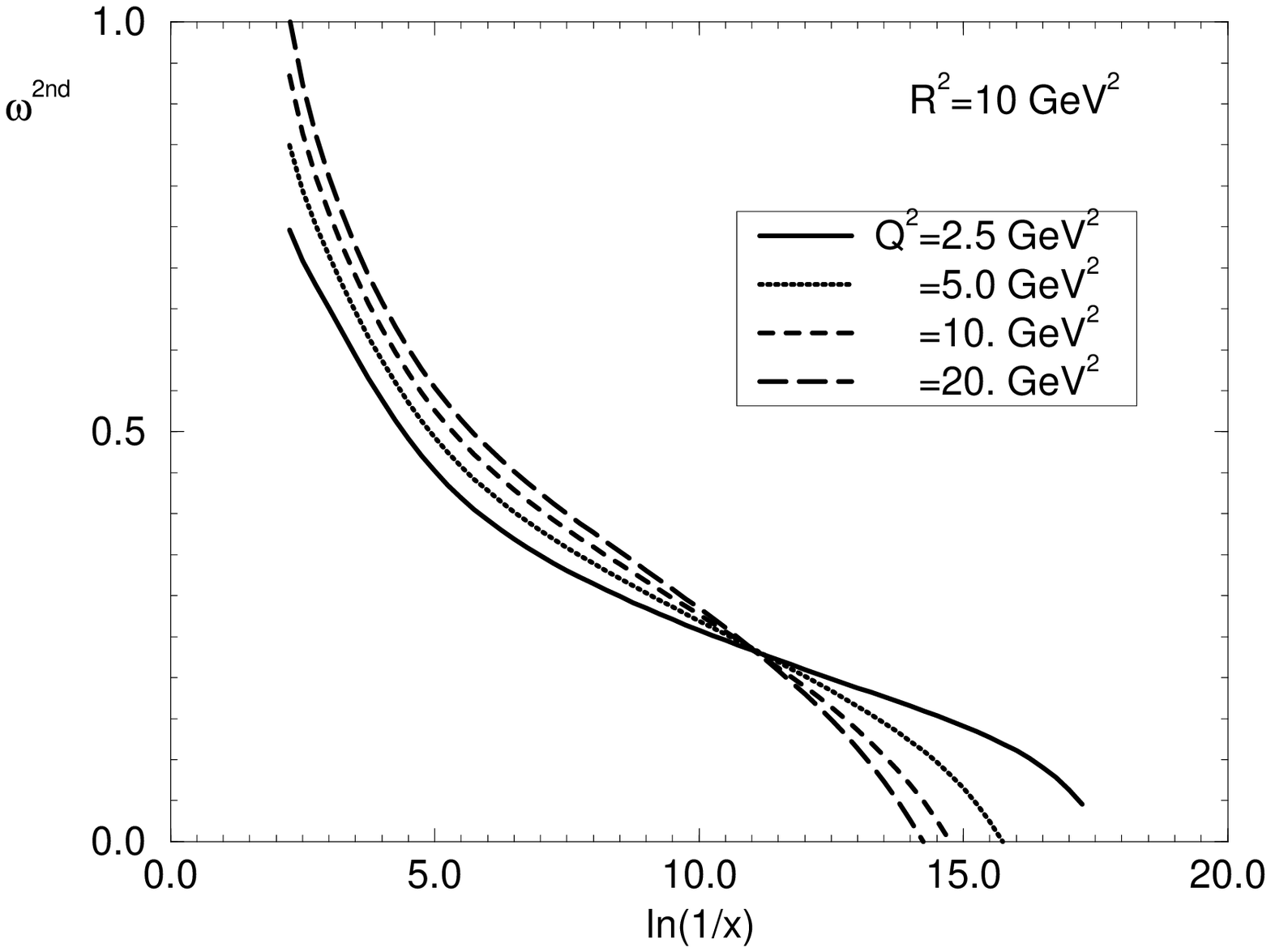,width=80mm} \\
\epsfig{file=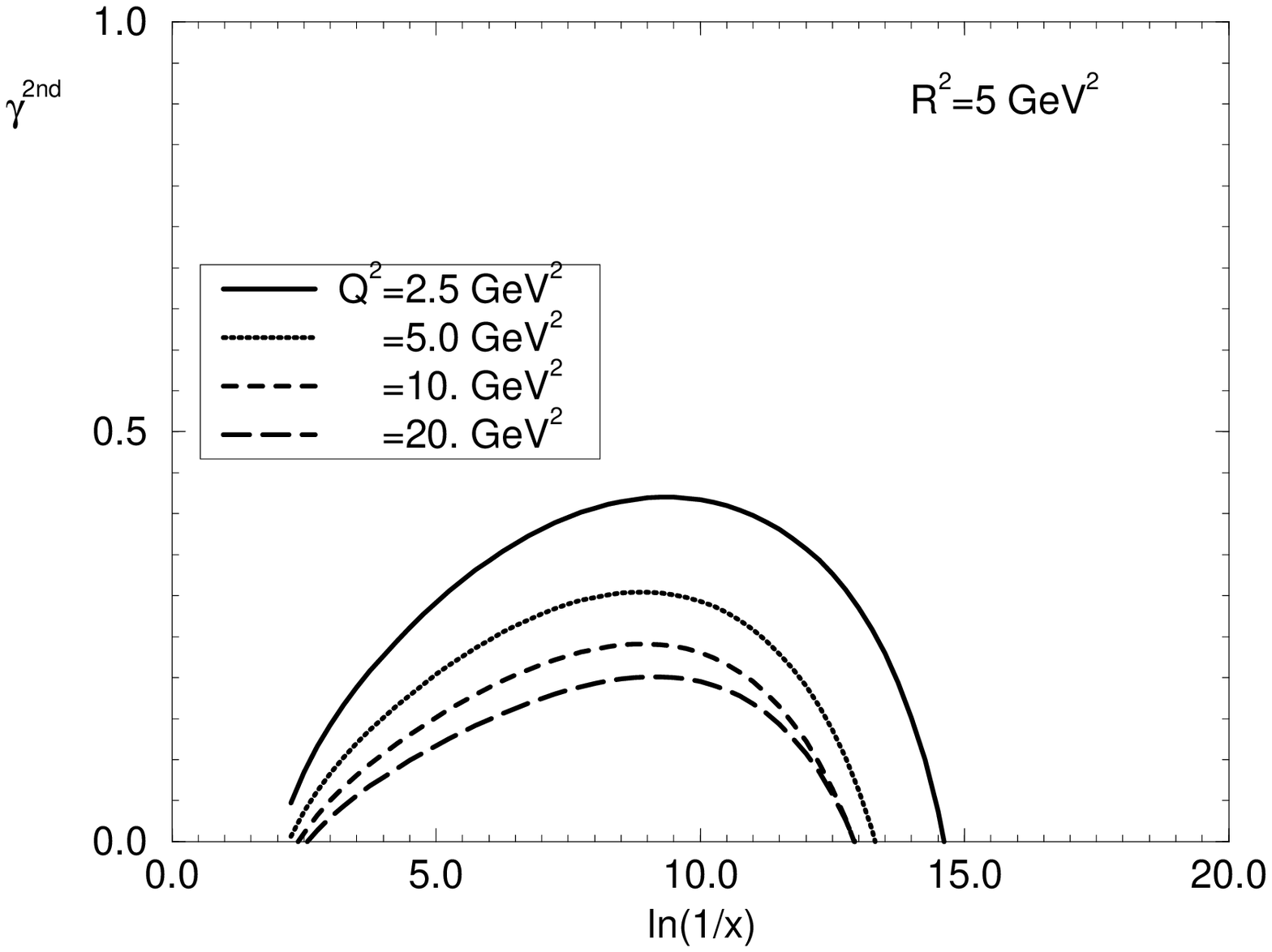,width=80mm} &\epsfig{file=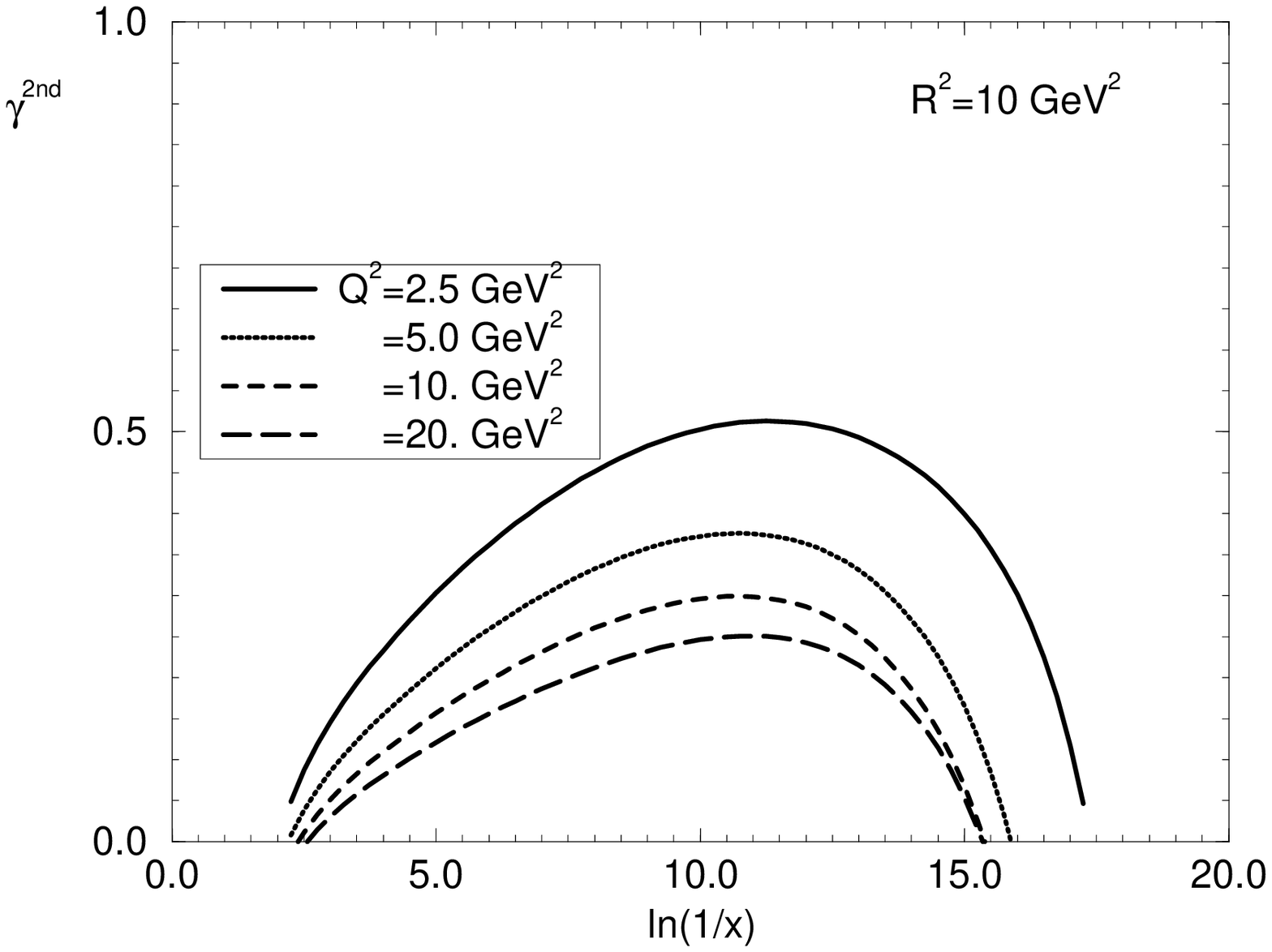,width=80mm} \\
\end{tabular}
\caption{{\em Second  
iteration calculations of $R_1$, $<\omega >$ and $<\gamma >$ as 
function of $y=ln(1/x)$ and $Q^2$ for two values of $R^2$. }}
\end{center}
\label{2ndit}
\end{figure}

Fig.\ref{2ndit} shows the second iteraction calculations for
 our three standard  observables: $R_1$, $< \omega >$
and $<\gamma >$. From the figure one can see
 that the second iteraction changes crucially
the values of these parameters. As it was already noted for the 
nuclear case\cite{AGL},
a remarkable feature
 is the crucial change of the value of the effective power $\omega(Q^2)
$ for the ``Pomeron" intercept which tends to zero at HERA kinematic region,
making possible the matching with ``soft" high energy phenomenology, even in 
the nucleon case.
It is also important to note that the second iteration makes
more pronounced all properties of the ratio $R_1$ and the anomalous dimension ( $<
\gamma>$). 

 However, all these features of the second iteration calculations 
we have to consider with many restrictions. Indeed, the only conclusion which
we can derive from our study is the fact that the correction to 
the Glauber (eikonal ) approach turns out to be large
in the region of small $x$ ( $x \,<\,10^{-2} $ ) 
 and becomes of the order of the first iteration for $x < 10^{-3}$. 
As it has been already discussed in \cite{GLR} \cite{AGL}, it occurs 
because the second
iteration gives a correction of the order
 of $\as ln(Q^2/Q_0^2) ln(1/x)$, which is close to unity  for the
DLA of pQCD ( for the DGLAP evolution  at small $x$).

Therefore, the iteration procedure cannot be an effective way to solve that
problem. We have to develop a 
different technique  to take into account rescatterings of  all the partons
 in the parton cascade which will be more efficient than the simple iteration
procedure for \eq{FINANS}. This is the subject of the next section.

\section{The general approach.}

\subsection{Why equation?}

We would like to suggest a new approach based on the new evolution equation
 to sum all SC. However, first of all we want to argue why an equation is
 better than any iteration procedure. To illustrate this point of view let us
differentiate the Mueller ( see \eq{MF} ) formula with respect to $y\,=\,\ln(1/x)$ and $
\xi\,=\,\ln Q^2$. It is easy to see that this derivative is equal to
\beq \label{DER}
\frac{\partial^2 x G(x, Q^2)}{\partial y\, \partial \xi}\,\,=\,\,
\frac{2 \, Q^2 }{\pi^2}\,\int d b^2_t \,\,\{\,\,1\,\,-\,\,e^{-\,\frac{1}{2}\,\s(x,
r^2_{\perp}\,=\,\frac{1}{Q^2})\,S(b^2_t)}\,\,\}\,\,.
\eeq
A nice property of \eq{DER} is that all quantities in \eq{DER} enter
at  small distances,
therefore it is under theoretical control. Of course,
 we cannot  avoid all difficulties
just changing  the solution procedure. Indeed, the nonperturbative
effects coming from the large distances are still important but they are all
 hidden in the boundary and initial conditions for the equation. Therefore,
an equation is a good ( correct ) way to separate  what we know ( small
distance contribution) from that we don't ( large distance contribution).

\subsection{ The generalized evolution equation.}

We suggest the following way to take into account the interaction of all
partons in a parton cascade with the target.
Let us differentiate the $b_t$-integrated Mueller
formula of \eq{14} in $y \, = \,\ln (1/x)$ and $ \xi = \ln(Q^2/Q^2_0)$.
 It gives
\beq \label{102}
\frac{\partial^2 x G ( y,\xi)}{\partial y \partial \xi}\,\,=\,\,
\frac{2 \,Q^2 R^2 }{ \pi^2}\,\,\,\left\{\,\, C \,\,+ \,\,ln(\kappa_{G} ( x', Q^2 )) \,\,+\,\,
E_1 (\kappa_{G} ( x', Q^2 ))  \right\} \,\,\,,
\eeq
where $\kappa_G (x,Q^2)$ is given by
\beq \label{KAPPA}
\kappa^{DGLAP}( x,Q^2) \,\,=\,\,\frac{N_c \as \pi }{2 Q^2 R^2 }\,x G^{DGLAP}(x,Q^2)\,\,.
\eeq
The expression (\ref{102}) can be rewritten in the form ( for fixed $\as$ )
\beq \label{103}
\frac{\partial^2 \kappa_G( y,\xi)}{\partial y \partial \xi}\,\,+\,\,
\frac{\partial \kappa_G(y, \xi)}{\partial y}\,\,=\,\,
\frac{ N_c\, \as}{\pi}\,\, \left\{\,\, C \,\,+ \,\,ln(\kappa_{G}) \,\,+\,\,
E_1 (\kappa_{G})  \right\} 
\,\,\equiv\,\,F(\kappa_G)\,\,.
 \eeq

Now, let us consider the expression of \eq{103} as an equation for  $\kappa_G$.
This equation has the following nice properties:

1.  It  sums all contributions of the order $ (\,\as\,y\,\xi\,)^n$ 
absorbing them in $x G (y,\xi)$, as well as all contributions of the order
of $\kappa^n$.
Therefore, this equation solves the old problem, formulated in Ref.\cite{GLR},
and 
 for $N_c\,\rightarrow \,\infty $ \eq{103} gives the complete
solution to our problem, summing all SC;

2 .The solution of this equation matches with the solution of the DGLAP
 evolution equation in the DLA of perturbative QCD at $\kappa\,\rightarrow \,0$;

3.  At small values of $\kappa$ ( $\kappa\,<\,1$ ) 
 \eq{103} gives the GLR equation. Indeed, 
for small $\kappa$ we can expand the r.h.s of \eq{102} keeping only the
 second term. Rewriting the equation through the gluon structure function
 we have
\beq \label{GLR}
\frac{\partial^2 x G( y,\xi)}{\partial y \partial \xi}\,\,=\,\,
\frac{\as N_c}{\pi}\,x G(x,Q^2)\,\,-\,\,\frac{\as^2 \,N^2_c}{8\,Q^2
\,R^2 }\,
( x G(x,Q^2))^2
\,\,,
\eeq
which is the GLR equation \cite{GLR} with 
 the coefficient in front of the second term 
 calculated by Mueller and Qiu \cite{MUQI}.

4. For $\as y \xi \,\approx\,1$ this equation gives the Glauber ( Mueller ) 
formula, that we have discussed in detail.

%5. This equation almost coincide with the equation that L.Mclerran
%with collaborators \cite{MCLER} for heavy ions derived from quite 
%different approach
%and with different technique. We are sure that {\it almost} will
%disappear when they will do more careful averaging over transverse 
%distances.

Therefore, the great advantage of this equation in comparison
 with the GLR one is the fact that it describes the region of large $\kappa$
and provides the correct matching both with the GLR equation and with the
Glauber ( Mueller ) formula.

%
%
%   BT DEPENDENT EQUATION
%   
%
%

We propose also an equation for the $b_t$-dependent gluon distribution.
 Let us consider the $b_t$-dependent Mueller
formula \eq{BMF}. Differentiating this expression
  in $y \, = \,\ln (1/x)$ and $ \xi = \ln(Q^2/Q^2_0)$ we obtain
\beq \label{1041}
\frac{\partial^2 x G ( y,\xi, b_t)}{\partial y \partial \xi}\,\,=\,\,
\frac{2 \,Q^2}{ \pi^3}\,\,\{\,\,1\,\,-\,\,e^{- \kappa_G(x,Q^2, b_t)}\,\,\}\,\,,
\eeq
where $\kappa_G (x,Q^2, b_t)$ is given by
\beq \label{BKAPPA}
\kappa_G^{DGLAP}( x,Q^2, b_t ) \,\,=\,\,\frac{N_c \as \pi^2 }{2 Q^2 }\,
x G^{DGLAP}(x,Q^2, b_t)\,\,.
\eeq
As it was discussed before, expression (\ref{1041}) can be rewritten in 
the form 
(for fixed $\as$)
\beq \label{1051}
\frac{\partial^2 \kappa_G( y,\xi, b_t)}{\partial y \partial \xi}\,\,+\,\,
\frac{\partial \kappa_G(y, \xi, b_t)}{\partial y}\,\,=\,\,
\frac{ N_c\, \as}{\pi}\,\,\,\{\,\,1 \,\,-\,\,e^{ - \kappa^{DGLAP}_G (x,Q^2,b_t)
}\,\,\}\,\,\equiv\,\,F_{b}(\kappa_G^{DGLAP})\,\,.
 \eeq
Now, let us consider the expression (\ref{1051})
 as the equation for  $\kappa_G$.
It means that we propose the following equation
\beq \label{NEQ}
\frac{\partial^2 \kappa_G( y,\xi, b_t)}{\partial y \partial \xi}\,\,+\,\,
\frac{\partial \kappa_G(y, \xi,b_t)}{\partial y}\,\,=\,\,
 \frac{N_c\, \as}{\pi}\,\,\,\{\,\,1 \,\,-\,\,e^{ - \kappa_G (x,Q^2,b_t)}\,\,\}
\,\,\equiv\,\,F_{b} (\kappa_G(x,Q^2,b_t))\,\,.
 \eeq
This expression is an equation for $\kappa ( y,\xi, b_t)$. The $b_t$-dependent
gluon distribution can be obtained from the solution of the above equation
using the expression (\ref{BKAPPA}).

 We claim that \eq{103} and \eq{NEQ} solve the problem of summing all Feynman 
diagrams in the following kinematic region:
\beq \label{PARNEQ1} 
\as \,\ln(1/x)\,\ln(Q^2/Q^2_0)\,\,\approx \,\,1 \, ,
\eeq 
\beq \label{PARNEQ2}
\as\,\ln(1/x)\,\,<\,\,1\,\,,\,\,\,\,\,\,\,\,\,\,\as\,\ln(Q^2/Q^2_0)\,\,<
\,\,1\,\, ,
\,\,\,\,\,\,\,\,\,\,
\as\,\,\ll\,\,1 ,
\eeq
%\beq \label{PARNEQ3}
%\kappa\,\,\leq\,\,1\,\, ,
%\eeq
\beq \label{PARNEQ4}
\as\,\kappa\,\,\leq\,\,1\,\,.
\eeq
The first two equations specify that we are doing our calculations in the 
double log approximation of pQCD while the third one introduces the new 
parameter $\kappa$ (see Ref.\cite{GLR}) which  takes into account the 
parton-parton interaction inside of the perturbative cascade. To 
illustrate how this set of parameters works let us consider an  
example simpler than the deep inelastic scattering with the proton, namely, the 
deep inelastic scattering with the meson built with heavy quarks (onia)
\cite{MU94}.
The mass of heavy quarks ($m_Q$) sets the scale for the distance in this 
interaction, and we have $r_{\perp}\,\,\propto\, \frac{1}{m_Q\,\as(m^2_Q)}$.
Therefore, we can safely apply pQCD for this problem.

\begin{figure}
\begin{center}
\begin{tabular}{c c}
\epsfig{file=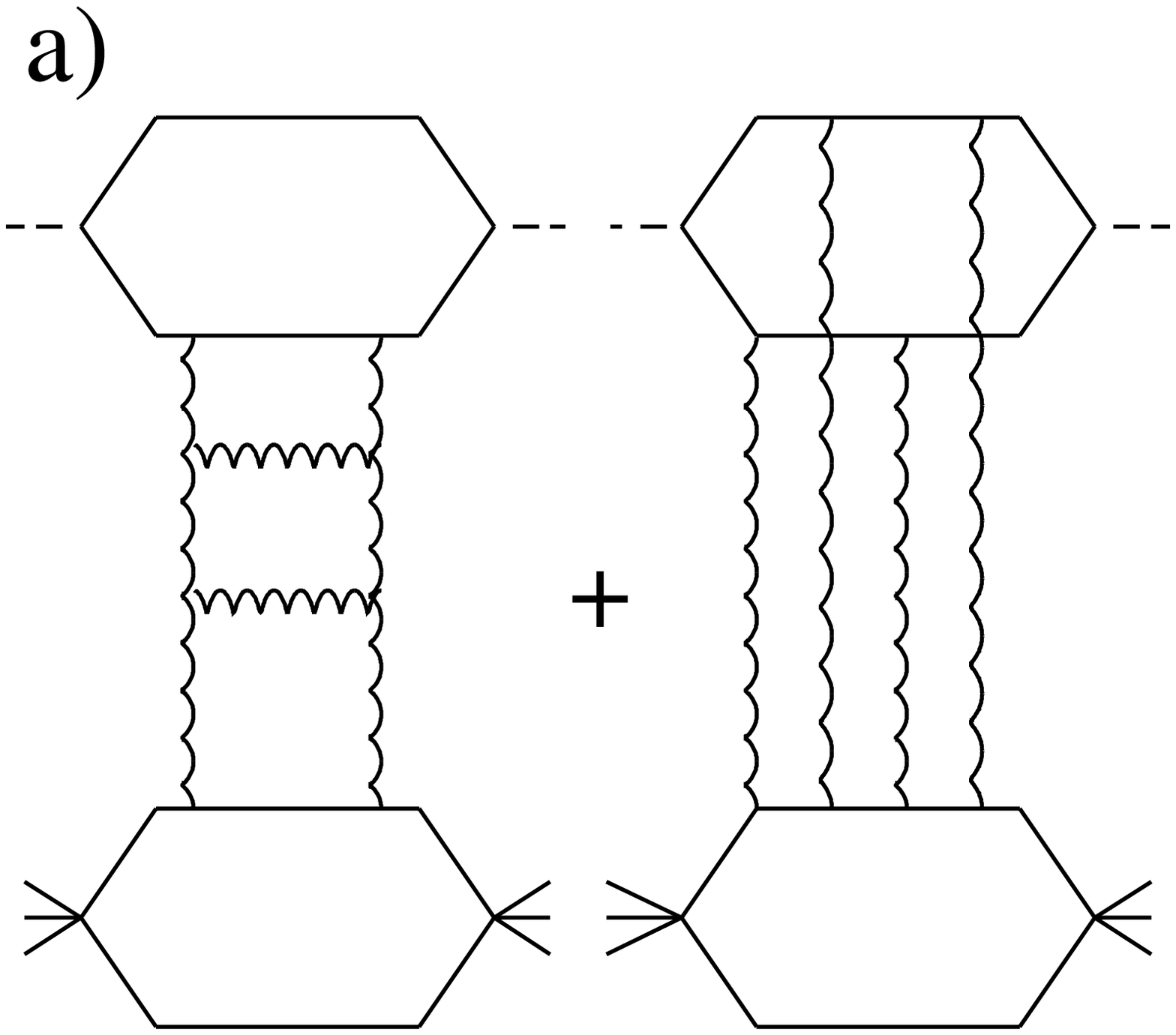,width=70mm}& \epsfig{file=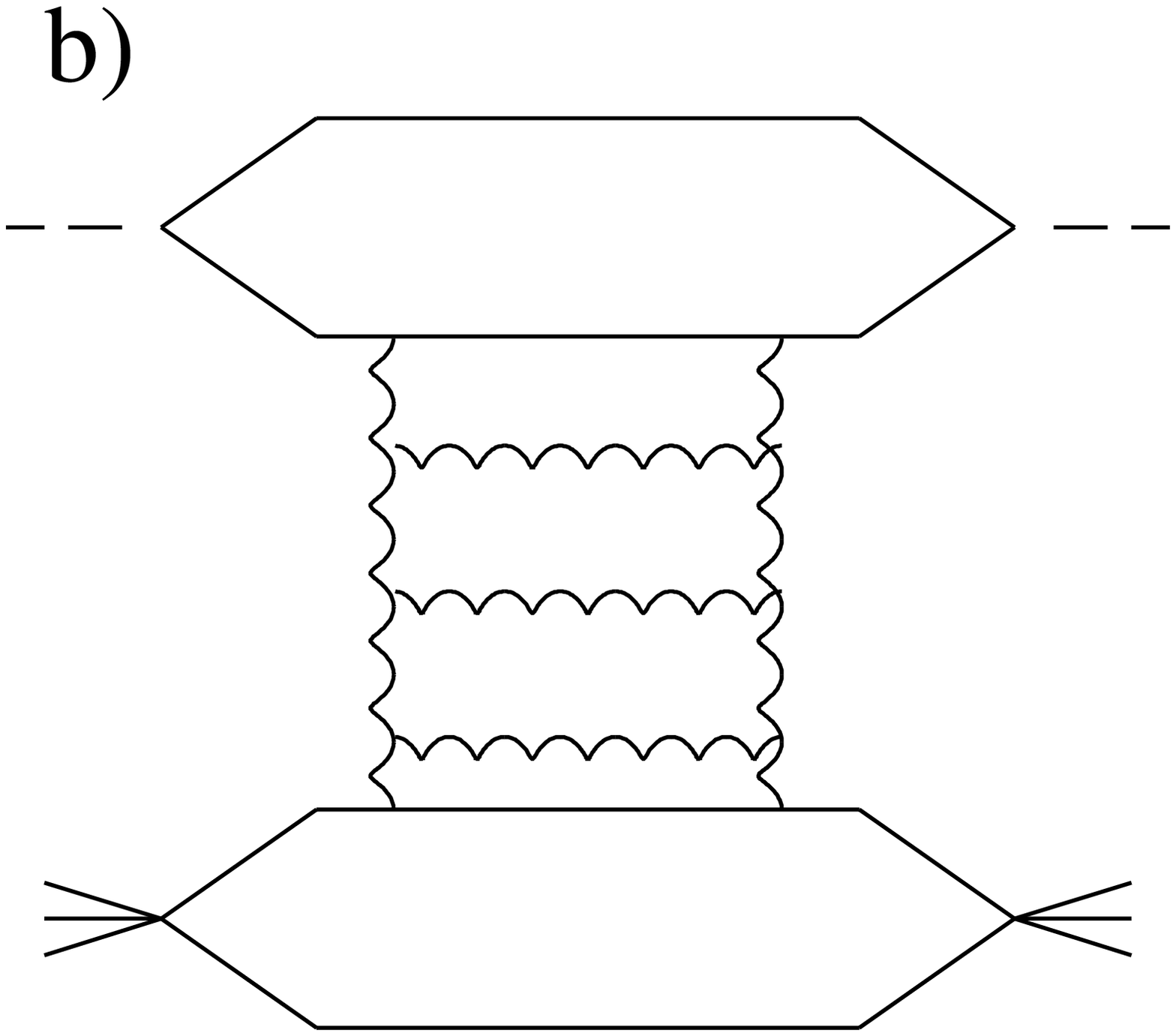,width=70mm}\\
\epsfig{file=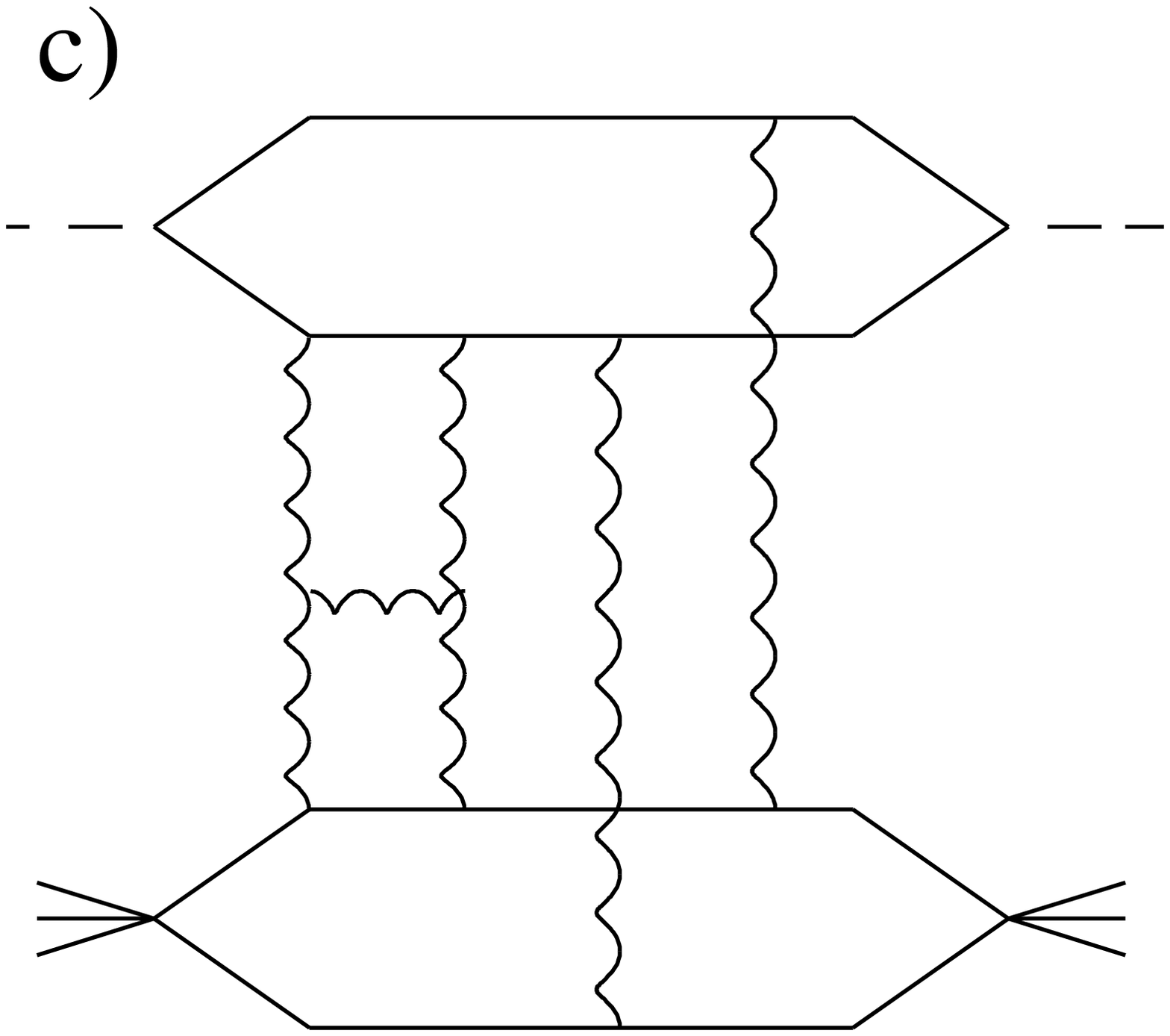,width=70mm} & \epsfig{file=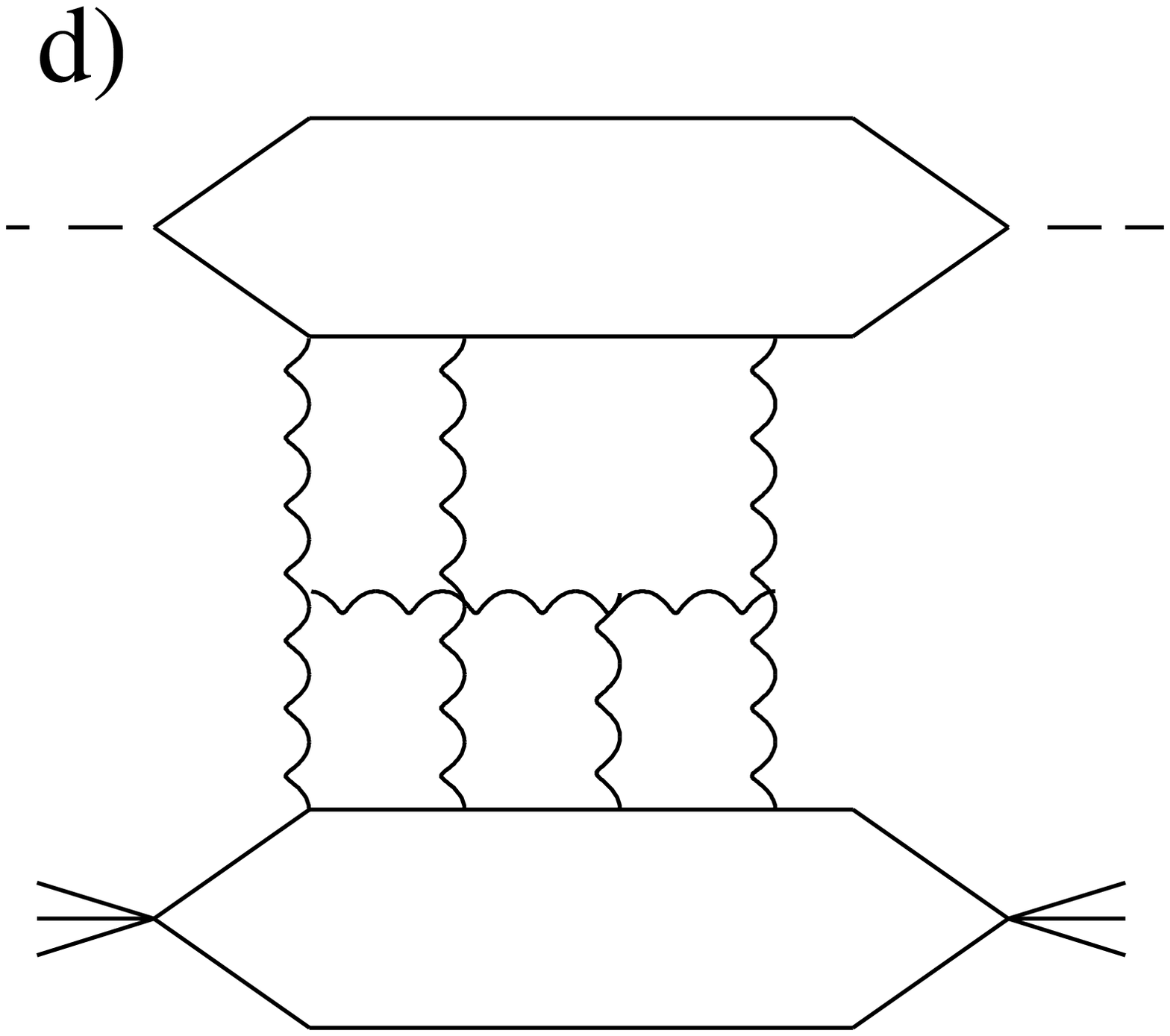,width=70mm}
\\
\end{tabular}
\caption{ {\em \label{NEQD} Diagrams of the deep inelastic scatering
with heavy quarks (onia) in the  order $\as^4(m^2_Q)$ (a) and
$\as^5(m^2_Q)$ (b - d) 
 of pQCD.}}
\end{center}

\end{figure}

\begin{figure}
\begin{center}
\begin{tabular}{c c}
\epsfig{file=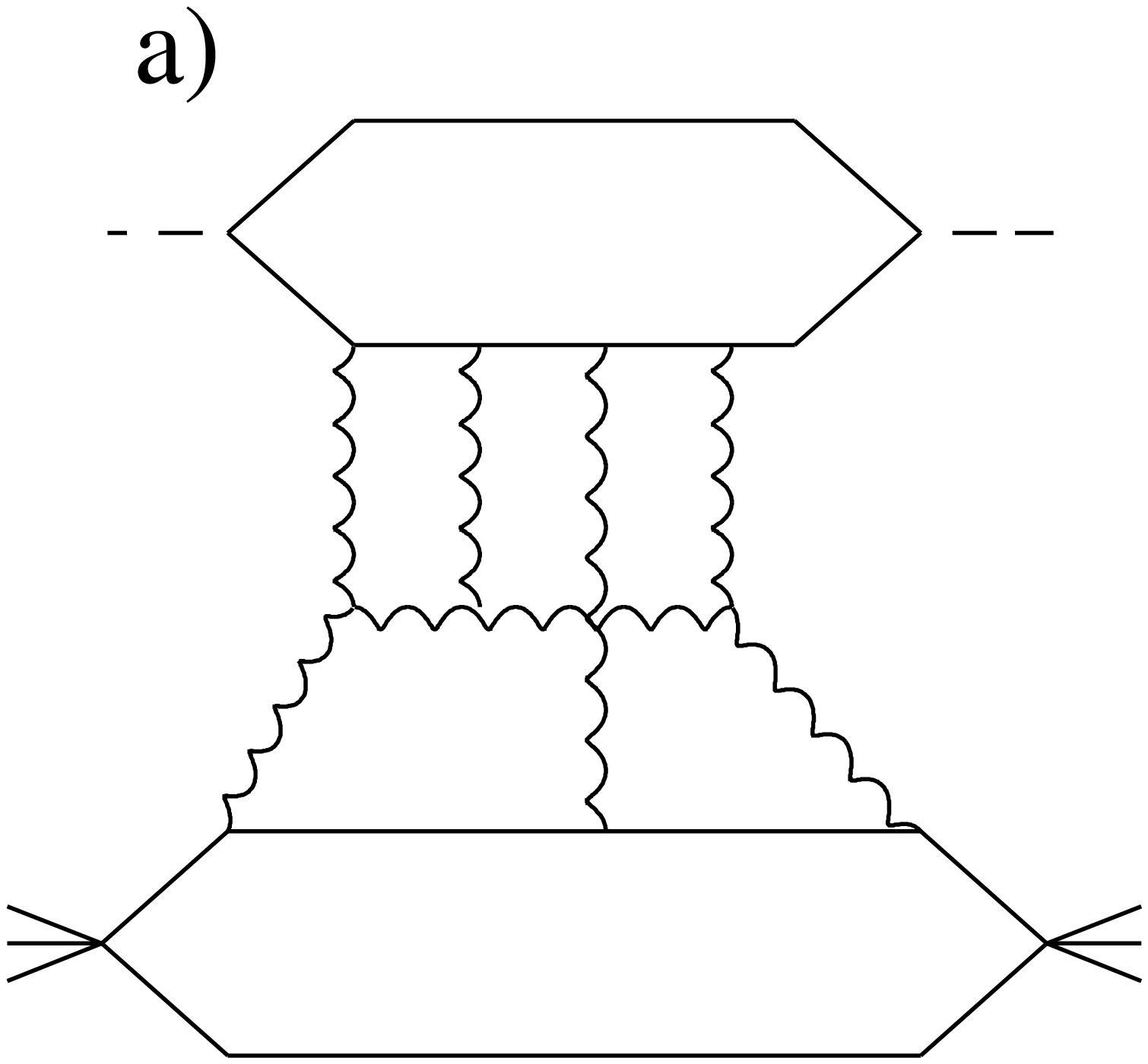,width=70mm} &\epsfig{file=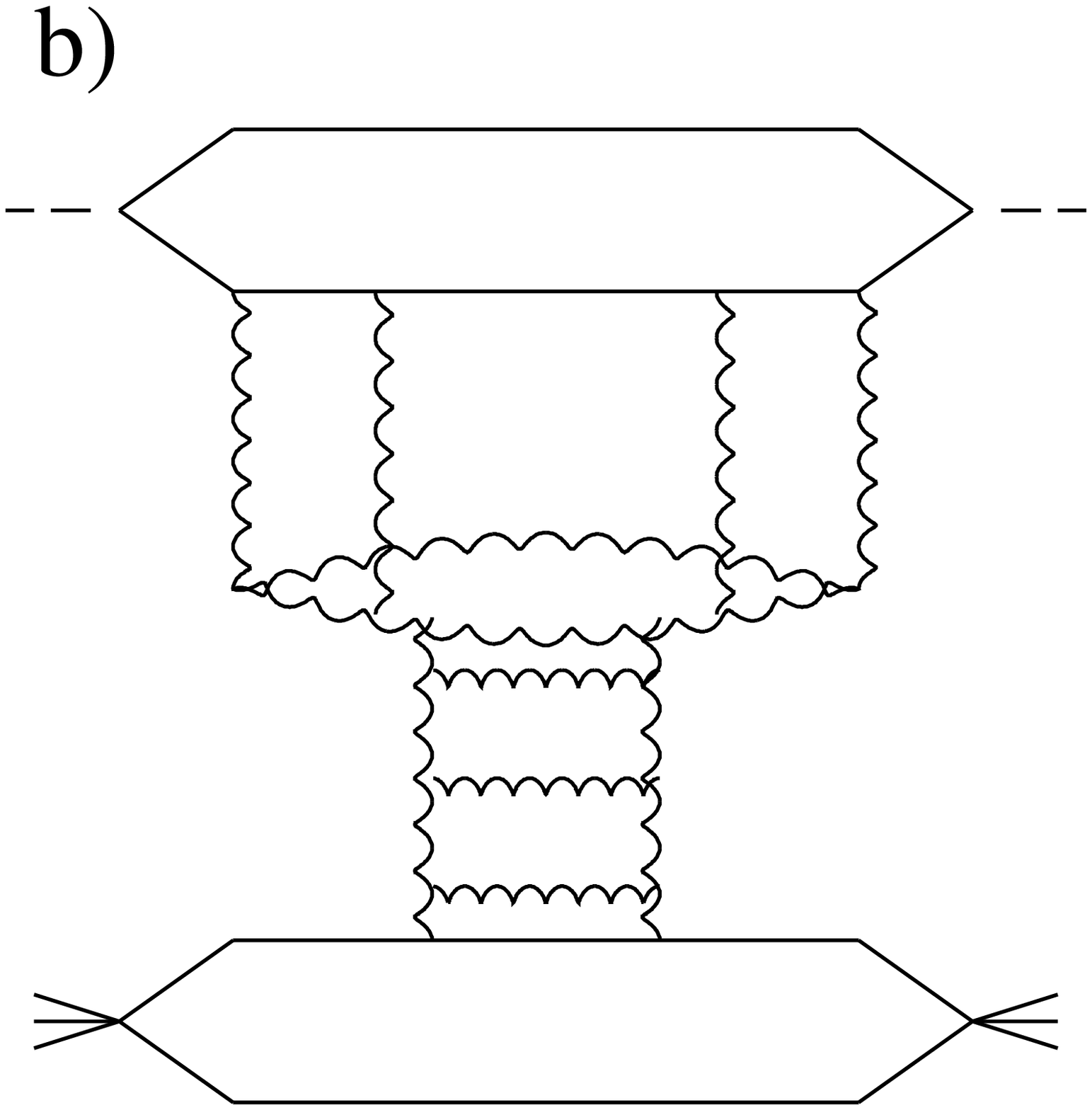,width=70mm} \\
\end{tabular}
\caption{ \label{NEQD1}{\em  The enhanced diagrams of the deep inelastic
scatering
with heavy quarks (onia) (not included in \protect\eq{103}).}}
\end{center}

\end{figure}

At order $\as^4(m^2_Q)$ of pQCD, using the set of parameters mentioned
above, the DIS can be described by the diagrams in Fig.\ref{NEQD}a. The first
diagram corresponds to the DGLAP evolution equations while the second one is the
Glauber - Mueller rescattering in the Born approximation. Going to the
next order of pQCD, one can see that the emission of one extra gluon
can be reduced: (i)  to the next order correction of the DGLAP evolution
equations (Fig.\ref{NEQD}b); (ii) to the emission of the extra gluon with
the rapidity close to the photon and its interaction with the target
( see Fig.\ref{NEQD}c ) which has been taken into account in the
generalized \eq{103} and (ii) the emission of the extra gluon with the
rapidity close to the target one ( see Fig.\ref{NEQD1}a). The last one is
the first from the more general class pictured in Fig.\ref{NEQD1}b ( so
called enhanced diagrams) that has not been included in \eq{103}.
However, in Ref.\cite{GLR}  it has been shown that integration over $k^2$
as well as over rapidity $y'$ ( see Fig.\ref{NEQD1}b ) concentrated at $y'$
close to the target rapidity and they give a small contribution of the
order $\as$ in comparison with the diagrams shown in Fig.\ref{NEQD}.
Therefore, we can neglected such contribution. However, there is a danger
in such an assumption, because it was made relying on perturbative
estimates while \eq{103} is  a nonperturbative one. Nevertheless, the
enhanced diagrams reveal themselves a class of diagrams with quite a
different topology which should be discussed and calculated separately. We
are going to do this job in our further publications. We must admit that
only such study will be able to establish the real kinematic region where
we can trust \eq{103}.


Eq. (\ref{103}) is the second order differential equation in partial
derivatives and we need two initial ( boundary ) conditions to specify the
solution.  The first one is, at fixed $y$ and $Q^2
\,\rightarrow \,\infty$,
 $$ \kappa\,\,\rightarrow\,\,\frac{N_c \,\as\,\pi
\,}{ 2\,Q^2\,R^2} \,x G^{DGLAP} (x,Q^2)\,\,.  $$ 
The second one we can fix
in the following way: at $ x = x_0 \,\,(y = y_0)$ which is small, namely,
in the kinematic region where $\as y \xi\, \leq\, 1$
 \beq \label{ic}
\kappa\,\,\rightarrow\,\,\kappa_{in}\,\,=\,\, \frac{N_c \,\as\,\pi }{
2\,Q^2\,R^2} \,x G (x,Q^2)\,\,, \eeq 
where $x G$ is given by the modified Mueller
formula ( see \eq{FINANS}).  Practically, we can take $x_0 \,=\,10^{-2}$,
since the corrections to the MF are small at this value of $x=x_0$. However,
as we have discussed, we cannot believe in any treatment of the SC at
large distances. Therefore, we will take $Q_0^2 \, = \, 1 GeV^2 $ as 
the initial value of $Q^2$ in the
modified Mueller formula. Thus, the 
boundary condition 
(\ref{ic}) will be considered only for $Q^2\,\geq\,Q^2_0\,=\, 1\,GeV^2 $. 
 For $Q^2\,< \,Q^2_0$ we have to specify the initial conditions and we will
discuss this point later.

The  boundary condition for the $b_t$-dependent equation (\ref{NEQ}) is fixed
in a similar form, namely, we  suggest the following initial condition for
$Q^2\,\geq Q^2_0$  where  the value of $\kappa$ at $x = x_0$
calculated
from the $b_t$ dependent gluon distribution

 \beq \label{bic}
\kappa\,\,\rightarrow\,\,\kappa_{in}\,\,=\,\, \frac{N_c \,\as\,\pi^2 }{
2\,Q^2} \,x G (x,Q^2, b_t)\,\,. \eeq

 \subsection{The asymptotic solution.}

 First
observation is the fact that \eq{103} has a solution which depends only
 on $y$. Indeed, one can check that $\kappa \,=\,\kappa_{asymp}(y)$ is
the solution of the following equation
\beq \label{105}
\frac{d \kappa_{asymp}}{d y}\,\,=\,\,F(\,\kappa_{asymp}\,)\,\,.
\eeq
The solution to the above equation is
\beq \label{106}
\int^{\kappa_{asymp}(y)}_{\kappa_{asymp}(y=y_0)}\,\,
\frac{ d \kappa'}{F(\kappa')}
\,\,=\,\,y - y_0\,\,.
\eeq

It is easy to find the behavior of the solution to \eq{106} at large value
of $ y$ since $F(\kappa)\,\,\rightarrow\,\,\bar \as \ln \kappa $ at
large $\kappa$ ( $\bar \as = \frac{N_c}{\pi}\,\as$ ). It gives
\beq \label{107}
\kappa_{asymp}\,\,\rightarrow \,\,\bar \as y \,\ln(\bar  \as y)\,\,\,\,\,\,
at\,\,\,\,\,\,\,\bar \as y \,\,\gg\,\,1\,\,.
\eeq
At small value of $y$, $F(\kappa)\,\,\rightarrow\,\,\bar \as \kappa$ and
we have
\beq \label{108}
\kappa_{asymp}\,\,\rightarrow\,\,\kappa_{asymp} ( y = y_0 )
\,\,e^{\bar \as ( y - y_0)} \,\,.
\eeq
 
We claim this  solution is the asymptotic solution to \eq{103}. To prove
this we have to consider the stability of the asymptotic solution. It
means, that we solve our general equation looking for the solution in the
form
\beq \label{STAB}
\kappa\,\,=\,\,\kappa_{asymp}(y)\,\,+\,\,\Delta \kappa(y,\xi - \xi_0)\,\,,
 \eeq
where $\Delta \kappa$ is small ($ \Delta \kappa\,\ll\,\kappa_{asymp}$) but
an arbitrary function at $\xi \,=\,\xi_0$. We have to prove that \eq{103}
 will not lead to big $\Delta \kappa$ ($\Delta
\kappa\,\gg\,\kappa_{asymp}$) at large $\xi$.

The following linear equation can be written for $\Delta \kappa$
\beq \label{109}
\frac{\partial^2 \Delta \kappa( y, \xi  )}{\partial y\,\partial \xi}\,\,+\,\,
\frac{\partial \Delta \kappa( y, \xi)}{\partial y}\,\,=\,\,\frac{d F(\kappa)}{
d \kappa}\,|_{\kappa = \kappa_{asymp}(y)}\,\,\Delta \kappa (y, \xi)\,\,.
\eeq
In Ref.\cite{AGL} was proven, that the solution of \eq{109} is much smaller
 than $\kappa$, since $\frac{d
F(\kappa)}{d \kappa}\,\,\rightarrow\,\,0$ at large $y$.

%
%
%  bt asymptotic equation.
%
%
The $b_t$-dependent equation also has an asymptotic solution.
Indeed, one can check that $\kappa \,=\,\kappa_{asymp}(y,b_t)$ is
the solution of the following equation
\beq \label{asyb}
\frac{d \kappa_{asymp}}{d y}\,\,=\,\,F_{b_t}(\,\kappa_{asymp}\,)\,\,,
\eeq
which can be rewritten as
\beq \label{asyb1}
\int^{\kappa_{asymp}(y,b_t)}_{\kappa_{asymp}(y=y_0,b_t)}\,\,
\frac{ d \kappa'}{F_{b_t}(\kappa')}
\,\,=\,\,y - y_0\,\, .
\eeq
This equation has the solution
\beq \label{SOL106}
\kappa_{asymp}(y,b_t)\,\,=\,\,\ln\{1\,\,+\,\,(\,e^{\kappa_0(y =
y_0,b_t)}\,\,-\,\,1\,)\,e^{\frac{N_c \as}{\pi}\,(y\,-\,y_0)}\,\}\,\,.
\eeq
It is easy to find from \eq{SOL106} the behavior of the solution of
\eq{asyb} at large values
of $ y$. It gives
\beq \label{asyb2}
\kappa_{asymp}(y;b_t)\,\,\rightarrow \,\,\bar \as y \,\,\,\,\,\,
\mbox{at} \,\,\,\,\,\,\,\bar \as y \,\,\gg\,\,1\,\,.
\eeq
At small values of $y$,
we have
\beq \label{asyb3}
\kappa_{asymp}(y,b_t)\,\,\rightarrow\,\,\kappa_{asymp} ( y = y_0,b_t )
\,\,e^{\bar \as ( y - y_0)} \,\,.
\eeq

Therefore, the asymptotic solution has a chance to be the solution of
 equation (\ref{103}) ((\ref{NEQ}) for $b_t$-dependent case) in the 
region of very small $x$. To prove that,  
we need to solve the equation 
in the wide kinematic region starting with the initial condition.
 We managed to do this only in the semiclassical approach.

\subsection{The semiclassical approach.}

In the semiclassical approach the solution of eq.
(\ref{103}) is supposed to be in the form
\bea
\kappa = e^S \, ,
\eea
where $S$ is a function with partial derivatives: $\frac{\partial S}{
\partial y} = \o $ and $\frac{\partial S}{\partial \xi} = \gamma $ 
which are smooth functions of $y$ and $\xi$.
It means that
\bea
\frac{\partial^2 S}{\partial \xi \partial y} \ll \frac{\partial S}{
\partial y}  \frac{\partial S}{\partial \xi} = \o \gamma \,\, .
\label{118}
\eea
Using eq.(\ref{118}), one can rewrite eq.(\ref{103}) in the form
\bea
\frac{\partial S}{\partial y} \frac{\partial S}{\partial \xi} +
\frac{\partial S}{
\partial y} = e^{-S} F(e^{S}) \equiv \Phi(S)
\label{119}
\eea
or
\bea
\o (\gamma + 1) = \Phi (S) \,\, .
\label{120}
\eea

 For an equation in the form
\bea
F(\xi, y, S, \gamma , \o ) = 0 \, ,
\label{121}
\eea
we can introduce the set of characteristic lines $ (\xi(y), S(y), \o (y),
\gamma (y) ) $, which satisfy a set of well defined equations 
(see, for example, Refs. \cite{Collins90}, \cite{Bartels91} for the method).
In Ref.  \cite{AGL} we developed a detailed calculation of the solution.
Here, we will concentrate on the $b_t$ dependent evolution. To solve the $b_t$ dependent equation in the semiclassical approach,
we need only to substitute the nonlinear term $F_{b_t}(\kappa)$, instead of 
$F (\kappa )$, in the expression (\ref{119}).
Using eq.(\ref{119}) and  eq.(\ref{120}), written for $\Phi_{b_t}$,
 we obtain the following set of equations for the characteristics
\bea
\frac{d \xi}{d y} = \frac{\Phi_{b_t} (S)}{(\gamma +1)^2}  \, ; \,\,\,\, 
\, \frac{d S}{d y} =   \frac{2 \gamma + 1}{(\gamma +1)^2}\Phi_{b_t} (S) \, ;
\,\,\,\, 
\frac{d \gamma}{d y} = \Phi'_{S;b_{t}} \frac{\gamma}{\gamma +1} \, \,,
\label{125}
\eea 
where $\Phi'_{S;b_{t}}\,=\, \frac{\partial \Phi_{b_t} }{\partial S}$.
The initial conditions for this set of equations we derive from eq.(\ref{ic}),
being
\bea
S_0 & = & ln \kappa_{in} (y_0, \xi_0) \nonumber \, ,\\
\gamma_0 & = & \left. \frac{\partial ln \kappa_{in} (y_0 , \xi )}
{\partial \xi} \right|_{\xi = \xi_0} \, .
\label{129}
\eea
The DGLAP equation for $\kappa$ is obtained taking the term in 
the curly in r.h.s of 
\eq{103} equal to $\kappa$, which gives $\Phi = \frac{\as Nc}{\pi}$ and 
$\Phi'_{S, b_{t}}\,=\, 0$ in \eq{125}.
The main properties of these equations have been considered in Ref.\cite{AGL}
analytically. Here, however,
 we restrict ourselves mostly to the numeric solution
of these equations.

  For $Q^2 \,\leq\,Q^2_0$ we  cannot specify the initial condition.
To find characteristics which correspond to such small virtualities, we put a
boundary condition at $Q^2 = Q^2_0$, namely
$\kappa_{bound}\,=\,\kappa(y,\xi = \xi_0)$. However, this boundary
condition cannot be arbitrary  at $x \,\rightarrow \,0$, but it should
satisfy the equation. On the
other hand we wish to incorporate in this boundary condition everything
that has been known on the parton densities experimentally. We suggest the
following boundary condition:
\beq \label{BOUND}
\kappa_{bound}\,\,=\,\,\kappa^{GRV}(Q^2=Q^2_0,x;b_t)\,\,\,\,\,at\,\,\,\,x\,\geq
\,x_0\,\,;
\eeq
$$
\kappa_{bound}\,\,=\,\,\kappa_{asymp}\,\,\,\,\,at\,\,\,\,x\,\leq\,x_0\,\,,
$$
where the value of $x_0$ we find from the equation 
$$
\kappa^{GRV}(Q^2 = Q^2_0,x = x_0;b_t)\,\,=\,\,1\,\,.
$$
Using \eq{BOUND} we can specify the boundary condition for our set of
equations, namely
\beq \label{BOUND1}
S_0(\xi = \xi_0,b_t)\,\,=\,\,\ln( \kappa_{bound}(y_0,\xi_0;b_t)\,\,;
\eeq
$$
\o_0(\xi = \xi_0)\,\,=\,\,\frac{\partial \ln
(\kappa_{bound}(\xi_0,y;b_t)}{\partial y}|_{y = y_0}\,\,.
$$
Using the boundary conditions of \eq{BOUND1}, it is better to rewrite
\eq{125} in the form:
\beq \label{BOUND2}
\frac{d \xi}{d y}\,\,=\,\,\frac{\o^2}{\Phi_{b_t}(S)}\,\,;\,\,\,\,
\frac{d S}{d y}\,\,=\,\,2\,\o\,+\,\frac{\o^2}{\Phi_{b_t}(S)}\,\,;\,\,\,\,
\frac{d \o}{d y}\,\,=\,\,\o^2\,\frac{\Phi'_{S;b_{t}}(S)}{\Phi_{b_t}(S)}\,\,.
\eeq 
 We would like to emphasize that our suggestion ( see \eq{BOUND} ) is
still rather arbitrary since the equation for $x_0$ has no deep
theoretical arguments. Actually, what we know theoretically is the fact
that $\kappa_{bound}\,\rightarrow \,\kappa_{asymp}$ at $\kappa^{GRV}\,
\geq\,1 $. This is the reason why we decided to restrict ourselves to the
solution of the equation for the trajectories that started at $y = 5$ with
the values of $Q^2 \,\geq \,Q^2_0\,=\,1\,GeV^2$. We are  studing
other trajectories on a separate paper.

\begin{figure}[hptb]
\begin{center}
\begin{tabular}{ c  c}
\psfig{file=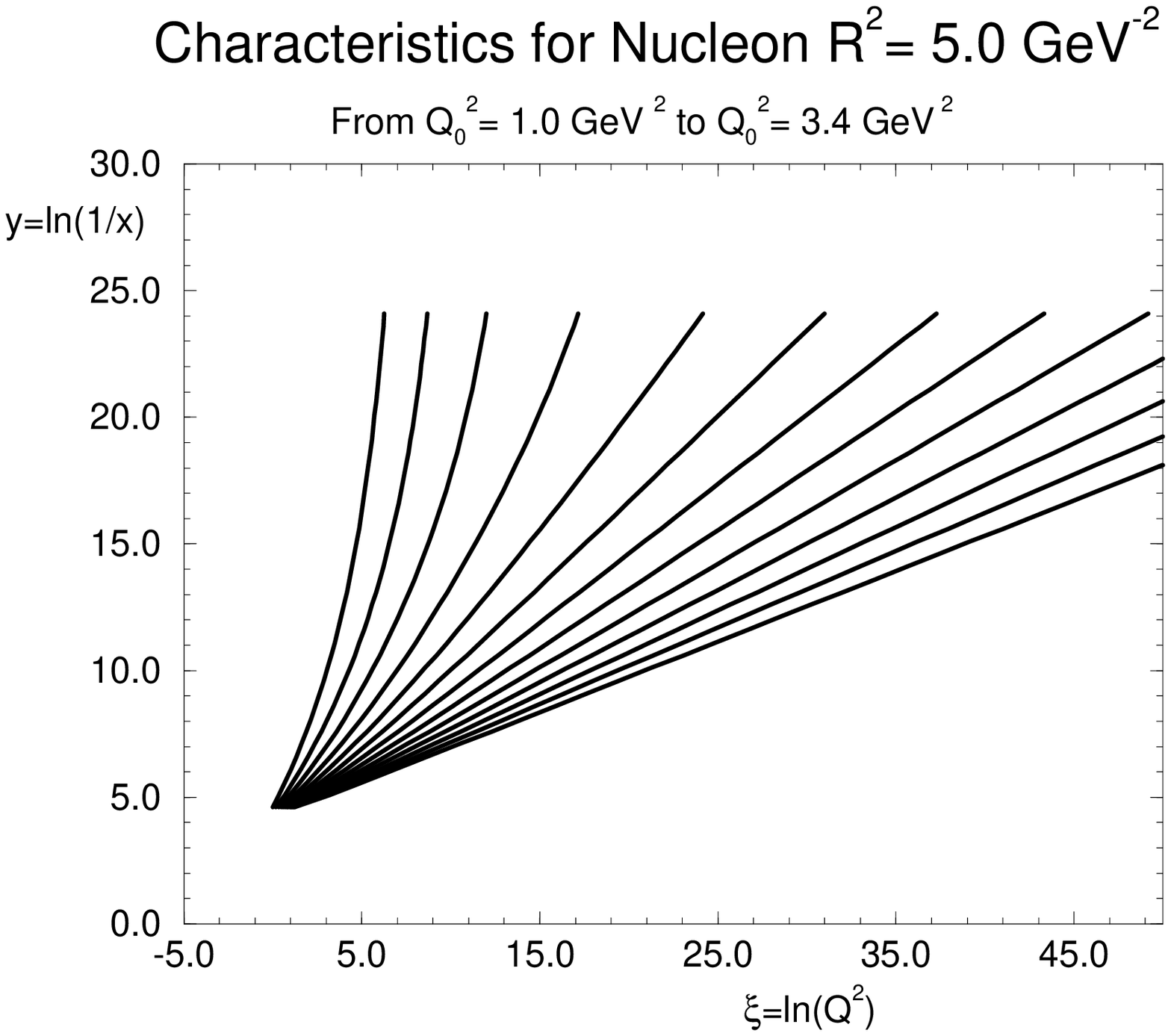,width=70mm} & \psfig{file=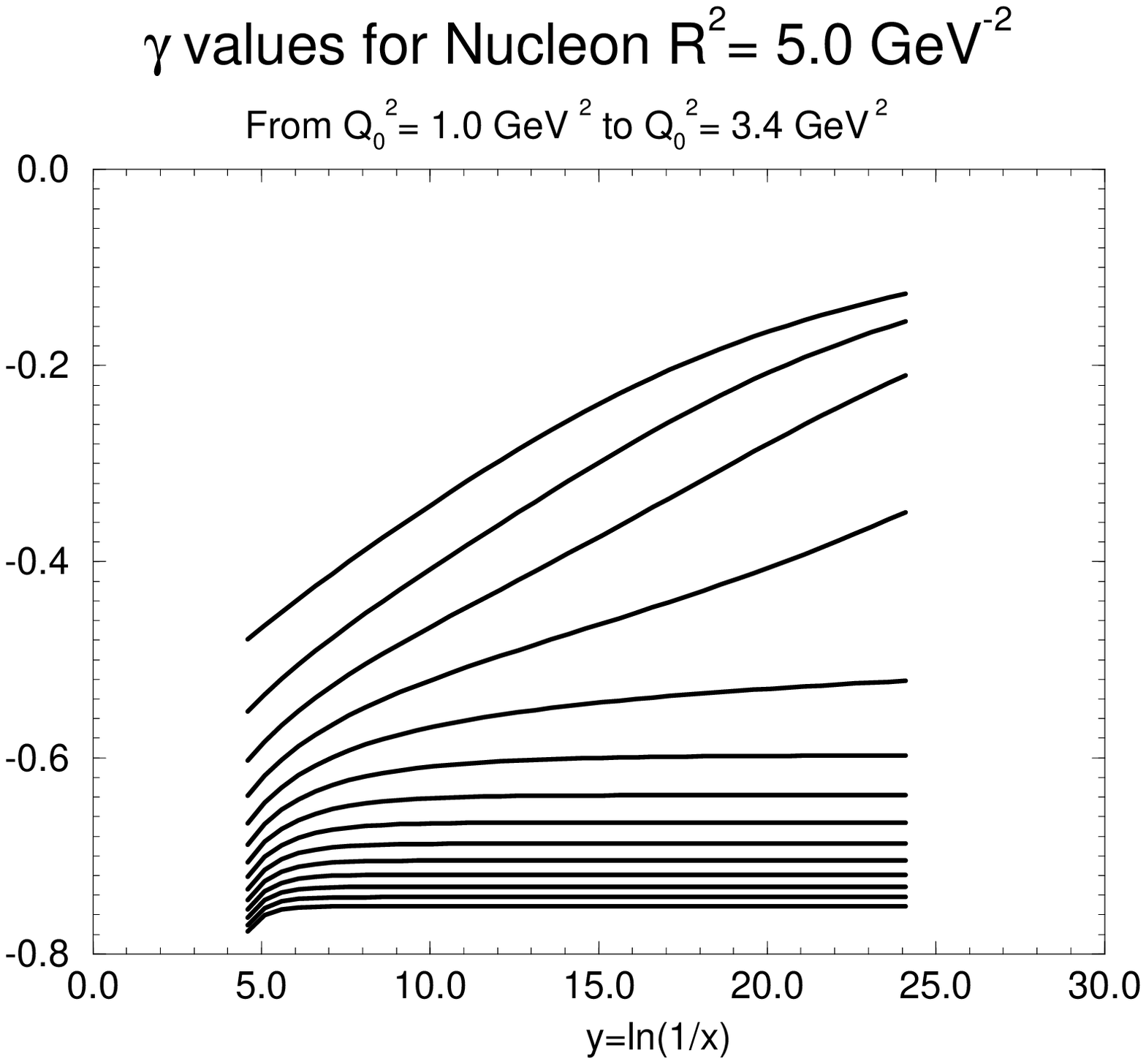,width=70mm}\\
\end{tabular}
\end{center}
\caption{ \em The trajectories (a) and the $\gamma$ values (b) for the solution of 
the generalized
 evolution equation for nucleon with $R^2 = \, 5 \, GeV^{-2}$.}
\label{scn}
\end{figure}

\begin{figure}[hptb]
\begin{center}
\begin{tabular}{ c c}
\psfig{file=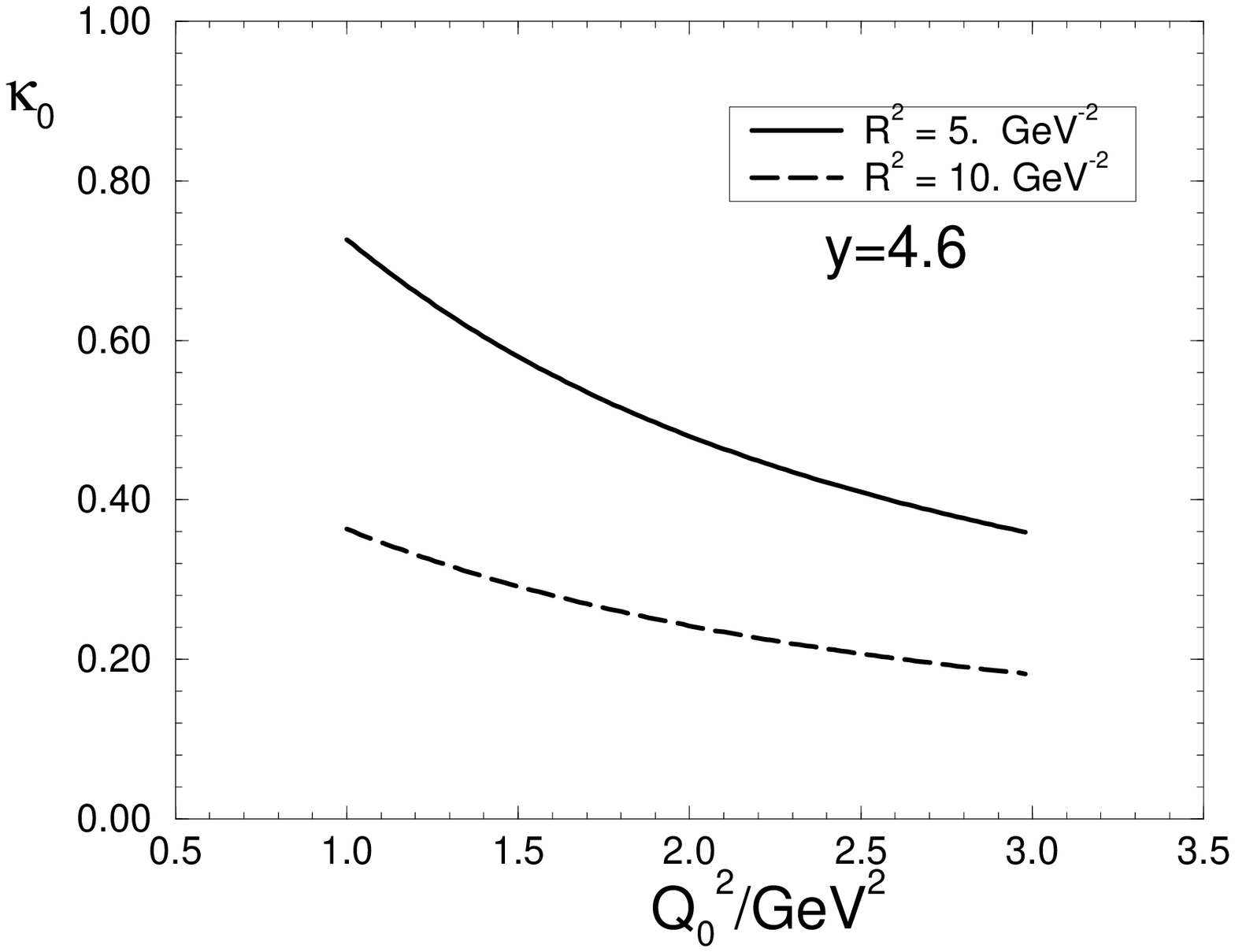,width=70mm} & \psfig{file=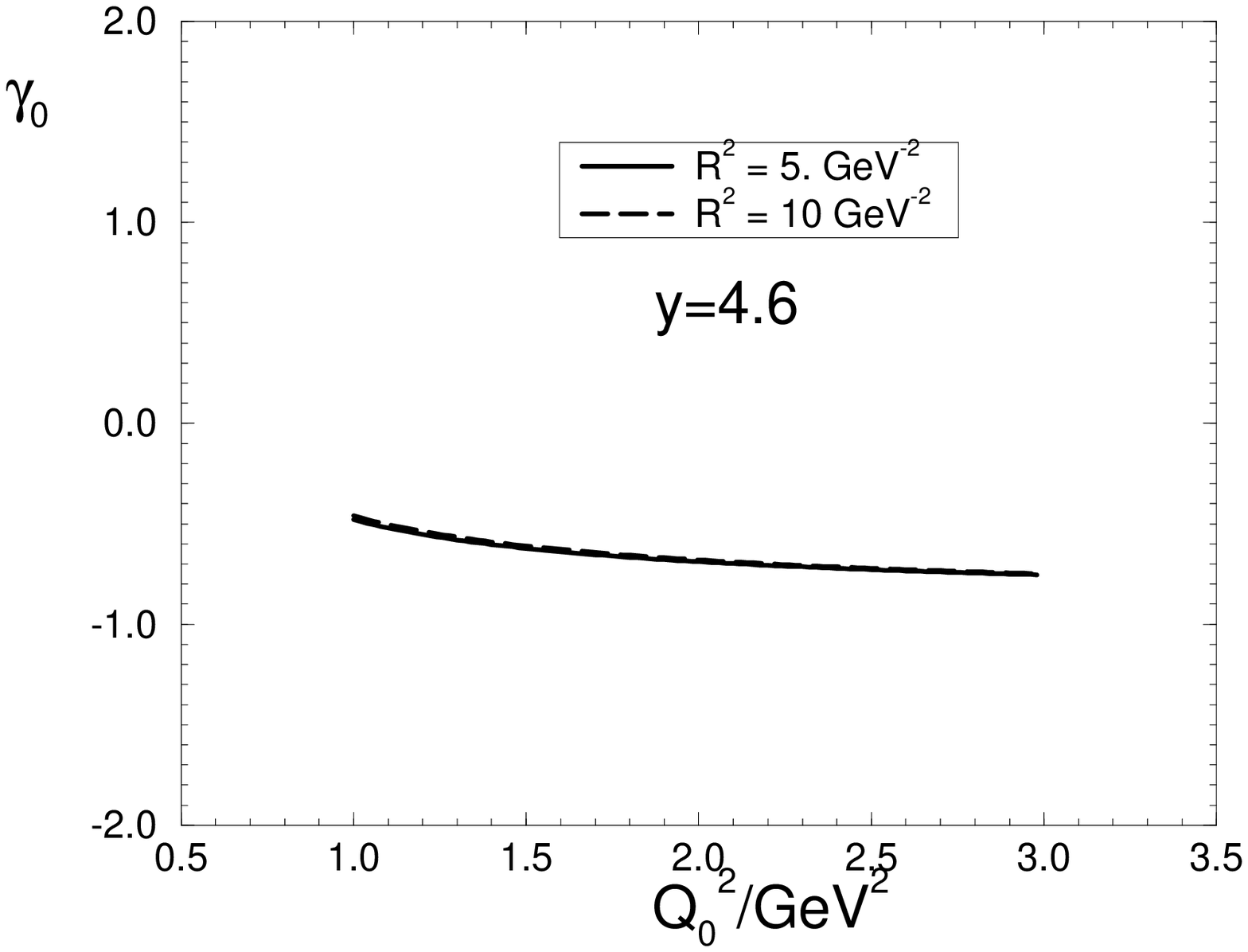,width=70mm}\\
\end{tabular}
\end{center}
\caption{\em The initial values $\kappa_0$ and $\gamma_0$ for $y=4.6$
as a function of $Q^2_0$.}
\label{sca}
\end{figure}

For the numerical solution we use the 4th order Runge - Kutta method
to solve our set of equations with the initial distributions of  \eq{129}.
 For a short notation we will use AGL to our nonlinear evolution equation
 \eq{103} and AGL$_{b_t}$ for the $b_t$ dependent nonlinear evolution equation
\eq{NEQ}.
Fig.\ref{scn} shows the general features of the solution.
Fig.\ref{scn}a shows the characteristic
curves in the $y$ v.s. $\xi$ plane for  different initial conditions.
Fig.\ref{scn}b shows the
 evolution of the respective $\gamma$
values with $y$.  The initial
conditions, plotted in Fig.(\ref{sca}), are calculated for $y=4.6$ ($x\, \approx \, 10^{-2}$)
and $Q^2_0$ from $1 \, GeV^{2}$ to $3.4 \, GeV^{2}$ with
$R^2 \, = \, 5 \, GeV^{-2}$.  When $\gamma$ goes to zero as $y$ grows, 
the nonlinear effects play an important
role.  The respective
trajectory tends to a vertical line, and the AGL solution tends to the
asymptotic one. When $\gamma$ goes to a constant, the AGL solution
tends to the DGLAP one.


\subsection{Comparison between different approaches to the SC} 
\begin{figure}[hptb]
\begin{center}
\begin{tabular}{ c c}
\psfig{file=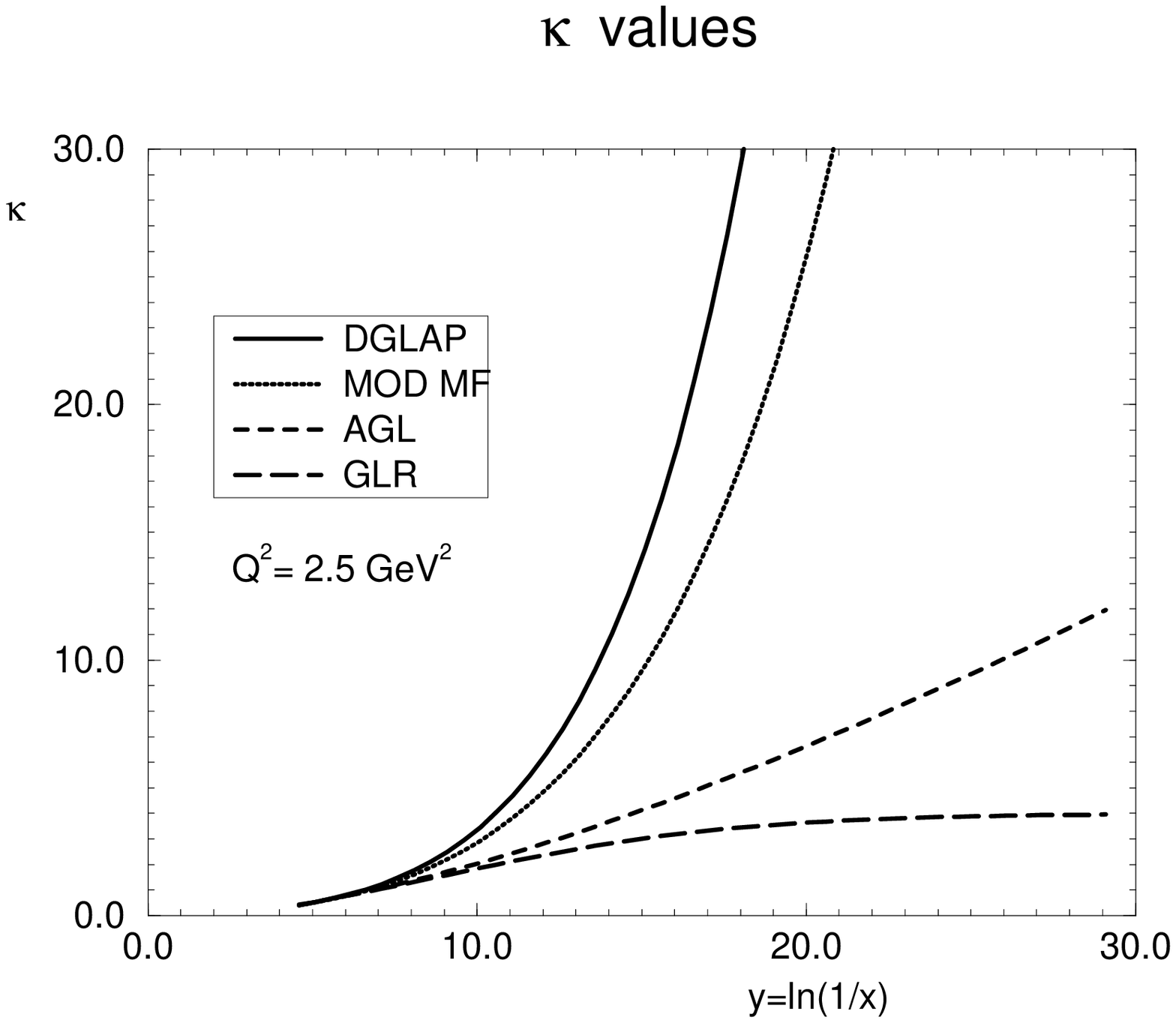,width=70mm} & \psfig{file=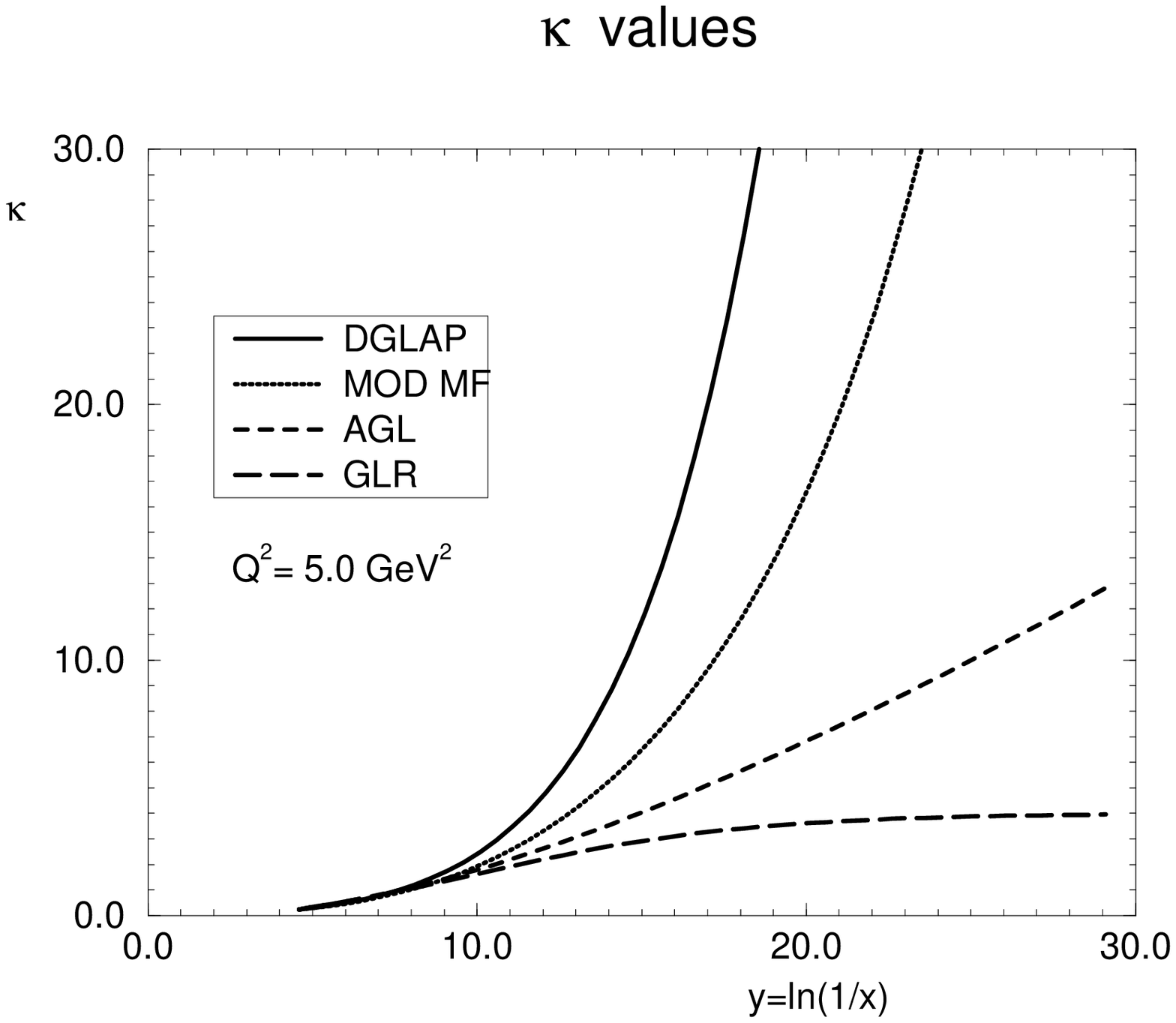,width=70mm}\\
\psfig{file=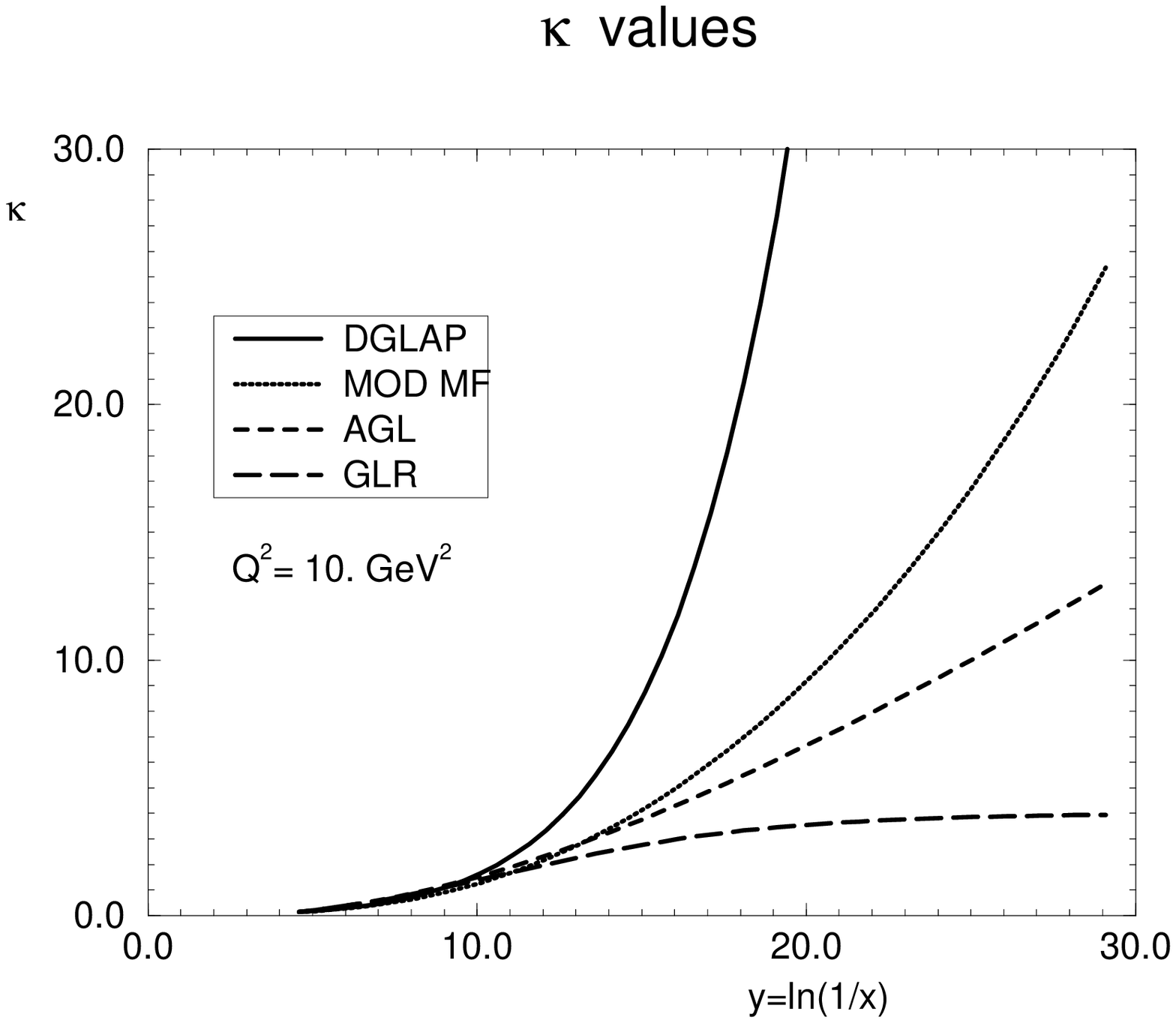,width=70mm} & \\
\end{tabular}
\end{center}
\caption{\em The $\kappa$ values solutions of the  nonlinear equation (AGL) and
the GLR equation. The values of $\kappa$ calculated from the modified
 Mueller Formula for
the gluon distribution (MF)
and from the  DGLAP gluon distribution is also shown.}
\label{aglkap1}
\end{figure}

Using the characteristics  and the
initial conditions we are able to
find the solution of AGL equation as a function of $y$ and $\xi$. For
a given point $( \, y_1, \, \xi_1 \, )$, we vary the initial condition
until we find the characteristic which passes throught the point.  From the
characteristic we can reconstruct the solution. The same procedure
can be used for GLR equation. In Fig.\ref{aglkap1} we compare
the  $\kappa$ values obtained from the solution of the  
nonlinear equation (\ref{103}) (AGL) and
the GLR equation. The values of $\kappa$ calculated from the modified 
Mueller Formula ( MOD MF)
and from the DGLAP evolution equation (GRV distribution) are also shown. The results are
presented as functions of $y$ for several values of $Q^2$. The initial
condition is the same for all equations. 

We can see from Fig.\ref{aglkap1} that the SC predicted by
 the MOD MF, the AGL equation and
by GLR equation have the same order of magnitude at HERA kinematic
 region ($5 \, < \, y \, < 10$). For very small $x$ ($y \, > \, 10$)
 the results are quite different. 
The values of $\kappa$ obtained from the  AGL equation suffer a sizeable reduction when 
compared with those obtained from the modified Mueller formula. 
The AGL solution increases monotonically with $y$, but much less then the
Mueller formula.
The GLR equation gives much stronger SC. In fact, the GLR equation predicts 
saturation at very small $x$ values. 

\begin{figure}[hptb]
\begin{center}
\psfig{file=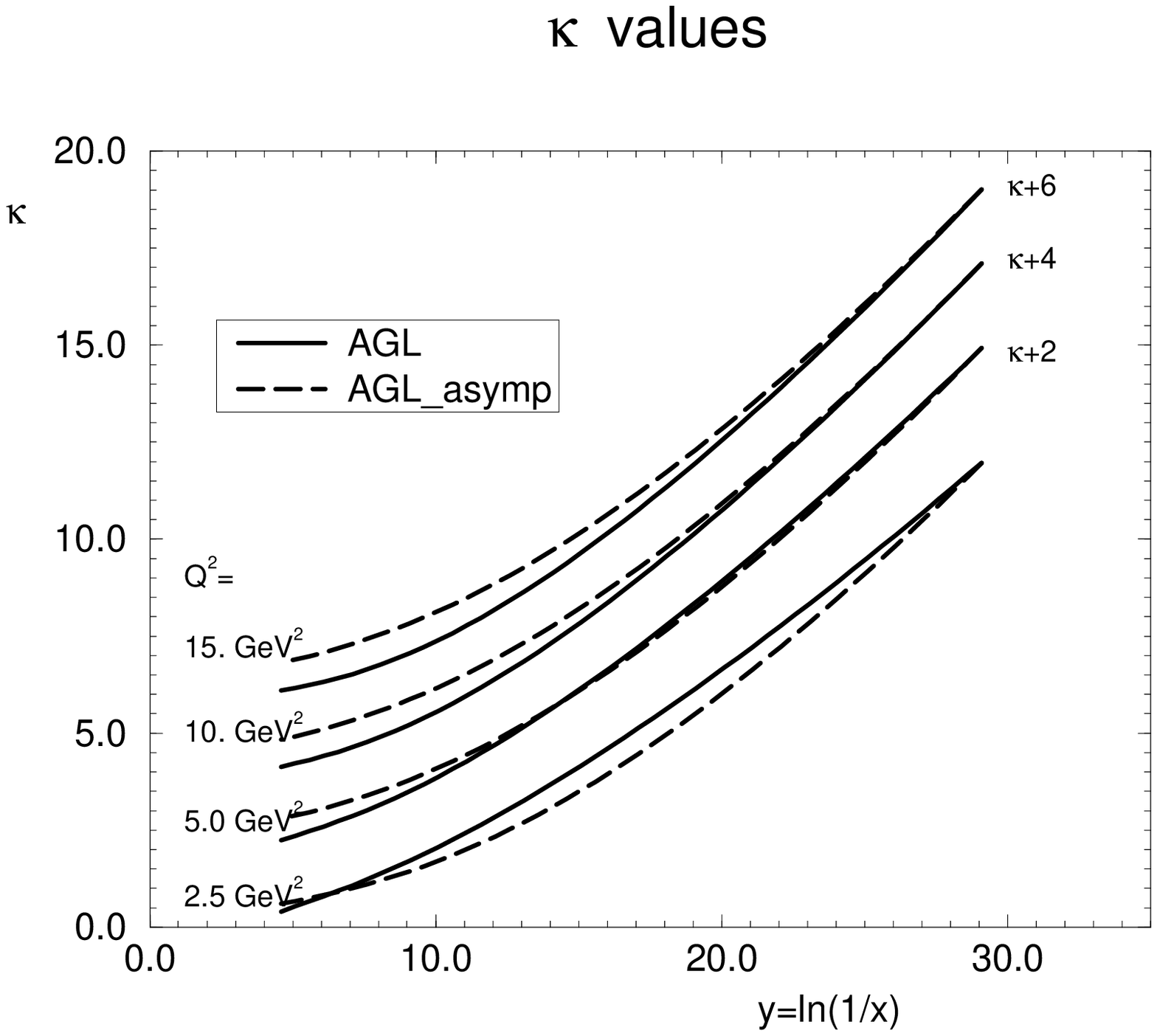,width=120mm}
\end{center}
\caption{\em The $\kappa$ values solutions of the  nonlinear equation 
(AGL) compared
with the backwards evolution of the asymptotic equation.}
\label{aglasym1}
\end{figure}

In Fig.\ref{aglasym1}, we present the solution of the AGL equation 
compared with the AGL$_{asymp}$ results. For each value of $Q^2$, the asymptotic 
equation is solved backwards from the value of $\kappa$, solution of
AGL equation (we took $y\, = \, 29.1$). We can see from the figure that the 
asymptotic solution crosses the AGL solution twice for $Q^2 \, \le \, 5 \, GeV^2$. 
We suppose it occurs because the AGL solution have not yet reached its 
asymptotic value for that $y$.

We would like to draw your attention to the fact that our asymptotic
solution turns out to be quite different from the GLR one. The GLR solution
in the region of very small $x$ leads to saturation of the gluon density
\cite{Collins90,Bartels91,BALE}. Saturation means that $\kappa$ tends
to a constant in the region of small $x$. The solutions of
\eq{103} approach the asymptotic solution  at $ x\,\rightarrow\,0$, which
 does not depend on $Q^2$, but exhibits sufficiently strong dependence
 of $\kappa$  on
$x$ ( see Fig.\ref{aglasym1} ), namely $\kappa \,\propto
\,\as \ln (1/x)\ln\ln(1/x)$.
The absence of saturation does not contradict any physics since gluons
are bosons and it is possible to have a lot of bosons in the same cell of
 the phase space. We should admit that A. Mueller first came to the same
 conclusion using his formula in Ref.\cite{MU90}.

\begin{figure}[hptb]
\begin{center}
\psfig{file=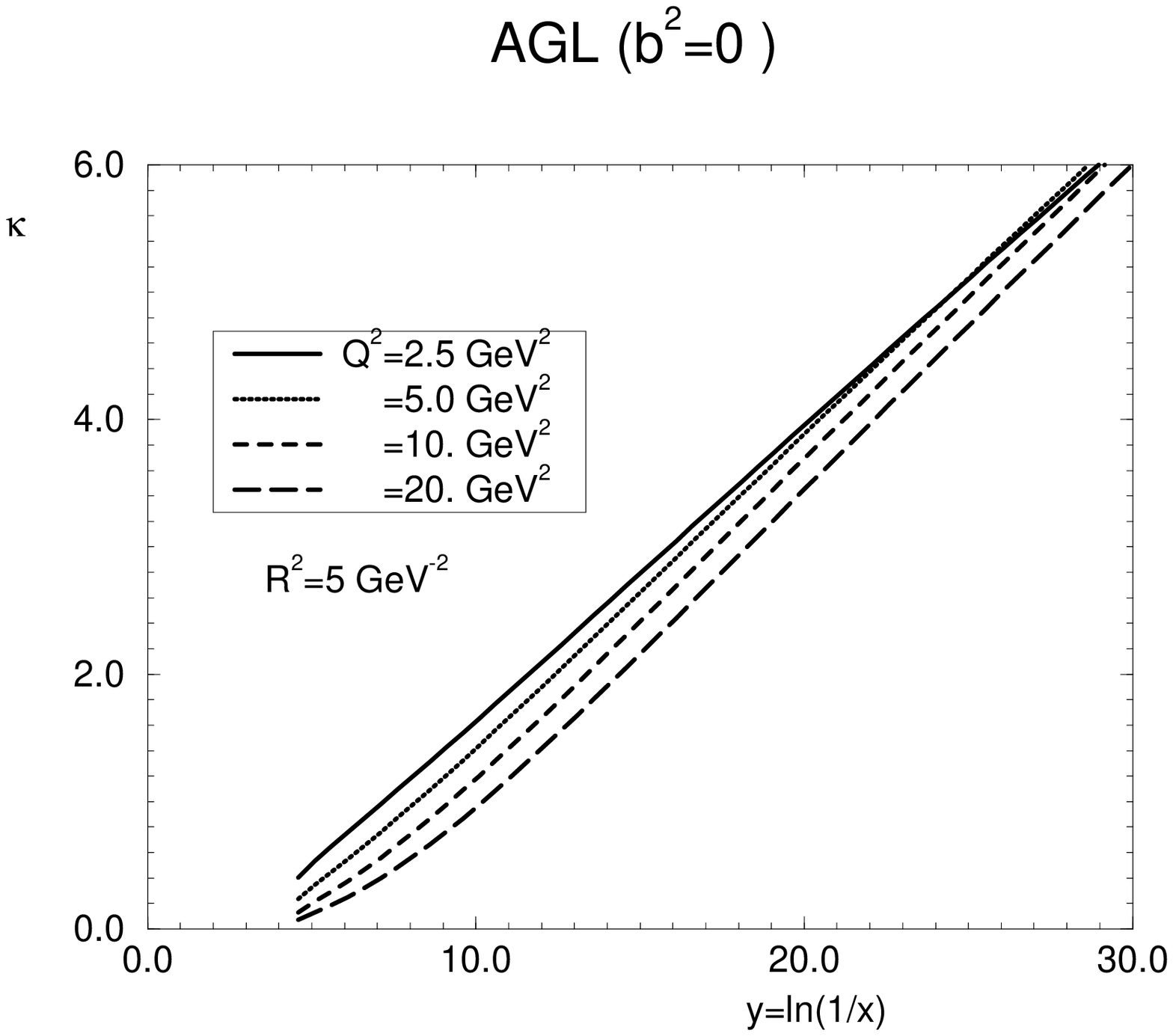,width=120mm}
\end{center}
\caption{\em The $\kappa$ values solutions of the  $b_t$-dependent 
nonlinear equation (AGL$_{b_t}$) calculated for $b_t^2 = 0$.}
\label{aglbt1}
\end{figure}

 To complete our discussion, we present in Fig.\ref{aglbt1} the solutions of
the   $b_t$-dependent 
nonlinear equation (AGL$_{b_t}$) calculated at $b_t^2 = 0$ for different values for
$Q^2$ and $R^2 = 5 \, GeV^{-2}$. One can see that the scale of the SC is
much bigger for $b_t$ =0 than for integrated over $b_t$  case. The
solutions approach the straight line
behaviour for $y$ close to its initial value. It occurs because the AGL$_{b_t}$
equation reaches its asymptotic behaviour for $\kappa (b_t)$ close to the
initial condition. Therefore, we can use the asymptotic solution as a good
approximation to the general solution of the equation at small values of
$b_t$.

\subsection{The $F_2$ structure function from the general approach.}

\begin{figure}
\begin{tabular}{c c}
\psfig{figure=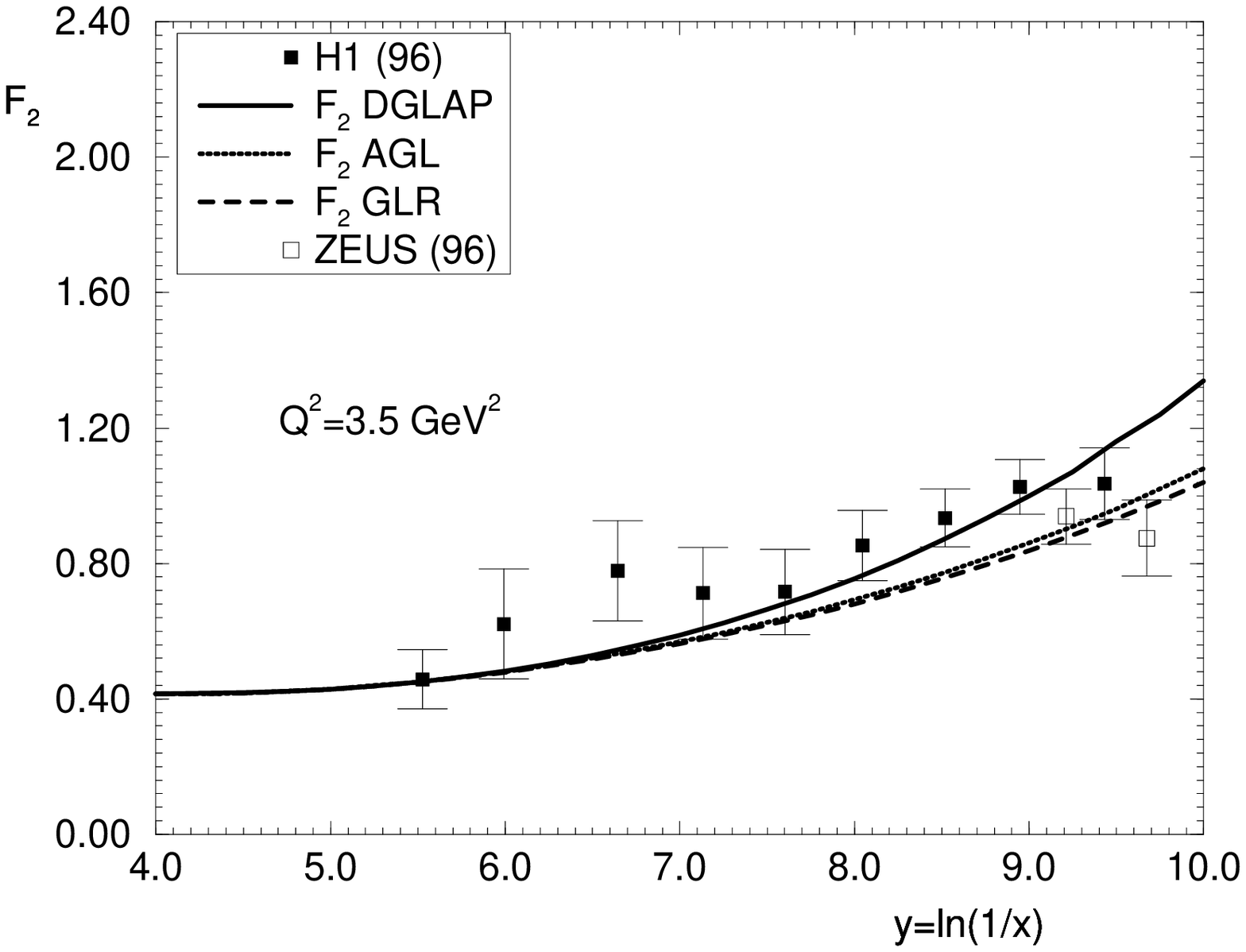,width=80mm}& \psfig{figure=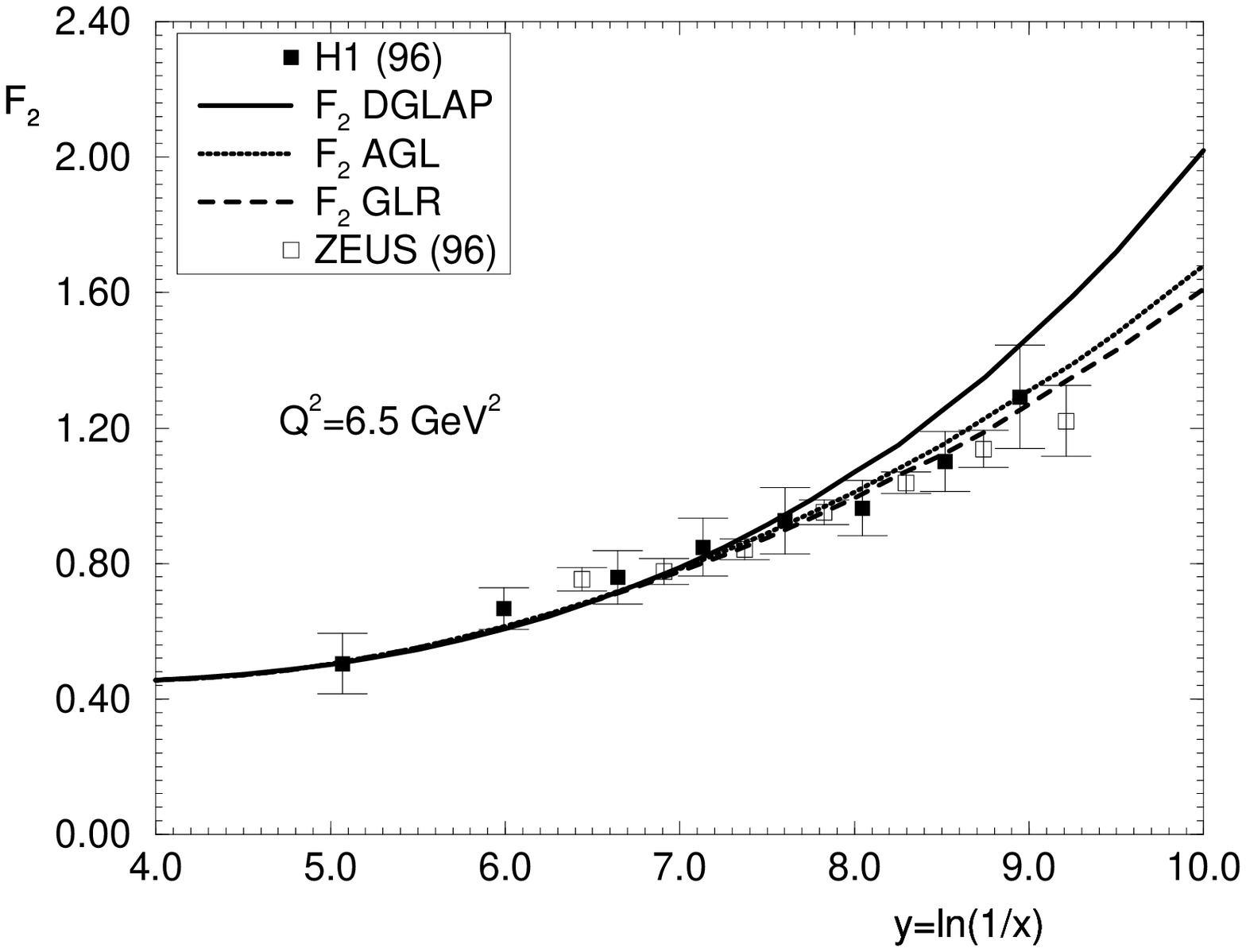,width=80mm}\\
\end{tabular}
\caption{\em $F_2$ calculated from different
gluon distributions using the expression (\protect\ref{f2g}) for the sea
component. }
\label{f2nuc14}
\end{figure}

In order to obtain the $F_2$ structure function from the general approach,
we calculate the valence and charm components of $F_2$ as described in
the section \ref{disf2}. We suppose that 
the sea component  is generated mainly from the scale violation mechanism. 
Thus, the sea component of $ F_2$ can be
 calculated  by the  simplified formula
\bea
F_2^{SEA}(x, Q^2) & = & F_2(x, Q^2_0 ) \, + \, \frac{1}{2 \pi} \sum^{N_f}_{1}\,\,Z^2_f
\,\int^{\ln Q^2}_{\ln Q^2_0} \as (Q'^2, N_f) d(\ln Q'^2) \nonumber \\
&  & \int^{1-x}_{0} \, dz \,\, [z^2 + (1-z)^2 ] \{ \frac{x}{1-z}
G (  \frac{x}{1-z}, Q'^2 ) \} \,\, ,
\label{f2g}
\eea
where $F_2(x, Q^2_0 )$ is the value of the sea component of $F_2$ at 
$Q^2=Q_0^2$. In this approach, we include the SC in the gluon
distribution in the  expression (\ref{f2g}). 
In Fig.\ref{f2nuc14} we show the results of $F_2$ with the sea component
calculated from different gluon distributions using expression 
(\ref{f2g}) for the sea component. The valence, charm, and the initial
value of the sea component (at $Q^2 \, = \, Q_0^2= 1 \, GeV^2$) are
calculated from the GRV distributions. We can see from the figure that
the SC predicted by the AGL and GLR equations are negligible for
$y \, < \, 7$. For $y \, > \, 7$, the SC start to become  important.
This result can be understood if we look at Fig.\ref{aglkap1}. The
$\kappa$ values given by the DGLAP, AGL and GLR equations are very close
for $Q^2 < 10 \, GeV^2 $ and $y\, < \, 7$, i.e.,  negligible SC are predicted
in this region. The results presented in Fig.\ref{f2nuc14} are consistent with that
presented in Fig.\ref{f2nuc9}.

%\begin{figure}
%\centerline{\psfig{figure=f2nuc.eps,width=100mm}}
%\caption{\it  The SC for $F_2$  calculated from
 %the Glauber formula (\protect\eq{f21}) (dotted line), and from
%the $F_2^{DLA}$ expression(\protect\eq{f2g}) using the gluon distribution
%from Galuber approach (dashed line). 
% We use  \protect$Q^2_0 \, = \, 0.4 \,\, GeV^2$ and $R^2= 5 \, GeV^{-2}$.}
%\label{f2nuc15}
%\end{figure}
%
%In Fig.\ref{f2nuc15} we compare the $F_2$ structure
%function calculated from the GRV distribution, given by \eq{f2grv},
%with SC for $F_2$. These SC are calculated in two different ways: 1)
%from the modified $F_2$ Glauber formula (\eq{f21}); and 2) from
%the modified gluon distribution \eq{FINANS} using expression (\ref{f2g}).
%As we can see, the DLA formula gives a consistent result for SC for $F_2$.
%

\section{Conclusions.}
In this paper we re-analyse the situation of the shadowing corrections
in QCD for proton deep inelastic structure functions. As have been
discussed in the introduction we view the physical interpretation of HERA
data as controversial. Here, we are going to summarize the result of
our analysis of the value of the SC, listing the questions that we asked
ourselves and the answers we got in the paper.

{\it 1.  Can we provide reliable estimates for the value of the SC in
QCD?}

Our answer is yes and, to answer this question, we reconsidered the
Glauber
- Mueller approach for the SC in DIS \cite{MU90} and suggested the
modified Mueller formula ( see \eq{FINANS}) to estimate the scale of the SC.
We consider these estimates as estimates from below for two
reasons: (i) in the eikonal (Glauber - Mueller ) approach only the fastest
partons interact with the target and, therefore, the real value of the SC 
should be bigger; (ii) we cannot trust the pQCD approach at large distances
which enter  into our formulae, in spite of the fact that the Mueller formula
gives the infrared stable answer for the deep inelastic structure
functions. We introduce a scale $r^2_0\,=\,\frac{1}{Q^2_0}$ and calculate
only the SC from the distances  $r_{\perp} \,\leq \,r_0$. We chose $Q_0 =
1 \,GeV$, relying on the fact, that all available parameterizations of the
deep inelastic structure functions describes the data down to photon
virtuality $Q^2 = Q^2_0 = 1 \,\, GeV^2$. We checked that the modified Mueller
formula gives the small corrections to $F_2 (x,Q^2)$ at large values of
$Q^2$ as well as to the momentum sum rules. We believe that we are able to
give a reliable estimate of the minimal value of the possible SC to the
parton densities.

{\it 2. What is the value of the SC?}

Using the modified Mueller formula we obtained the estimates for the value
of the SC given in Figs.(\ref{r1n}), (\ref{omn}) and (\ref{gmn}).
 These figures show 
that the SC are essential
in HERA kinematic region  to gluon  densities, especially
to $<\g >$ and $<\o>$. We calculated also the SC for $b_t$
distributions and showed that the SC at $b_t = 0$ turn out to be much
bigger than the  integrated over the $b_t$ parton densities.  One can see
from Figs.(\ref{rb0}) that $< \g(b_t=0)>$ is smaller 
that $\g = \frac{1}{2}$ for
 $Q^2$ smaller then $10 \, GeV^2$  and $<\o(b_t=0)>$ reaches 
the value of the intercept
of the soft Pomeron making possible the matching of soft and hard
processes. Therefore, our answer to the question formulated in the title
of this subsection is that the value of the SC is big enough to be seen 
in the gluon structure function at
 HERA kinematic region. However, the SC to $F_2(x,Q^2)$ turns out to be
rather
small. This is a reason why the SC has not been observed and the whole issue
 of the SC has been hidden in uncertainties of the value of the gluon
 structure function in current attempts to describe the HERA data
 in the framework of the DGLAP evolution.

We would like to point out that all numbers
were calculated using the GRV parameterization for the gluon structure
function and our reasoning for the choice of the GRV parameterization is
(i) it describes the experimental data  for $Q^2 \geq 1 GeV^2$; (ii) the
initial virtuality in the GRV parametrization is very low ( $Q^2_0 = 0.3
GeV^2$) and, therefore, the DLA of pQCD works better than in the other
parameterizations ( we recall that the Mueller formula for the SC was
proven in the DLA). However, we are planning to calculate the value of the
SC using other parameterization in future. It remains also for  further
publications the important question how the value of the SC depends on the
choice of the initial parton distribution at fixed value of $x$.

{\it 3. Where is the BFKL Pomeron?}

Our answer is that the BFKL Pomeron is hidden under the SC. We argue this
point in subsection 2.9. Here, we want to repeat that this statement
depends crucially on the choice of the gluon correlation radius $R^2$ and
we argued that HERA data gives $R^2 = 5 \,GeV^{-2}$. This result means
that the break of the operator product expansion appears rather as the
result of the SC than the BFKL dynamics ( see Ref.\cite{MU96} ).
We also argued that the BFKL Pomeron should not be seen in all other
processes even in the inclusive production of the jet in DIS which was
specially proposed to test the BFKL Pomeron\cite{HOTSPOT}, since for
all these processes $R^2$ is smaller than for the proton target.
We have to search for new processes to recover the BFKL dynamics.
 It is interesting to notice that the fact that the BFKL Pomeron is
hidden
under the SC is seen directly from Figs.\ref{fig.1} and \ref{Fig.3}
by comparing two curves:
$\kappa\, = \,1$ and $< \gamma > \,=\,\frac{1}{2}$. One can notice that
the first curve is to the right of the second one. It means that
decreasing the value of $Q^2$ at any fixed $x$ we meet first the curve $
\kappa\,=\,1$ where the SC become essential and only for smaller value of
$Q^2$  we approach the
curve $< \gamma >\,=\,\frac{1}{2}$ where the BFKL dynamics could enter 
in the
game.

{\it 4. How well works the Glauber - Mueller approach?}

We showed that the corrections to the Glauber - Mueller approach are big
and it can give only estimates from below for the value of the SC ( see 
section 3 ).

{\it 5. Can we develop a general approach in QCD to calculate the SC?}

Our answer is that we can. We suggest the new evolution equation
and discussed its theoretical accuracy. We argued that this equation solves
the problem of the SC summing up all diagrams of the order $( \as \ln ( 1/x)
\ln  (Q^2/Q^2_0))^n$ and $ (\kappa)^n$ considering $\as \ln(1/x) \ll 1$,
 $\as \ln(Q^2/Q^2_0) \ll 1 $, $\as \ll 1$ and $\kappa \ll 1$, where
 $\kappa = \frac{3 \pi \as}{2\,Q^2 R^2} x G(x,Q^2)$. This equation
allows
us to calculate the value of the SC in a wider kinematic region than the
GLR equation and study the parton density to the left of the critical
line of the GLR equation ( $<\g> = 1/2$). 

{\it 6. Do the parton densities reach saturation?}

Our answer is no. We solve the new evolution equation and found out that
the gluon density $xG(x,Q^2)$ tends to the asymptotic behaviour
( $x G(x,Q^2) \,\rightarrow xG^{asym}(x,Q^2) = 2 R^2 Q^2 \ln
(1/x)\ln\ln(1/x)$). We discuss the more detailed comparison of the
solution to the new evolution equation with the GLR one in subsection 4.3.

{\it 7. Can we match the deep inelastic cross section with the cross
section of the real photoproduction?}

We proved that the solution of the new evolution equation gives the mild
behaviour of the  parton densities  with energy (asymptotically,
they tend to $\ln (W)$ ). It gives us a hope to describe the matching
between soft and hard processes taken into account the SC using the new
evolution equation. We suppose to do this in our further publications.

{\it 8. What can be a practical use of our approach?}

We were impressed by the fact that  HERA data on $F_2(x,Q^2)$ can be
fitted by the simple formula \cite{BUCH}:
$$
F_2(x,Q^2)\,\,=\,\,m \log\frac{Q^2}{Q^2_0}\,\log\frac{x_0}{x}\,\,,
$$
where $m = 0.377$, $x_0 = 0.074$ and $Q^2_0 = 0.5 GeV^2$.
Such a parameterization leads to the gluon structure function \cite{BUCH}
$xG(x,Q^2)\,=\,3 \log(x_0/x)$, which energy ($x$) dependence is just the
same as for the solution to our evolution equation. Therefore, we can
interpret our results as the theoretical justification of the Buchmuller
- Haidt parameterization. Our physical idea sounds as follows:
everything has happened at $x$ of the order of $10^{-2}$, at such values
of $x$ the parton densities have reached their asymptotic values and 
we see only asymptotic behaviour of the parton densities at smaller values
of $x$. We suppose to work out this idea in the nearest future. 

{\it 9. Has everything been done?}

Of course, the answer is no. We have a lot of problems to solve.
First, as we have mentioned, we need to estimate the values of the SC
using other parameterizations of the parton densities but not
only the GRV one. We have to study more carefully the dependence of the
value of the SC on the initial parton distributions and to check how
stable are our results. Second, we have to find the complete 
solution for the running
$\as$. The experience with the GLR equation shows that the
running $\as$ leads to important properties of the SC  as the
appearance of the critical line and so on. Third, we need to take into
account that the partons in the parton cascade can interact not only with
the target but with the other partons (so called enhanced diagrams).
We can find out the region of the applicability of the new equation only
after solving this problem. We suppose to do this as the next step in our
approach.

We believe that our paper is only  the first step in reconsidering the
whole issue of the SC in DIS. Actually, we showed here, that the new
parameter that governs the size of the SC is $\kappa$ and the value of the
SC is not negligible in HERA kinematic region. This conclusion makes
rather suspicious
the current interpretation of HERA data on the basis of the DGLAP
evolution equations without the SC  and much work is
needed to clarify physics of DIS that is behind  HERA data.

While completing this paper, our attention was drawn  to three recent
papers \cite{INA} \cite{GLMSL} \cite{FSR} which deal with the Glauber -
Mueller approach to the SC for the deep inelastic structure function
$F_2(x,Q^2)$
for nuclei and for the hard diffractive production. The results obtained
in
these paper 
are close to ours.

{\bf Acknowledgements:} 
E. Levin wish to thank E. Gotsman and U. Maor for stimulating discussions
as well as all participants of the theory seminar in HEP department of the
Tel Aviv university for encouraging criticism and interest. 
 Work partially
 financed by  CNPq, CAPES and FINEP, Brazil.

\end{document}

%% file: defi.tex
 1
\def\as{\alpha_{\rm S}}

\def\citenum#1{{\def\@cite##1##2{##1}\cite{#1}}}
\def\citea#1{\@cite{#1}{}}
\def\as{\alpha_{\rm S}}

\def\eps{\epsilon}
\def\g{\gamma}

\def\o{\omega}
\def\O{\Omega}

\def\pa{\partial}

\def\s{\sigma}

\def\({\left(}
\def\){\right)}

\def\citenum#1{{\def\@cite##1##2{##1}\cite{#1}}}
\def\citea#1{\@cite{#1}{}}

\def\l1vt{\vec{l_{1\perp}}}

\def\rt{r_{\perp}}
\def\bt{b_{\perp}}
\def\rt2{$r^2_{\perp}$}
\def\bt2{$b^2_t$}

\def\jol1{$J_0(\,l_{1\perp}\,r_{\perp}\,)$}

\def\citea#1{\@cite{#1}{}}

\def\beq{\begin{equation}}
\def\eeq{\end{equation}}
\def\bea{\begin{eqnarray}}
\def\eea{\end{eqnarray}}

\def\eq#1{{Eq.~(\ref{#1})}}

\def\bbbz{{\mathchoice {\hbox{$\sf\textstyle Z\kern-0.4em Z$}}
{\hbox{$\sf\textstyle Z\kern-0.4em Z$}}
{\hbox{$\sf\scriptstyle Z\kern-0.3em Z$}}
{\hbox{$\sf\scriptscriptstyle Z\kern-0.2em Z$}}}}
\def\npb#1#2#3{    {\it Nucl. Phys. }{\bf B#1} (19#2) #3}
\def\plb#1#2#3{    {\it Phys. Lett. }{\bf B#1} (19#2) #3}
\def\prd#1#2#3{    {\it Phys. Rev. }{\bf D#1} (19#2) #3}

\def\zpc#1#2#3{    {\it Z. Phys. }{\bf C#1} (19#2) #3}

\def\sjnp#1#2#3{   {\it Sov. J. Nucl. Phys. }{\bf #1} (19#2) #3}

\def\l{\lambda}

%% file: ncln1.bbl
\begin{thebibliography}{99}

\bibitem{HERA} 
ZEUS collaboration, M. Derrick et.al.:{\it Zeit. Phys.} {\bf C65} (1995) 379;\\
H1 collaboration, T. Ahmed et.al: \npb{439}{95}{471}; \\
H1 collaboration, S. Aid et al.: \npb{470}{96}{3}

\bibitem{MRS}
A.D. Martin, R.G. Roberts and W.J. Stirling: \plb{306}{93}{145}.
\bibitem{CTEQ}
 CTEQ Collaboration, H.L.Lai et al.:\prd{51}{95}{4763}.

\bibitem{GRV}
M. Gluck, E. Reya and A. Vogt: \zpc{53}{92}{127}.
M. Gluck, E. Reya and A. Vogt: \zpc{67}{95}{433}.

\bibitem{GLR} L. V. Gribov,
  E. M. Levin and M. G. Ryskin: {\it Phys.Rep.} {\bf 100} (1983) 1.

\bibitem{FRST}  A.L. Ayala, M.B. Gay Ducati and E.M. Levin: \plb{388}{96}{188}

\bibitem{MUQI}
A.H. Mueller and J. Qiu: {\it Nucl. Phys.}  {\bf B268} (1986) 427.

%\bibitem{HERAPSI}
%H1 Collaboration, S.Aid et al.: DESY 96-037, March 1996.

\bibitem{BFKL}
 E.A. Kuraev,
  L.N. Lipatov and V.S. Fadin: {\it Sov. Phys. JETP} {\bf 45}
        (1977) 199 ;
\,\,Ya.Ya. Balitskii and L.V. Lipatov:{\it Sov. J. Nucl. Phys.} 
{\bf 28} (1978)
822;\,\,L.N. Lipatov: {\it Sov. Phys. JETP} {\bf 63} (1986) 904.


\bibitem{LR87}
E.M. Levin and M.G.Ryskin:\sjnp{45}{87}{150}.

\bibitem{MU90}
A.H. Mueller:\npb{335}{90}{115};

\bibitem{AGL}
A.L. Ayala, M.B. Gay Ducati and E.M. Levin: CBPF-FN-020/96, hep-ph 
9604383, April 1996; to appear in {\it Nucl. Phys. B}.

\bibitem{RY} 
M.G. Ryskin: \zpc{57}{93}{89}.

\bibitem{FIVE}

S. Brodsky et al: \prd{50}{94}{3134}.
\bibitem{DGLAP}
V.N. Gribov and L.N. Lipatov:{\it Sov. J. Nucl. Phys.} {\bf 15} (1972)
438; L.N. Lipatov: {\it Yad. Fiz.} {\bf 20} (1974) 181; G. Altarelli and
G. Parisi:{\it Nucl. Phys.} {\bf B126} (1977) 298; Yu.L. Dokshitser:{\it
Sov. Phys. JETP} {\bf 46} (1977) 641.


\bibitem{AGK}
V.A.Abramovski, V.N. Gribov and O.V. Kancheli: \sjnp{18}{73}{308}.

\bibitem{r20} 
M. Abramowitz and I.A. Stegun: {\it ``Handbook of Mathematical Functions"},
Dover Publication, INC, NY 1970.

\bibitem{VENEZ} G.Veneziano: \plb{52}{74}{220},\npb{74}{74}{365}; G.
Marchesini and G. Veneziano: \plb{56}{75}{271}; M. Ciafaloni, G.
Marchesini and G. Veneziano:\npb{98}{75}{472}.

 \bibitem{HT} J.Bartels:\zpc{60}{93}{471},\plb{298}{92}{589};
E.M. Levin, M.G. Ryskin and A.G. Shuvaev: \npb{387}{92}{204}.

\bibitem{LIP} L. N. Lipatov: Hep-ph/9610276, Octorber, 1996.

\bibitem{JAR}T. Jaroszewicz:\plb{116}{82}{291}.

\bibitem{MW}G. Marchesini and B. Webber: \npb{349}{91}{617};
E.M. Levin, G. Marchesini, M.G. Ryskin and B. Webber: \npb{357}{91}{167}.

\bibitem{MU96} A.H. Mueller: \plb{396}{97}{251}.

\bibitem{LEREN} E.M. Levin: \npb{453}{95}{303}.

\bibitem{HOTSPOT} A.H. Mueller: {\it Nucl. Phys. B ( Proc. Suppl.)}{\bf
18C}(1991) 125; J. Bartels, A. DeRoeck and M. Loewe: \zpc{54}{92}{635};
J. Kwiecinski, A.D. Martin, P.J. Sutton:
\plb{287}{92}{254},\prd{46}{92}{921};W-K. Tang:\plb{278}{91}{363};J.
Bartels:{\it J. Phys.}{\bf G19}(1993)1611; E. Laenen and E. Levin:{\it J.
Phys.}{\bf G19}(1993)1582.

\bibitem{LALE}
E. Laenen and E. Levin: {\it Ann. Rev. Nucl. Part.} {\bf 44} (1994) 199.

\bibitem{EKL} 
R.K. Ellis, Z. Kunst and E. M. Levin:\npb{420}{94}{517}.

\bibitem{BARY} 
J. Bartels and M.G. Ryskin: DESY - 96 - 238, hep-ph 9612226. 

\bibitem{SOFTPOMERON} A. Donnachie and P.V. Landshoff: \plb{185}{87}{403}, 
\npb{311}{89}{509};
 E. Gotsman, E. Levin and U. Maor:
\plb{353}{95}{526}.

\bibitem{GRV95}
M. Gluck, E. Reya and A. Vogt: \zpc{67}{95}{433}.

\bibitem{MU94} 
A.H. Mueller: \npb{415}{94}{373}.

\bibitem{Collins90} 
J.C. Collins and J. Kwiecinski: \npb{335}{90}{89}.

\bibitem{Bartels91} 
J. Bartels, J. Blumlein and G. Shuler:\zpc{50}{91}{91}.

\bibitem{BALE}
J. Bartels. and E. Levin: \npb{387}{92}{617}.


\bibitem{BUCH} W. Buchm\"uller and D. Haidt: Hep-ph/9605428, May, 1996.

\bibitem{INA}
Zheng Huang, Hung Jung Lu and Ina Sarcevic: AZPH - TH - 97 - 07, hep - ph
/9705250.

\bibitem{GLMSL}
E.Gotsman,E. Levin and U. Maor: TAUP 2405/97, hep - ph/9701226.
 
\bibitem{FSR}
L. Frankfurt, W. Koepf and M. Strikman: OSU - 97 - 0201, hep - ph
/9702216.

\end{thebibliography}
